\documentclass[11pt,a4paper]{article}
 \pdfoutput=1
\usepackage{jcappub}
\usepackage{amsmath}
\usepackage{amsfonts,color}
\usepackage{amssymb,float}
\usepackage{mathtools}
\usepackage[utf8]{inputenc}
\usepackage{url}
\usepackage{graphicx}
\usepackage{refstyle}
\usepackage{wasysym}
\usepackage{tabularx}
\usepackage{accents}
\usepackage{graphicx}
\usepackage{color}
\usepackage[dvipsnames]{xcolor}
\usepackage{hyperref}

\hypersetup{
    colorlinks=true,
    linkcolor=blue,
    filecolor=magenta,      
    citecolor=red
}

\usepackage{subfigure} 
\usepackage{hyperref} 

\makeatletter
\long\def\dddddot#1{%
  {\mathop {#1}\limits ^{\vbox to-1.4\ex@ {\kern -\tw@ \ex@ \hbox {\normalfont .....}\vss }}}%
}
\long\def\multidots#1#2{%
  \count@=0
  {{\mathop {#2}\limits ^{\vbox to-1.4\ex@ {\kern -\tw@ \ex@ \hbox {\normalfont %
  \loop%
  \ifnum#1>\count@%
  .%
  \advance\count@ by1%
  \repeat%
  }\vss }}}}%
}
\makeatother

\title{\boldmath A data-driven Reconstruction of Horndeski gravity via the Gaussian processes}

\author[a]{Reginald Christian Bernardo,}
\author[b,c]{Jackson Levi Said}

\affiliation[a]{National Institute of Physics, University of the Philippines Diliman, Quezon City 1101, Philippines}
\affiliation[b]{Institute of Space Sciences and Astronomy, University of Malta, Malta, MSD 2080}
\affiliation[c]{Department of Physics, University of Malta, Malta, MSD 2080}

\emailAdd{rbernardo@nip.upd.edu.ph}
\emailAdd{jackson.said@um.edu.mt}

\abstract{
\textbf{We reconstruct the Hubble function from cosmic chronometers, supernovae, and baryon acoustic oscillations compiled data sets via the Gaussian process (GP) method and use it to draw out Horndeski theories that are fully anchored on expansion history data. In particular, we consider three well-established formalisms of Horndeski gravity which single out a potential through the expansion data, namely: quintessence potential, designer Horndeski, and tailoring Horndeski. We discuss each method in detail and complement it with the GP reconstructed Hubble function to obtain predictive constraints on the potentials and the dark energy equation of state.}
}

\begin{document}

\maketitle
\flushbottom

\section{Introduction}
\label{sec:intro}

Flat $\Lambda$CDM has dominated as the cosmological concordance model since the discovery of the accelerating expansion of the Universe \cite{Riess:1998cb,Perlmutter:1998np} some decades ago. It mimics numerous observations in a wide variety of cosmological and astrophysical settings \cite{dodelson2003modern,Clifton:2011jh}. It does this despite the necessity of large portions of the model requiring particle physics beyond the standard model as well as future solutions to foundational problems such as fine-tuning issues, the horizon and coincidence problems, and the cosmological constant problem \cite{RevModPhys.61.1,Bull:2015stt}. In this background, there have been several interesting proposals to replace $\Lambda$CDM ranging from dynamical dark energy \cite{Sahni:1999gb, Sahni:2006pa, Copeland:2006wr}, extended gravity \cite{Capozziello:2011et}, to beyond general relativity (GR) \cite{Clifton:2011jh}, and others.

In recent years, observations of some cosmological parameters appear to feature growing discrepancies with early Universe predictions, based on vanilla $\Lambda$CDM, and late time cosmology-independent measurements, and is fast becoming a central issue in $\Lambda$CDM cosmology \cite{DiValentino:2020zio,DiValentino:2020vvd,Aghanim:2018eyx,Bernal:2016gxb}. The value of the Hubble parameter at current times $H_0$ has encapsulated one part of this cosmological tension with late time measurements taken from Cepheid calibrated type Ia supernovae events \cite{Riess:2019cxk} and strong lensing by distant quasars \cite{Wong:2019kwg} resulting in very high values of $H_0$, while early Universe predictions based on a $\Lambda$ cosmology produce a much lower value \cite{Aghanim:2018eyx,Ade:2015xua}. Some other measurements point to an even larger $H_0$ tension \cite{Riess:2020sih,Pesce:2020xfe,deJaeger:2020zpb}.

Further analysis of potential solutions within $\Lambda$CDM such as nonflat cosmologies \cite{DiValentino:2019qzk, Handley:2019tkm}, or more exotic contributions from particle physics beyond the standard model, may be done. However, the issue may ultimately require a reexamination of the gravitational contribution to $\Lambda$CDM, together with the underpinnings of general relativity (GR) \cite{Bull:2015stt,Copeland:2006wr}. The plethora of modifications to GR form a large landscape of theories on which to build cosmological model. On the other hand, many of these theories have been shown to be dynamically equivalent to a second order gravitational theory, in terms of metric tensor derivatives, provided a scalar field degree of freedom is allowed. In this context, by exploring Horndeski gravity \cite{Horndeski:1974wa}, which is the most general second-order theory of gravity that contains only one scalar field, we can investigate a large region on cosmology beyond $\Lambda$CDM.

Horndeski gravity encompasses many of the popular formulations of gravity that have been studied. For instance, despite being organically fourth order $f(R)$ gravity \cite{Sotiriou:2008rp, DeFelice:2010aj, Nojiri:2006gh, Hu:2007nk, Appleby:2007vb, Starobinsky:2007hu, Appleby:2009uf} can be transformed into a Horndeski class of models through an appropriate mapping \cite{Sotiriou:2008ve}. However, the recent measurement of the propagation speed of gravitational waves has severely constrained many of the promising branches of the Horndeski landscape \cite{TheLIGOScientific:2017qsa,Ezquiaga:2018btd}. Some of these theories include quartic and quintic Galileon models \cite{Nicolis:2008in,Deffayet:2009wt}, de-Sitter Horndeski \cite{Martin-Moruno:2015bda}, the Fab Four \cite{Charmousis:2011bf}, and purely kinetic coupled models~\cite{Gubitosi:2011sg}, among others. While many Horndeski gravity models continue to be viable, an interesting new avenue of Horndeski gravity has started to emerge in which the curvature associated with the mediation of gravity is replaced with teleparallel torsion \cite{Krssak:2018ywd}. This organically lower order form of gravity has been shown to produce a much larger space of Horndeski theories \cite{Bahamonde:2019shr,Bahamonde:2020cfv} which can be shown to revive some models within Horndeski gravity \cite{Bahamonde:2019ipm}.

In this work, we reconsider the range of viable Horndeski gravity models using Gaussian process regression (GP) which is a method of reducing noise as well as simulating new intermediary points in a data set. GP functions as a non-parametric reconstruction technique by employing a covariance function whose hyperparameters are fixed in a Bayesian approach so that the Kernel approximates the data set and can produce a smoothed continuous data set with respective uncertainties. Thus the more points in a data set, the more well constrained the hyperparameters will be. The only drawback of this approach is if clustering occurs in a data set then the hyperparameters may be optimized for only part of the data set which may translate into other regions being poorly reconstructed. GP has been widely used in cosmology ranging from reconstructing the value of $H_0$ as in Refs. \cite{Seikel2012, Shafieloo:2012ht, Seikel:2013fda, Yennapureddy:2017vvb, Gomez-Valent:2018hwc, Li:2019nux, Liao:2019qoc, Keeley:2020aym, Renzi:2020fnx, Colgain:2021ngq} where the contentious value of $H_0$ has been approximated using various compiled sources of expansion data, to the value of $f\sigma_8$ at current times as in Ref. \cite{Benisty:2020kdt}, and gravitational wave analysis \cite{Belgacem:2019zzu,Moore:2015sza,Canas-Herrera:2021qxs}.

More recently, GP has been applied to the inverse problem in extended models of gravity in which arbitrary classes of gravity are constrained through the prism of compiled data sets without assuming particular models \textit{a priori} which is a core problem in modified gravity. By allowing the gravitational sector Lagrangian to remain largely unprescribed, various studies have produced viable ranges that such models would need to satisfy. Consider Refs.~\cite{Briffa:2020qli,Cai:2019bdh,Ren:2021tfi} where background expansion data was used to produce restrictions in $f(T)$ gravity for medium redshift ranges, while in Ref.~\cite{LeviSaid:2021yat} the same goal was achieved but with growth data. This approach has also been used in the context of interacting models between dark energy and dark matter as in Ref.~\cite{Cai:2015zoa}. Another approach to reducing the space of large classes of gravitational theories is Ref.~\cite{Reyes:2021owe} in which viable paths to a Horndeski model are explored in terms of their predictions on cosmological parameters. The work is interesting because it delves into the classes of models that continue to be viable in the background of the speed of GW constraint.

In the present work, we trace down viable paths of Horndeski gravity in the context of the range of priors on $H_0$ which forms the base of the so-called Hubble tension. In Sec.~\ref{sec:horndeski_theory} we briefly review Horndeski gravity and the classes of models we will study later on. Sec.~\ref{sec:gaussian_process} then expands on some GP background together with an explanation of how various data sets were used with the GP approach to produce the Hubble diagram with the various prior choices. The core work on Horndeski gravity is contained in Sec.~\ref{sec:painting_horndeski} where each of the family of viable Horndeski classes is investigated through the GP reconstruction method, and where we also produce the equation of state for the effective dark energy component of the theory. Throughout the work, we assume a $(-1,+1,+1,+1)$ metric signature and use geometric units where $c = M_{\text{Pl}}^2 = 1$ where $M_{\text{Pl}}^2 = 1/\left(8 \pi G\right)$, unless otherwise stated.

\section{Horndeski theory}
\label{sec:horndeski_theory}

Horndeski gravity is one of the most prominent modifications of $\Lambda$CDM cosmology beyond GR since it is synonymous with many of the extensions or modifications of GR, at least in terms of it dynamical equations. Thus, Horndeski gravity brings together a wide swath of theories of gravity and makes their study much more accessible for general analysis. The base of theory emerges from the Lovelock theorem \cite{Lovelock:1971yv} in which the Einstein-Hilbert action is found to be a unique theory that produces second-order field equations. However, on adding a single scalar field, Horndeski gravity produces a much richer plethora of models that spans the entire range of Lagrangians that produce second-order equations of motion while only containing a single scalar field. It is worthwhile to mention that Horndeski gravity is constructed in a curvature-based context (through the Levi-Civita connection), and that other teleparallel proposals have been made in recent years \cite{Bahamonde:2019ipm,Bahamonde:2019shr,Bahamonde:2020cfv}.

Besides avoiding Ostrogradsky instability problems \cite{Kobayashi:2019hrl}, Horndeski gravity also only contains one extra (scalar) degree of freedom which is propagating \cite{Hou:2017bqj}, which may appear as a massive or massless gravitational wave. Horndeski gravity features a number of equivalent representations, which in our case we present through the action \cite{Kobayashi:2019hrl}
\begin{equation}
	\mathcal{S}_{\rm H} = \int {\rm d}^4 x \left(\mathcal{L}_2 + \mathcal{L}_3 + \mathcal{L}_4 + \mathcal{L}_2\right) + \mathcal{S}_{\rm mat}(\psi,g_{\mu\nu})\,,
\end{equation}
where $\psi$ represents the matter fields in the matter action $\mathcal{S}_{\rm mat}$ and $g_{\mu\nu}$ is the metric tensor, and
\begin{align}
	\mathcal{L}_2 &= G_2\left(\phi,\,X\right)\,,\\
	\mathcal{L}_3 &= -G_3\left(\phi,\,X\right)\Box \phi\,,\\
	\mathcal{L}_4 &= G_4\left(\phi,\,X\right) R + G_{4,X}\left[\left(\Box\phi\right)^2 - \left(\nabla_{\mu} \nabla_{\nu} \phi\right)\left(\nabla^{\mu} \nabla^{\nu} \phi\right)\right]\,,\\
	\mathcal{L}_5 &= G_5\left(\phi,\,X\right)G_{\mu\nu}\nabla^{\mu} \nabla^{\nu}\phi\nonumber\\
	&- \frac{1}{6}G_{5,X}\left[\left(\Box\phi\right)^3 - 3\left(\Box \phi\right)\left(\nabla_{\mu}\nabla_{\nu}\phi\right)\left(\nabla^{\mu}\nabla^{\nu}\phi\right) + 2\left(\nabla^{\mu}\nabla_{\alpha}\phi\right) \left(\nabla^{\alpha}\nabla_{\beta}\phi\right) \left(\nabla^{\beta}\nabla_{\mu}\phi\right) \right]\,,
\end{align}
in which $X = - \frac{1}{2}\nabla_{\sigma}\phi \nabla^{\sigma}\phi$ is the kinetic term associated with the scalar field, $\Box\phi = \nabla_{\mu}\nabla^{\mu}\phi$ is the d'Alembertian operator, $G_{\mu\nu}$ is the Einstein tensor, and $G_i\left(\phi,\,X\right)$ are arbitrary functions of the scalar field and its kinetic term.

Of particular importance is the impact of Lovelock's theorem on this formulation of Horndeski gravity. Notice that only the Ricci scalar appears as a purely gravitational scalar while the Einstein tensor appears as a coupling with the scalar field derivative terms. Thus the Lovelock theorem renders a finite Lagrangian for Horndeski theory. Another interesting feature of Horndeski gravity is its subclasses such as Brans-Dicke theory \cite{Brans:1961sx} which occurs for $G_3 = 0 = G_5$, $G_2 = 2\omega X/\phi$ and $G_4 = \phi$, $f(R)$ gravity \cite{Sotiriou:2008rp} when $G_3 = 0 = G_5$, $G_2 = f(\phi) - \phi f'(\phi)$ and $G_4 = f'(\phi)$, and GR for the choice where $G_2 = 0 = G_3 = G_5$ and $G_4 = 1/2$.

The recent binary neutron star merger which was recorded in the multimessenger events in which GW170817 \cite{TheLIGOScientific:2017qsa} together with an electromagnetic counterpart GRB170817A \cite{Goldstein:2017mmi} were used to severely constrain the speed of GW propagation to within one part in $10^{15}$. The result of this is a Horndeski theory that is drastically reduced in potential models \cite{Ezquiaga:2017ekz}. This is described by a smaller class of viable actions
\begin{equation} \label{eq:action}
    \mathcal{S}_{\rm H - c} = \int d^4 x \sqrt{-g} \left( f \left( \phi \right) R + K \left( \phi, X \right) - G \left( \phi, X \right) \Box \phi  \right) + \mathcal{S}_{\rm mat}(\psi,g_{\mu\nu}) 
\end{equation}
where we refer to $f$, $K$, and $G$ as the conformal, $k$-essence, and braiding potentials, respectively. This theory continues to encompasses a large swath of alternative gravity theories that have been widely studied, including $f(R)$ gravity \cite{Sotiriou:2008rp,DeFelice:2010aj,Nojiri:2006gh}, generalized Brans-Dicke theories \cite{Brans:1961sx}, covariant Galileons \cite{Nicolis:2008in,Deffayet:2009wt}, quintessence \cite{Tsamis:1997rk}, and kinetic gravity braiding \cite{Deffayet:2010qz, Kobayashi:2011nu}, among others \cite{Kase:2018aps}.

In this work, we restrict our attention to particular subclasses of the viable Horndeski gravity models in Eq.~(\ref{eq:action}) where we single out particular classes formed by subclasses of the $\left( f, K, G \right)$ functions. In particular, we focus on the following subclasses:
\begin{enumerate}
\item Quintessence \cite{Tsamis:1997rk}:
\begin{equation}
f \left( \phi \right) = 1/2
\end{equation}
\begin{equation}
K \left( \phi, X \right) = X - V\left(\phi\right)
\end{equation}
\begin{equation}
G \left( \phi, X \right) = C 
\end{equation}
where $C$ is a constant. We refer to $V$ as the quintessence potential.
\item Designer Horndeski \cite{Arjona:2019rfn}:
\begin{equation}
f \left( \phi \right) = 1/2
\end{equation}
\begin{equation}
K \left( \phi, X \right) = K \left( X \right)
\end{equation}
\begin{equation}
G \left( \phi, X \right) = G \left( X \right) .
\end{equation}
\item Tailoring Horndeski \cite{Bernardo:2019vln}:
\begin{equation}
f \left( \phi \right) = 1/2
\end{equation}
\begin{equation}
K \left( \phi, X \right) = X - 2 \Lambda
\end{equation}
\begin{equation}
G \left( \phi, X \right) = G \left( X \right)
\end{equation}
where $\Lambda$ is a constant.
\end{enumerate}
In addition to the gravitational action (\ref{eq:action}), we consider a perfect fluid of energy density $\rho$ and pressure $P$ and assume spatial flatness for a homogeneous and isotropic cosmology. This produces Friedmann equations for each subclass which are presented in Sec.~\ref{sec:painting_horndeski}, where they are inverted so that the ensuing data sets can be used to constrain the values of the Lagrangian contributions.

\section{Determining the Hubble diagram using Gaussian process regression}
\label{sec:gaussian_process}

We devote this section to presenting a brief introduction to GP in Sec.~\ref{subsec:gp_recap} which is then applied directly to expansion data to obtain the Hubble diagram in Sec.~\ref{subsec:gp_application}. In this section, we discuss the assumptions and dependencies that this approach is built on and thus the context in which to interpret the results that follow.

\subsection{Gaussian process: A Brief Review}
\label{subsec:gp_recap}

The GP regression is a powerful tool that merges the idea of a kernel and Bayesian analysis to make meaningful predictions of a function $H$ from a given data set $\{ \left( z, H(z) \right) \}$ \cite{10.5555/971143, 10.5555/1162254}. Most notably, it is a non-parametric way of learning a function and so is a refreshing change of view in making cosmological predictions usually based on arbitrary parametrizations and Markov chain Monte Carlo (MCMC) methods \cite{Seikel2012, Shafieloo:2012ht, Seikel:2013fda}. In light of the existing tensions between early, i.e., during last scattering, and local cosmological observations, GP has also naturally emerged as a go-to approach in cosmology and is increasingly becoming a popular tool to make cosmological predictions without assuming a particular model of cosmology \cite{Shafieloo:2012ht, Colgain:2021ngq,Yennapureddy:2017vvb,Seikel2012,Seikel:2013fda, Benisty:2020kdt, Belgacem:2019zzu,Moore:2015sza,Canas-Herrera:2021qxs, Briffa:2020qli,Cai:2019bdh, LeviSaid:2021yat, Cai:2015zoa, Reyes:2021owe, Wang:2017jdm, Gomez-Valent:2018hwc, Zhang:2018gjb, Mukherjee:2020vkx, Aljaf:2020eqh, Li:2019nux, Liao:2019qoc, Busti:2014aoa, Cai:2015pia, Renzi:2020fnx}.

Consider an observation $\{ \left( z, H(z) \right) \}$ with uncertainties captured by the covariance matrix $C$. In order to reconstruct the function $H(z^*)$ at the coordinates $z^*$ via the GP, we first assume a kernel $K\left( z^* , \tilde{z}^* \right)$ that relates the function values at different coordinates $z^*$ and $\tilde{z}^*$ in the data set of observations. In terms of this kernel, the mean and the covariance of the GP reconstruction of the $n$th derivative of $H(z)$ at $z^*$ are given by
\begin{equation}
    \langle H^{* (n)} \rangle = K^{(n, 0)} \left( z^*, Z \right) \left[ K\left( Z, Z \right) + C \right]^{-1} H \left( Z \right)\,,
\end{equation}
and
\begin{equation}
    \text{cov} \left( H^{* (n)} \right) = K^{(n, n)} \left( z^*, z^* \right) - K^{(n, 0)} \left( z^*, Z \right) \left[ K\left(Z, Z\right) + C \right]^{-1} K^{(0, n)} \left(Z, z^*\right)\,,
\end{equation}
respectively, where $Z$ stands for the union of the redshifts of the measurements and $y^{(n, m)}$ refers to the $n$th derivative of a function $y$ with respect to its first argument and the $m$th derivative with respect to the second argument. This kernel depends on hyperparameters that will be trained using a Bayesian implementation in order to fit the observations. However, in contrast with parametric approaches, the hyperparameters do not specify the shape of the function, but rather, they describe its characteristic scales in the $z$ and $H$ directions, typically, as a length scale $l$ and an amplitude $A$. In particular, we consider the simplest, infinitely-differentiable, and arguably the most widely-used kernel in machine learning, known as the \textit{radial basis function} (RBF), also often referred to as the squared exponential kernel,
\begin{equation}\label{eq:kernel}
    K\left(z, \tilde{z}\right) = A \exp\left( - \dfrac{\left(z - \tilde{z}\right)^2}{ 2 l^2 } \right)\,.
\end{equation}
The hyperparameters $\left( l, A \right)$ are then selected by optimizing, or more consistently, marginalizing over the marginal likelihood $\mathcal{L} = p \left( H | Z, l, A \right)$. However, marginalization over the hyperparameters usually demands more computational power and can be impractically time consuming. Fortunately, in cosmology, and other applications, where the $\mathcal{L}$s are at least approximately Gaussian, optimization usually turns out to be a good approximation. In what follows, we will optimize over the hyperparameters to reconstruct the local Hubble function.

Two remarks are in order. First, the GP mean function is kept to zero in our reconstructions. This seems reasonable as we intend to obtain function values in regions encompassing the domain of the data points. Most importantly, this is in line with a data-driven, model-independent approach which would otherwise be compromised with, for example, a $\Lambda$CDM or another mean function. Second, it is often the case in GP applications that the reconstruction is tested for various kernels which lead to redundant results. For this reason, we pursue this work with only the most natural GP kernel (\ref{eq:kernel}), and instead investigate in detail the consequences of the choice of the kernel in a different paper \cite{Bernardo:2021mfs} where we show that the reconstructions of the Hubble function per kernel can be at most in mild statistical tension with each other.

Our implementation is based on the public codes \textit{GaPP} \cite{Seikel2012} and \textit{scikit} \cite{scikit-learn} for the GP and \textit{cobaya} \cite{2020arXiv200505290T} and \textit{getdist} \cite{Lewis:2019xzd} for calibration of the absolute magnitude $M$ via MCMC to be described shortly. The codes used in this paper also used $numpy$ \cite{2020NumPy-Array}, $scipy$ \cite{2020SciPy-NMeth}, $seaborn$ \cite{Waskom2021}, and $matplotlib$ \cite{4160265}. A friendly implementation of the computations in this work is communicated as a two-part \textit{jupyter} notebook \cite{jupyter} which can be freely downloaded \cite{reggie_bernardo_4810864}.

\subsection{Application to the Hubble expansion}
\label{subsec:gp_application}

We now apply the GP approach using the RBF kernel in Eq.~(\ref{eq:kernel}) to a number of combined Hubble data sources, which we use to reconstruct the $H(z)$ data. This is done using three core sources of $H(z)$ data, namely cosmic chronometers (CC), supernovae of Type Ia (SNe) and baryonic acoustic oscillation (BAO). In addition, we consider three priors on $H(z)$ that have been reported in the literature, namely the Riess prior which is $H_0^{\rm R19} = 74.03 \pm 1.42 \,{\rm km\, s}^{-1} {\rm Mpc}^{-1}$ \cite{Riess:2019cxk}, the Carnegie-Chicago Hubble prior $H_0^{\rm TRGB} = 69.8 \pm 1.9 \,{\rm km\, s}^{-1} {\rm Mpc}^{-1}$ \cite{Freedman:2019jwv}, and the latest value from the Planck collaboration (P18) $H_0^{\rm P18} =67.4 \pm 0.5 \,{\rm km\, s}^{-1} {\rm Mpc}^{-1}$ ($\Omega_{m0} = 0.3153 \pm 0.0073$ and $\Omega_\Lambda = 0.6847 \pm 0.073$) \cite{Aghanim:2018eyx}. These $H_0$ priors clearly represent the current Hubble tension and are considered in this analysis to shed more light to this intriguing puzzle.

Adding some background on these priors, the $H_0^{\rm P18}$ estimates is the result of using (flat) $\Lambda$CDM as a fiducial model and cosmic microwave background (CMB) data from the Planck mission to predict the late time value of $H_0$ \cite{Aghanim:2018eyx}. This contrasts with the cosmology independent estimates that come from local observations, where $H_0^{\rm R19}$ is the highest of these values and comes from long period observations of Cepheids in the Large Magellanic Cloud using the Hubble Space Telescope which significantly reduces the uncertainty in the measurement of $H_0$. We also consider measurements of the tip of the red giant branch (TRGB) where this turning point is used as a standard candle to predict values of $H_0^{\rm TRGB}$. While other prior values exist in the literature, these adequately represent the tension in the value of the Hubble constant.

On the data sets themselves, the CC data comprise the majority of the combined data set points with 29 (our of 40) points within the $z \lesssim 2$ range, and do not rely on any cosmological models \cite{Moresco:2016mzx, Moresco:2015cya, 2014RAA....14.1221Z, 2010JCAP...02..008S, 2012JCAP...08..006M}. For the SNe data set, we employ both the full Pantheon data set \cite{Scolnic:2017caz} and the compressed Pantheon compilation together with the CANDELS and CLASH Multi-cycle Treasury (MCT) data \cite{Riess:2017lxs}. This is done using the corresponding covariance matrix, and where only five of the six points in Ref.~\cite{Riess:2017lxs} are used since the sixth point is non-Gaussian. In particular, we incorporate the compressed SNe data in the analysis by promoting the $E(z)$ data points to $H(z) = H_0 E(z)$ using the $H_0$ priors and then feeding resulting $H(z)$ and the corresponding covariance matrix into the GP regression.

CC data depends on a differential ages technique between galaxies while the SNe data is compiled using Cepheid calibrated distance measurements for SNe event. Finally, we consider the BAO data from BOSS and eBOSS in Refs.~\cite{Alam:2016hwk, Bautista:2020ahg, Gil-Marin:2020bct, Tamone:2020qrl, deMattia:2020fkb, Neveux:2020voa, Hou:2020rse, Agathe:2019vsu, Blomqvist:2019rah}. While BAO data is not entirely independent from $\Lambda$CDM (due to the assumption of a fiducial radius of the sound horizon $r_s=147.78\,\mathrm{Mpc}$), they add value to the analysis as being data that emanates from the growth of large scale structure which has a strong bearing on the value of $H_0$. In particular, these measurements are made in terms of the comoving distance $d_M(z)/r_s$ and $d_H(z)/r_s$, where 
\begin{equation}
    d_M(z) = (1 + z) d_A(z)\,,
\end{equation}
and
\begin{equation}
    d_H(z) = c/H(z)\,,
\end{equation}
in which the Hubble expansion can be determined. Instead of relying on the sound horizon radius $r_s$, which consequently depends on a cosmological model which in this case is $\Lambda$CDM, we calculate the ratio $d_M(z)/d_H(z)$ from the BAO data set, cancelling out the dependence on $r_s$, and compliment this with a GP reconstructed angular-diameter distance $d_A(z)$ from the compressed Pantheon samples. This step assumes spatial-isotropy and the distance-duality relation, $d_A(z) = d_L(z)/(1 + z)^2$, where $d_A(z)$ and $d_L(z)$ are the angular-diameter distance and luminosity distance, respectively, and also relies on a calibrated SNe absolute magnitude. Arguably, both spatial-isotropy and the distance-duality relation can be considered as cosmological assumptions, but neither requires any hard-parametrization like $\Lambda$CDM and, most importantly, both assumptions can be supported just as well with any metric theories of gravity.

\begin{figure}[h!]
\center
\includegraphics[width = 0.5 \textwidth]{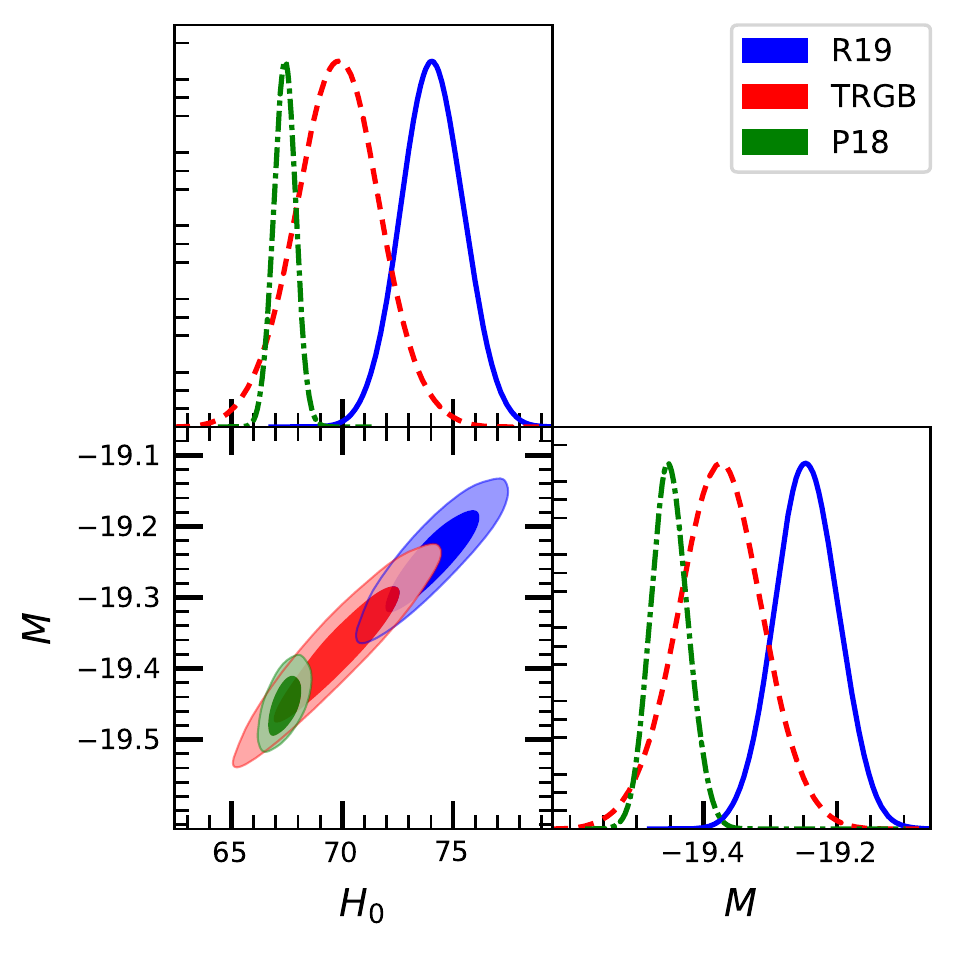}
\caption{The posteriors of the calibrated SNe absolute magnitude $M$ for each $H_0$ (in units of $\,{\rm km\, s}^{-1} {\rm Mpc}^{-1}$) prior. The mean and standard deviations are $M = -19.25 \pm 0.05, -19.38 \pm 0.06$, and $-19.45 \pm 0.03$ for the R19, TRGB, and P18 $H_0$ priors, respectively.}
\label{fig:M_calibration}
\end{figure}

Now, the full Pantheon data set was also considered where a cosmological model is necessary to obtain the Hubble data and thus calibrate the absolute magnitude $M$. Nonetheless, this quantity is related to the intrinsic brightness of a luminous object and so should not reflect, or at least, weakly reflect, the dynamics of cosmic expansion. This assertion is strongly supported by the fact that $M$ can be calibrated, provided an $H_0$ prior, with subpercent precision even when considering only the $z < 0.1$ redshifts (the first 11 of 40 points) in the compressed Pantheon samples \cite{Scolnic:2017caz}. The calibrated $M$ which will be used later for the GP reconstruction of $H(z)$ per $H_0$ prior is shown in Fig.~\ref{fig:M_calibration}.

\begin{figure}[h!]
\center
\includegraphics[width = 0.45 \textwidth]{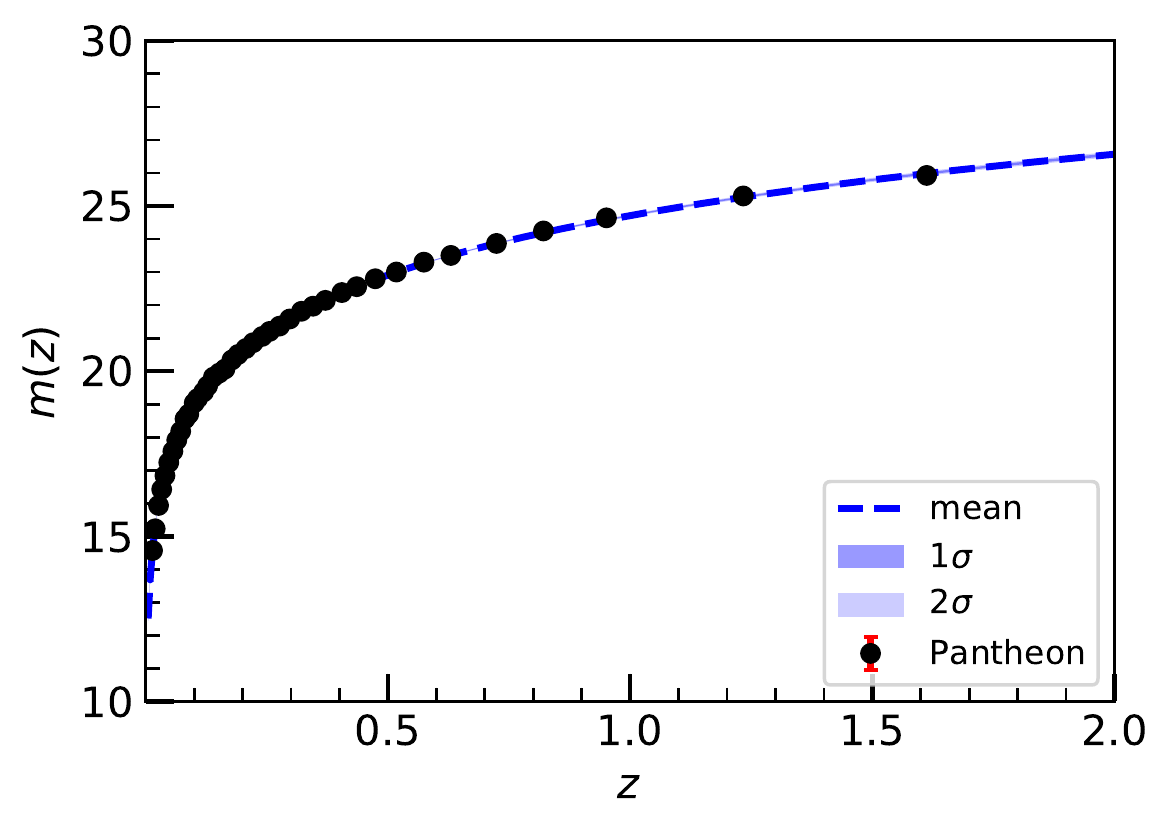}
\caption{GP reconstruction of the SNe apparent magnitudes as a function of the redshift $z$.}
\label{fig:pantheon_mz}
\end{figure}

The mean and standard deviations are $M = -19.25 \pm 0.05, -19.38 \pm 0.06$, and $-19.45 \pm 0.03$ for the R19, TRGB, and P18 $H_0$ priors, respectively. Lastly, the GP reconstructed Pantheon SNe apparent magnitudes is presented in Fig.~\ref{fig:pantheon_mz}.

We emphasize that the remarkable precision of this GP reconstruction of $m(z)$, achieved by performing the GP in $\log(z)$ (due to the drastic variance in the density of points across $z$), justifies and completes the outlined sound horizon-free construction of $H(z)$ from BAO.

In the present study, we perform a number of GP analyses using a combination of prior values. GP is a non-parametric reconstruction implementation but it does depend on the kernel hyperparameters. Given the level of precision in the discrepancy in $H_0$, we consider the RBF kernel in order to reduce any fine differences between these estimations of $H_0$.

\begin{figure}[ht]
\center
	\subfigure[ $\chi^2_\text{R19} = 26.2$ ]{
		\includegraphics[width = 0.45 \textwidth]{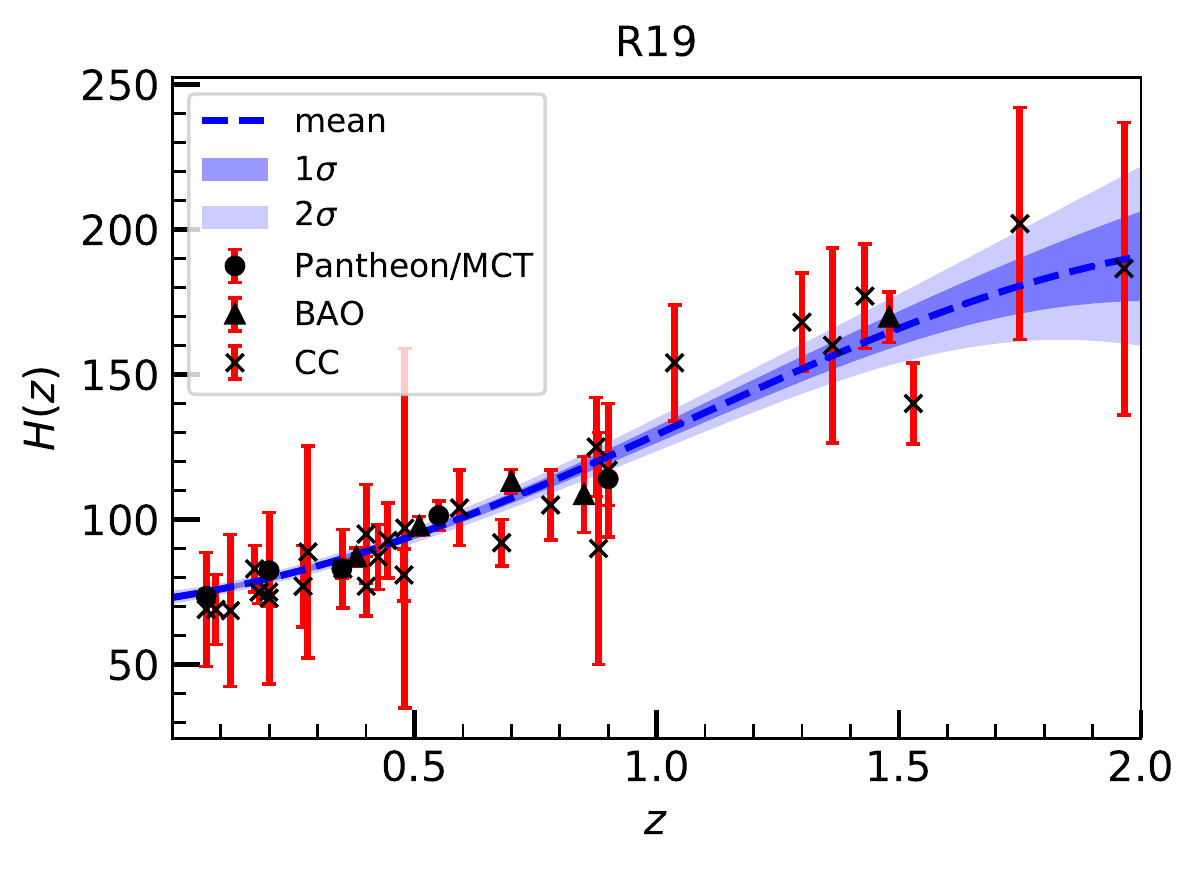}
		}
	\subfigure[ $\chi^2_\text{TRGB} = 19.8$ ]{
		\includegraphics[width = 0.45 \textwidth]{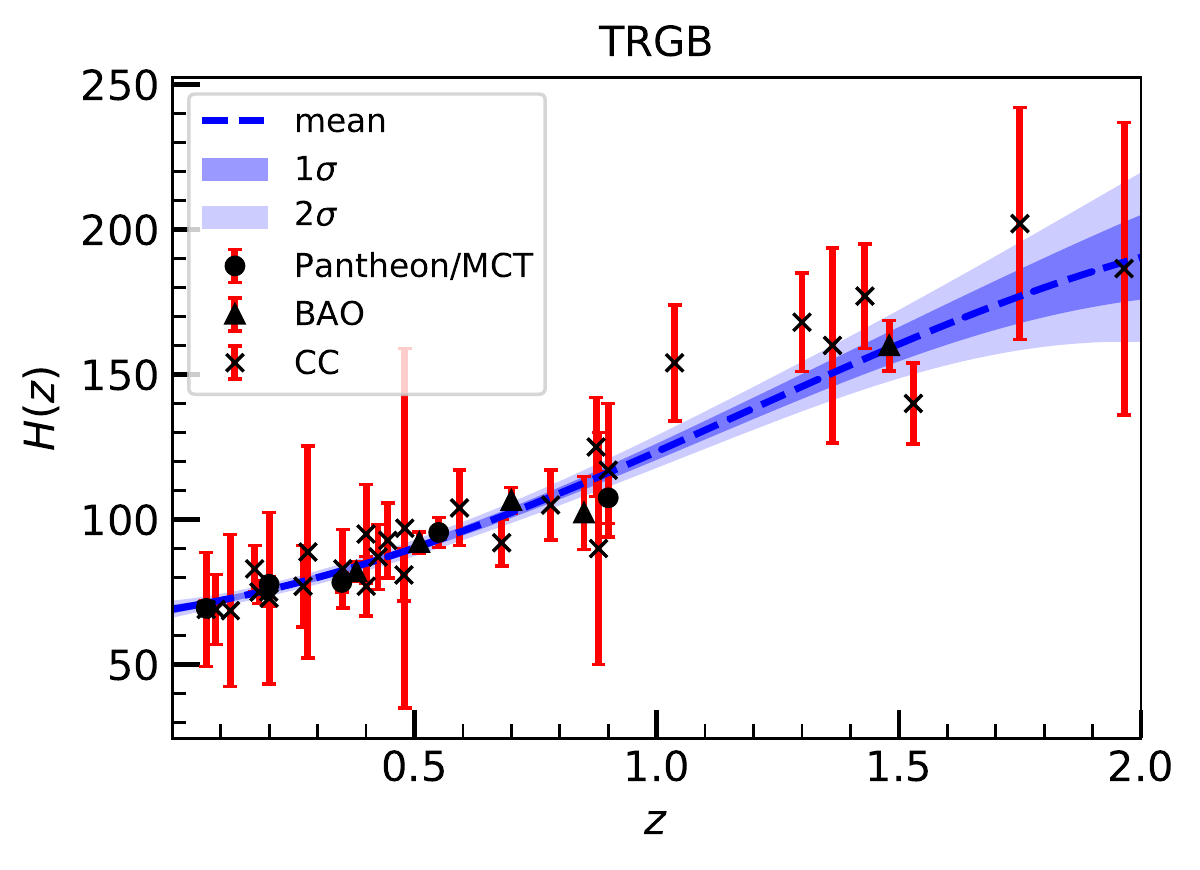}
		}
	\subfigure[ $\chi^2_\text{P18} = 25.8$ ]{
		\includegraphics[width = 0.45 \textwidth]{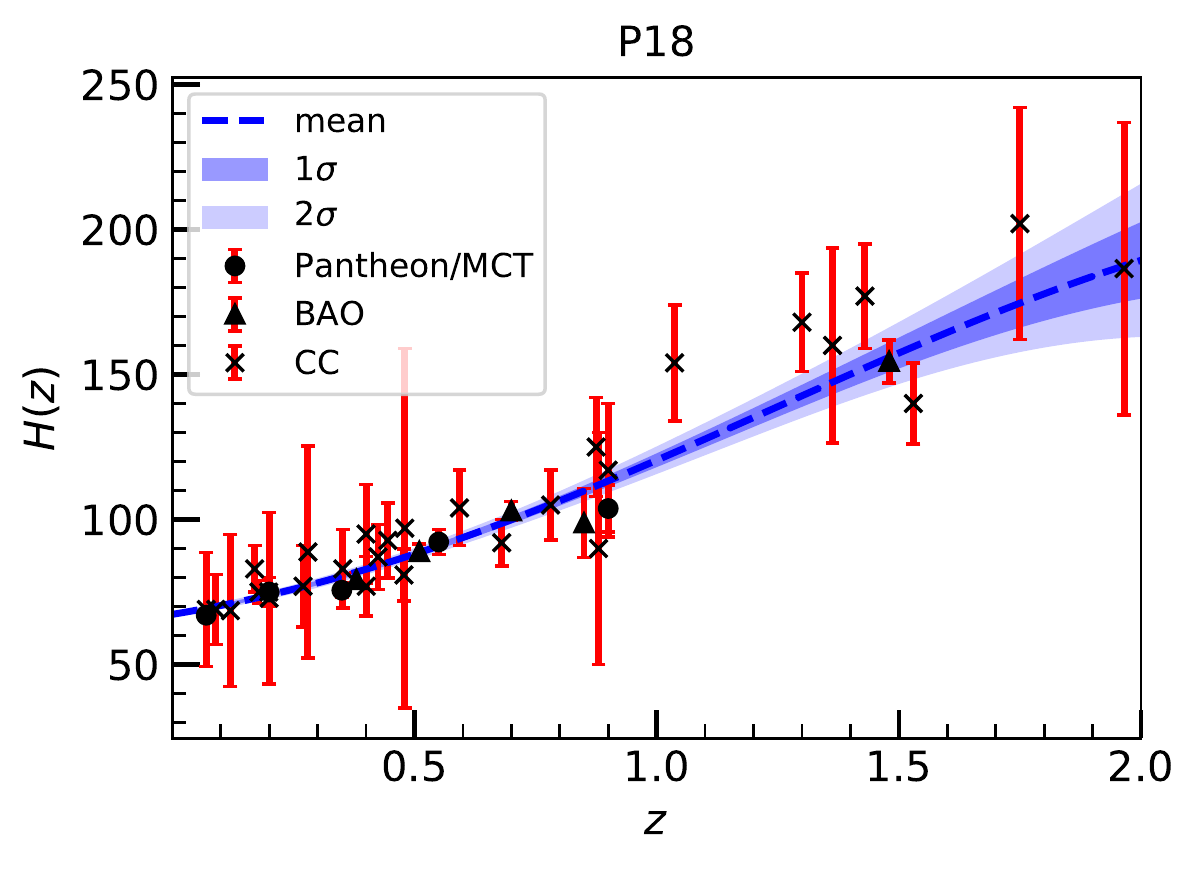}
		}
	\subfigure[ $\chi^2_{\Lambda\text{CDM}} = 28.5$ ]{
		\includegraphics[width = 0.45 \textwidth]{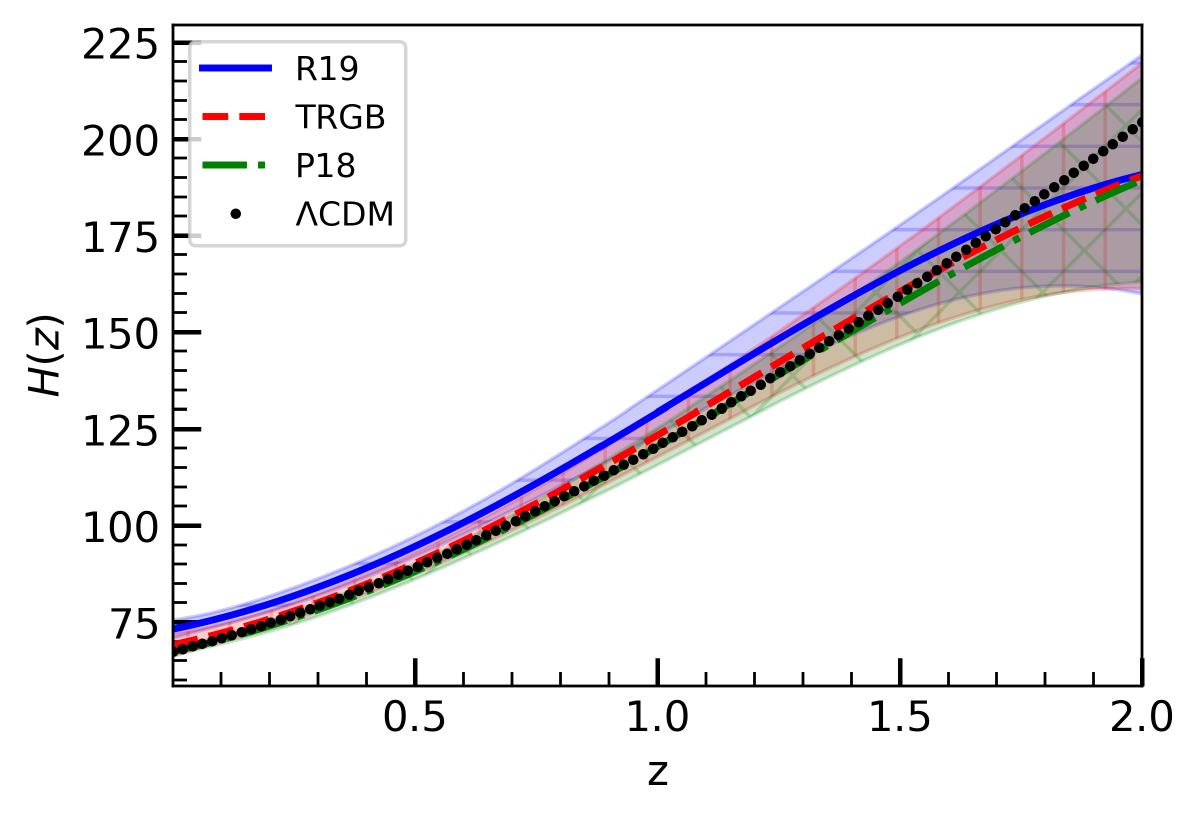}
		}
\caption{(a-c) GP reconstructed Hubble function $H(z)$ (in units of $\,{\rm km\, s}^{-1} {\rm Mpc}^{-1}$) given the full data set (Pantheon/MCT + BAO + CC) and $H_0$ prior (R19, TRGB, P18); (d) GP reconstructed Hubble function $H(z)$ for the for the full data set with different $H_0$ priors and the $\Lambda$CDM contour. The filled-hatched regions in (d) show the $2\sigma$ confidence intervals for each prior. Hatches used: $\left( '-' : H_0^\text{R19} \right)$ \cite{Riess:2019cxk}, $\left( '|' : H_0^\text{TRGB} \right)$ \cite{Freedman:2019jwv}, $\left( '\times' : H_0^\text{P18} \right)$ \cite{Aghanim:2018eyx}.}
\label{fig:Hz_rec}
\end{figure}

The reconstructed Hubble parameter is shown for the combined CC, Pantheon/MCT, and BAO data sets for each of the $H_0$ priors in Fig.~\ref{fig:Hz_rec}. For completeness, we also write down the $\chi^2$ likelihoods for each of these reconstructions and the one for $\Lambda$CDM (taking in the P18 constraints \cite{Aghanim:2018eyx}) in the subtitles of Fig.~\ref{fig:Hz_rec}, and where
\begin{equation}
    \chi^2 = \sum_i \frac{\left(H_{\rm obs}(z_i) - H_{\rm recon}(z_i)\right)^2}{\sigma_{\rm obs}(z_i)^2}\,,
\end{equation}
for each data set in the combination. We do this to quantity possible over fitting which is a common pathology of non-parametric reconstruction methods. Indeed, this shows up in the $\chi^2$-statistics ($< 40$, number of data points). Nonetheless, it can be seen that the constrained $\Lambda$CDM model based on the Planck 2018 release also overfits the joint data set. On the other hand, this overfitting by $\Lambda$CDM can in fact also be viewed as a remarkable coincidence considering that the Planck mission probes the physics of last scattering ($z \sim 1000$). The relative value of $\Delta \chi^2 = \chi^2_i - \chi^2_{\Lambda\text{CDM}}$ may then be taken to provide a rough degree of how well a GP reconstruction fits the data.

In Fig.~\ref{fig:Hz_rec}, we also notice that the low redshift behavior of the reconstructed Hubble diagram is not drastically different for the different prior selections. However, the uncertainties at high redshifts are impacted by this choice with R19 giving the largest uncertainties region. In fact, these reconstructions produce small uncertainty regions up to roughly $z \sim 1.5$ after which these regions start to grow due to the sparsity of observational data points in that region.

In the next section, the reconstructed $H(z)$ and its first derivative $H'(z)$ will be used to predict Horndeski Lagrangian potentials. In doing so, it is important to highlight that $H(z)$ and $H'(z)$ are naturally correlated together because of the GP reconstruction \cite{Seikel2012,Seikel:2013fda}. That is, at each redshift $z^*$, this is fleshed out by the covariance matrix
\begin{equation} \label{eq:cov_f0_f1}
    \text{cov} \left( H(z^*), H'(z^*) \right) = K^{(0, 1)}\left( z^*, z^* \right) - K\left( z^*, Z \right) \left[ K\left(Z, Z\right) + C \right]^{-1} K^{(0, 1)} \left( Z, z^* \right)\,.
\end{equation}
To consistently obtain a function $f \left(H (z^*), H'(z^*) , \theta \right)$ where $\theta$ stands for any additional parameters, with a variance $\text{var}\left(\theta\right)$, we draw a large number of samples, \textit{a la} Monte Carlo (MC), from the multivariate Gaussian distribution
\begin{equation}
\left(
\begin{matrix}
H  \\
H' \\
\theta
\end{matrix}
\right)
\sim 
\mathcal{N}
\left[  
\left(
\begin{matrix}
H  \\
H'  \\
\theta
\end{matrix}
\right)
,
\left(
\begin{matrix}
\text{cov} \left( H , H \right) & \text{cov} \left( H, H' \right) & 0 \\
\text{cov} \left( H, H' \right) & \text{cov} \left( H', H' \right) & 0 \\
0 & 0 & \text{var}\left( \theta \right)
\end{matrix}
\right)
\right]\,,
\end{equation}
and then take the mean, standard deviation, and other relevant statistical outputs of the resulting posterior distribution of the function $f$.

\section{Gaussian Processes Reconstruction of Horndeski Gravity}
\label{sec:painting_horndeski}

In this section, we discuss in detail three model-building implementations in Horndeski cosmology and complement this with the GP reconstructed Hubble function. This outputs meaningful predictions, mean and confidence intervals, of the Horndeski potentials and the corresponding dark energy equation of state.

\subsection{Quintessence potential}
\label{subsec:quintessence}

The simplest and perhaps most well-known potential construction method is in quintessence cosmology \cite{Tsamis:1997rk}. In this model, the Friedmann and scalar field equations are given by
\begin{equation}
\label{eq:Feq_quint}
3 H^2 = \rho + \dfrac{\dot{\phi}^2}{2} + V\left( \phi \right)\,,
\end{equation}
\begin{equation}
\label{eq:Peq_quint}
2 \dot{H} + 3 H^2 = -P -\dfrac{\dot{\phi}^2}{2} + V\left( \phi \right) \,,
\end{equation}
and
\begin{equation}
\label{eq:Seq_quint}
\ddot{\phi} + 3 H \dot{\phi} + V'\left(\phi\right) = 0\,,
\end{equation}
respectively. It can of course be shown that Eq.~(\ref{eq:Seq_quint}) follows from Eqs.~(\ref{eq:Feq_quint}) and (\ref{eq:Peq_quint}) as long as the perfect fluid's energy is thermodynamically conserved. Using the Friedmann equations, it can then be shown that the potential and kinetic terms of the scalar field are given by
\begin{equation}
\label{eq:V_quint}
V \left( \phi \right) = \dot{H} + 3 H^2 - \dfrac{\rho - P}{2}\,,
\end{equation}
and
\begin{equation}
\label{eq:phip2_quint}
\dot{\phi}^2 = -2 \dot{H} - \left( \rho + P \right)\,.
\end{equation}

Eq.~(\ref{eq:V_quint}) shows that the quintessence potential can be uniquely determined given a Hubble function and matter fields. The kinetic term $X = \dot{\phi}^2/2$ (through Eq.~(\ref{eq:phip2_quint})) can be similarly obtained. We additionally assume non-relativistic matter fields which take on the P18 value for the current matter density parameter \cite{Aghanim:2018eyx}.

\begin{figure}[h!]
\center
	\subfigure[ ]{
		\includegraphics[width = 0.45 \textwidth]{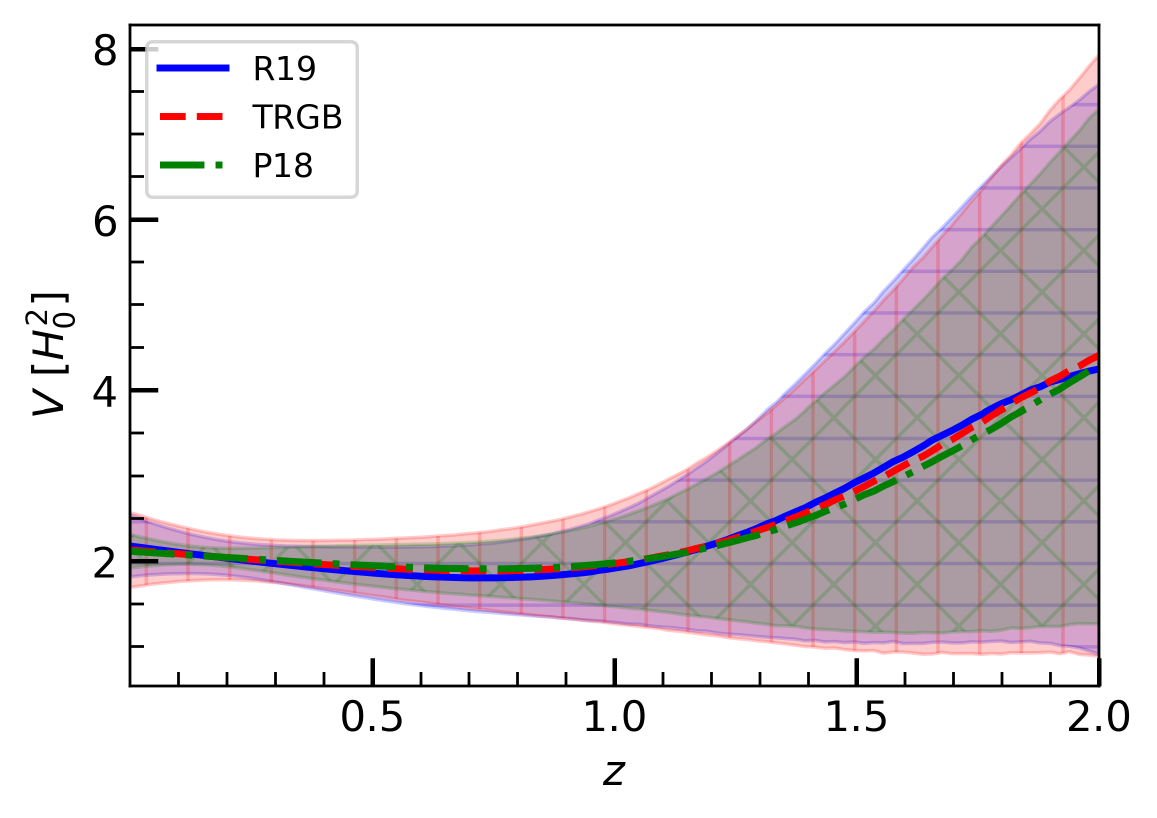}
		}
	\subfigure[ ]{
		\includegraphics[width = 0.45 \textwidth]{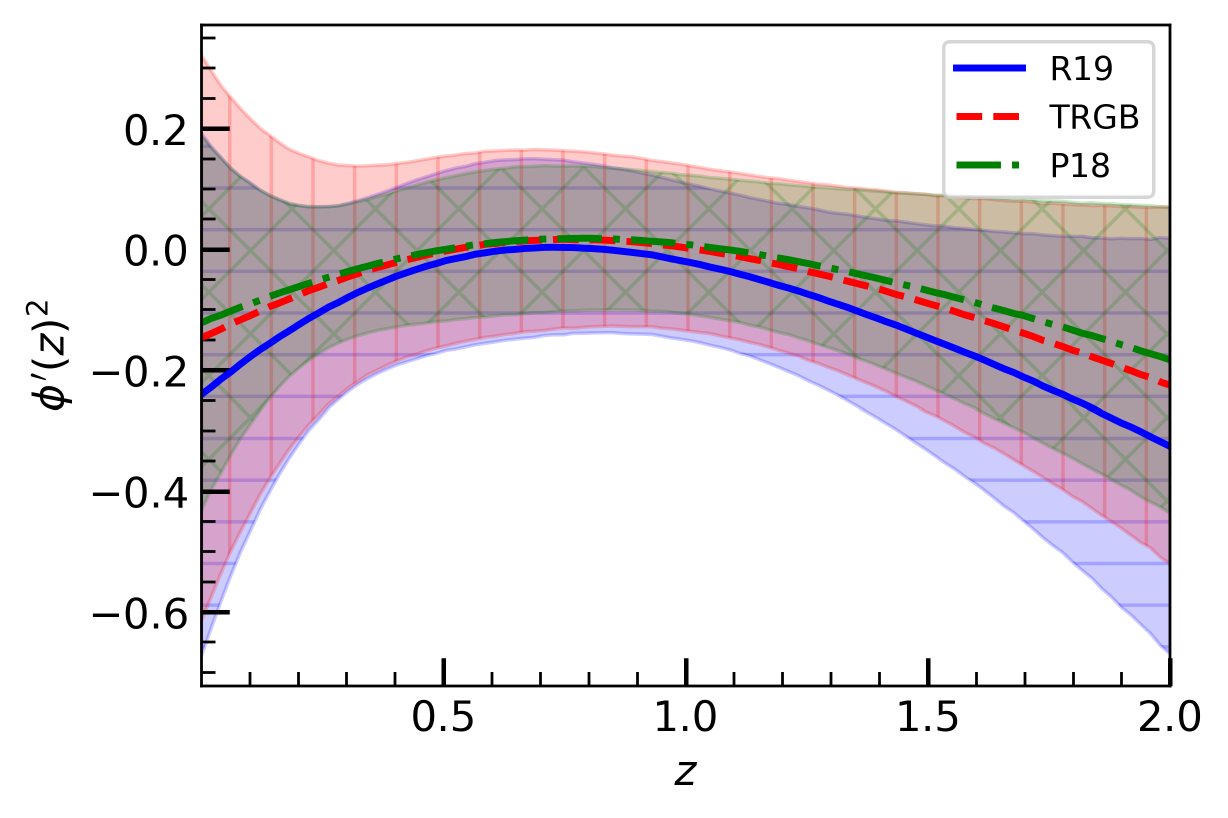}
		}
\caption{(a) Reconstructed quintessence potential $V(z)$ and (b) $\phi^\prime(z)^2$ for varying $H_0$ prior. The filled-hatched regions show the $2\sigma$ confidence intervals for each prior. Hatches used: $\left( '-' : H_0^\text{R19} \right)$ \cite{Riess:2019cxk}, $\left( '|' : H_0^\text{TRGB} \right)$ \cite{Freedman:2019jwv}, $\left( '\times' : H_0^\text{P18} \right)$ \cite{Aghanim:2018eyx}.}
\label{fig:V_phip2_quint}
\end{figure}

\begin{figure}[h!]
\center
\subfigure[ ]{
		\includegraphics[width = 0.45 \textwidth]{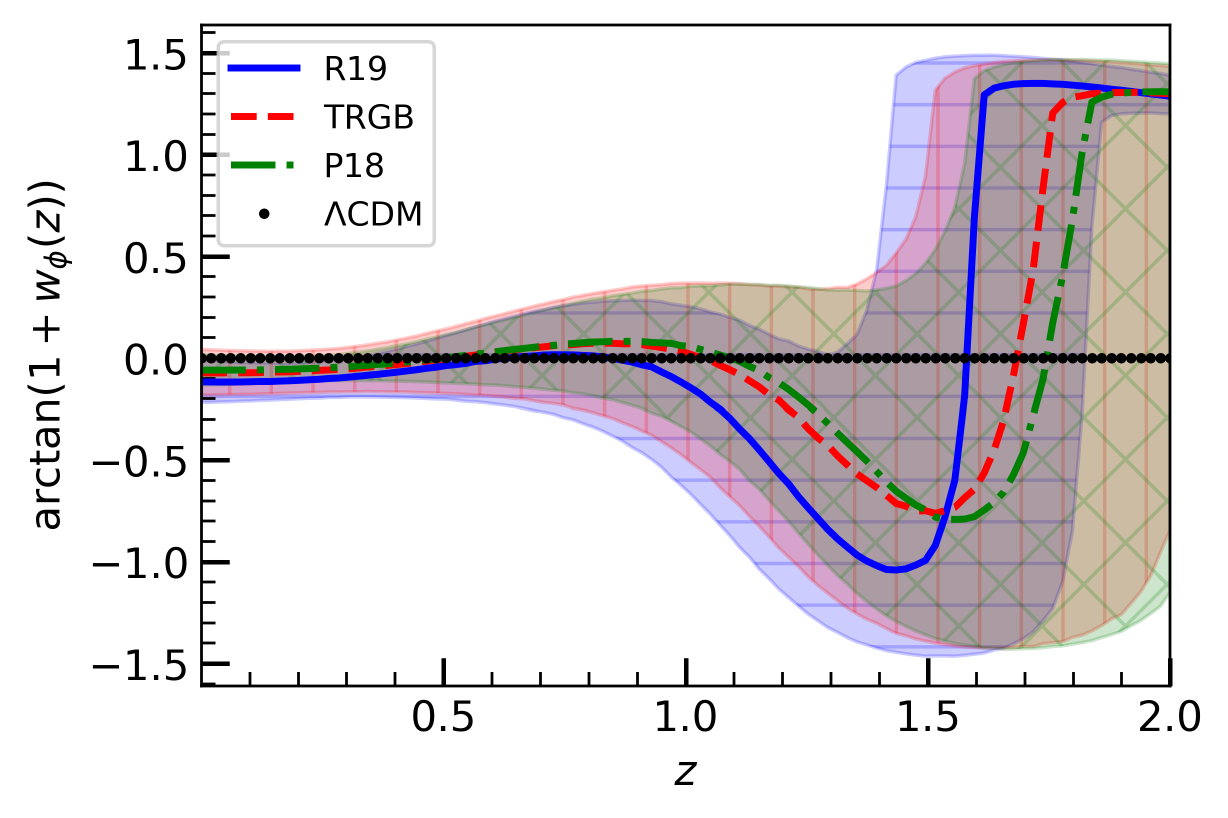}
		}
	\subfigure[ ]{
		\includegraphics[width = 0.4 \textwidth]{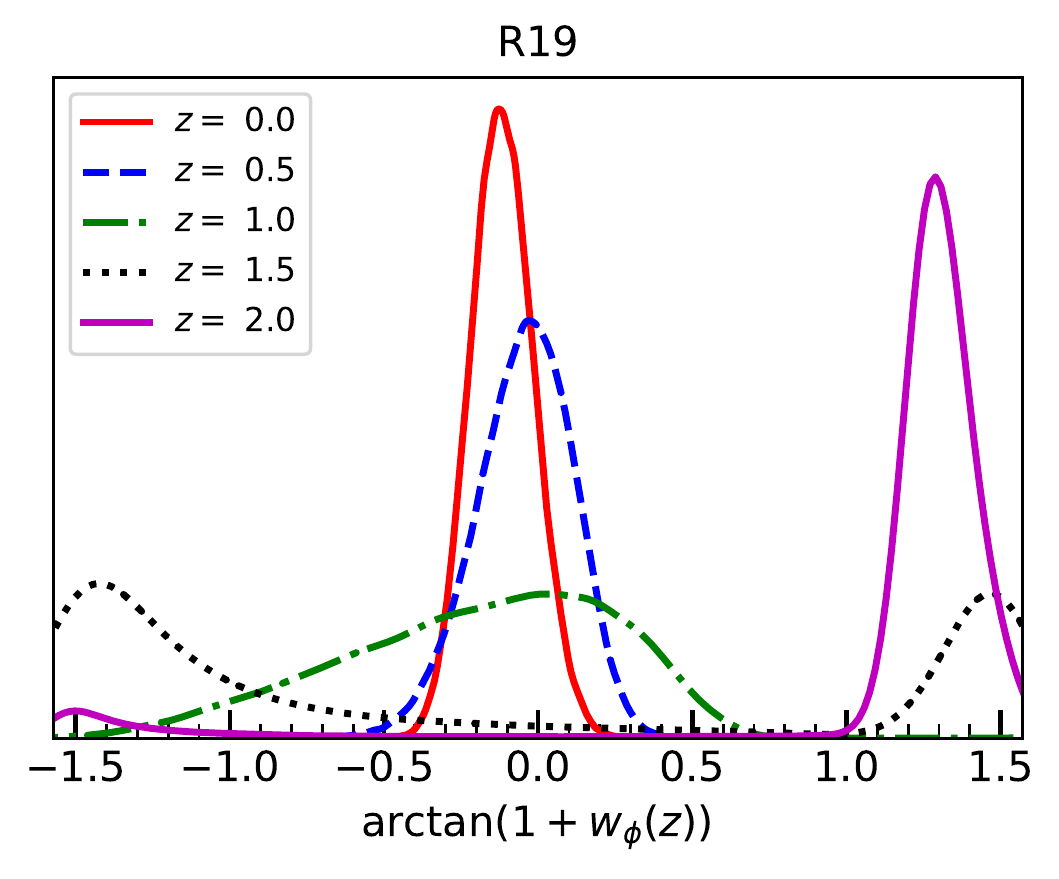}
		}
	\subfigure[ ]{
		\includegraphics[width = 0.4 \textwidth]{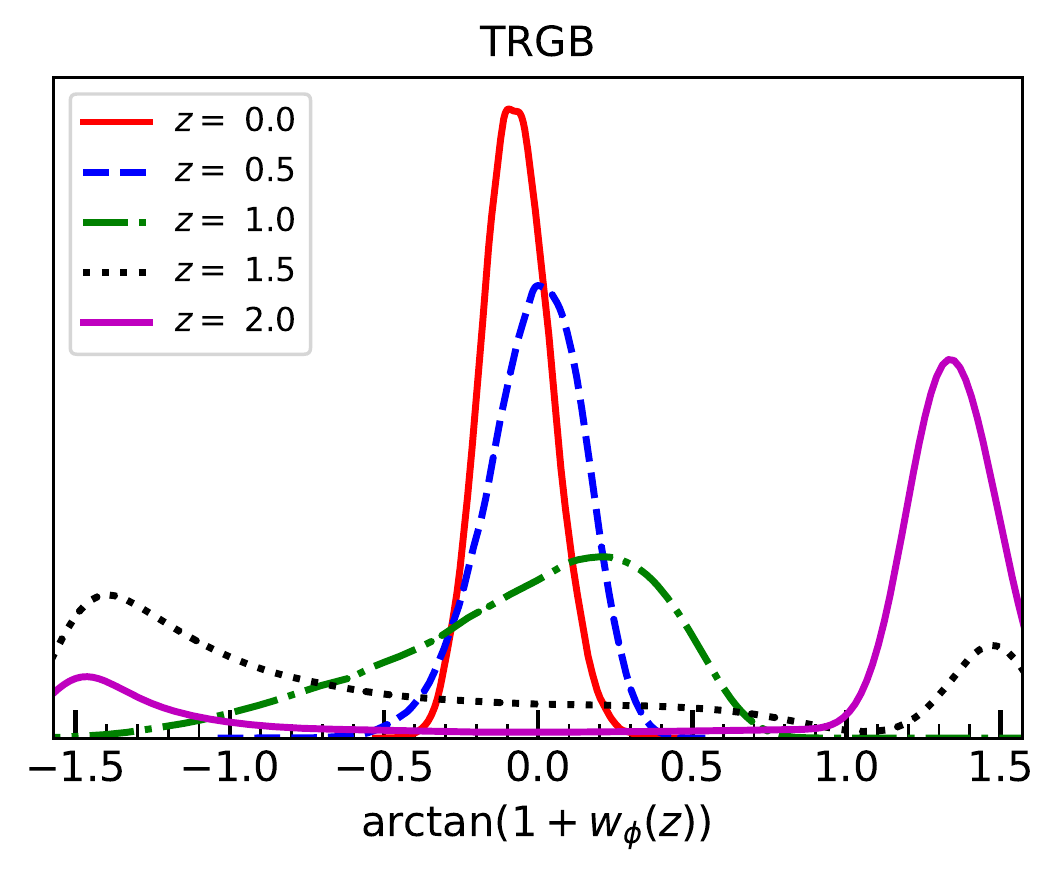}
		}
	\subfigure[ ]{
		\includegraphics[width = 0.4 \textwidth]{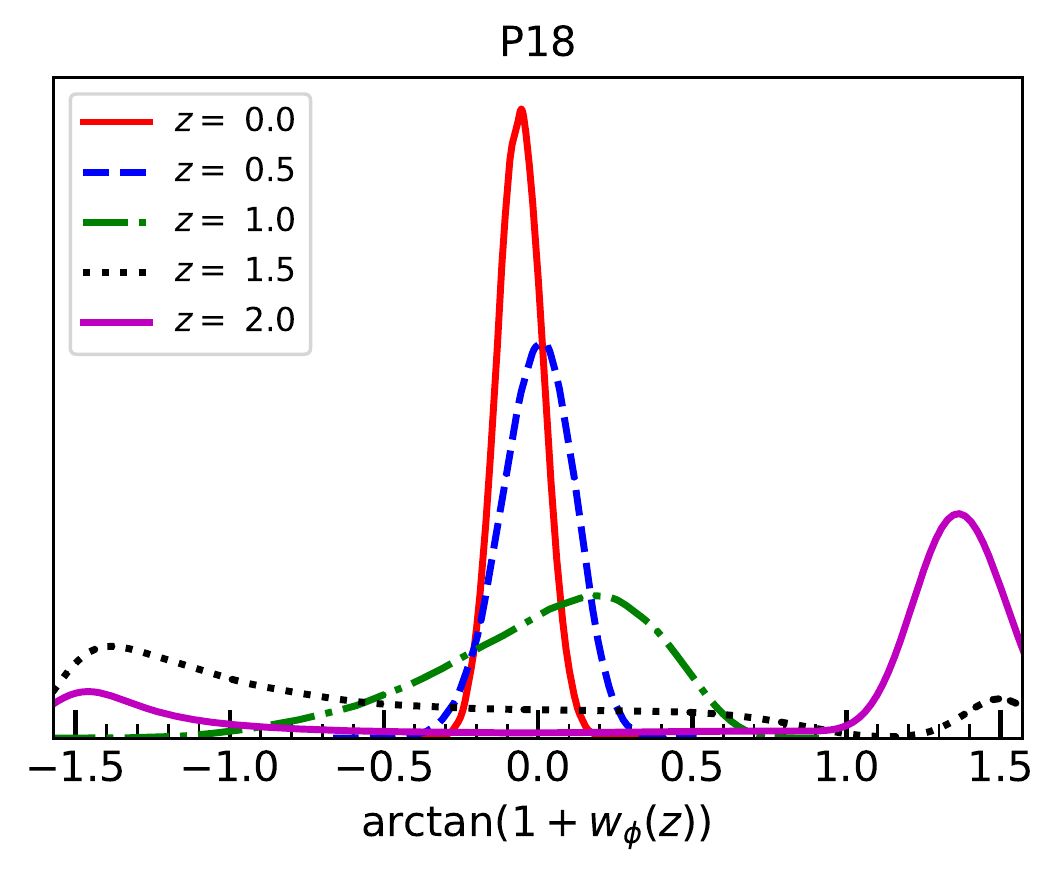}
		}
\caption{(a) The compactified dark energy equation of state, $\arctan\left(1 + w_\phi(z)\right)$, for quintessence for various $H_0$ priors as a function of the redshift. The solid, dashed, and dash-dotted lines represent the median of the distribution and the filled-hatched regions show the $34.1\%$ of the probability mass surrounding the median from both sides. Hatches used: $\left( '-' : H_0^\text{R19} \right)$ \cite{Riess:2019cxk}, $\left( '|' : H_0^\text{TRGB} \right)$ \cite{Freedman:2019jwv}, $\left( '\times' : H_0^\text{P18} \right)$ \cite{Aghanim:2018eyx}. Posteriors of the compactified dark energy equation of state at sample redshifts $z = 0, 0.5, 1, 1.5, 2$ are presented for the (b) R19, (c) TRGB, and (d) P18 $H_0$ priors, respectively.}
\label{fig:w_de_quint}
\end{figure}

We can now use the GP reconstructed Hubble function to predict the shapes of the $V$ and $\phi'(z)^{2}$, where $\dot{f} \left( t \right) = - (1 + z) H(z) f'(z)$ for an arbitrary function $f$ of time (or redshift). The results are shown in Fig.~\ref{fig:V_phip2_quint}.
Note that by expressing the dimensionful quantities, in this case, $V$, in units of $H_0$, the tensions which exist at $z = 0$ for the different $H_0$ priors can be alleviated in the construction. For example, in Fig.~\ref{fig:V_phip2_quint}, the mean values of $V$ and $\phi'(z)^2$ for any one of the priors always lie well within the $1\sigma$ contours of the two other priors. In this case, it can be seen that the potential and kinetic term traces out a particular shape regardless of the $H_0$ prior. This then constrains the quintessence potential. In principle, the explicit function $V\left(\phi\right)$ can also be obtained by integrating $\phi'(z)$ to obtain $\phi$ up to a constant. However, we find that the mean $\phi'(z)^2$ is mostly-negative throughout the late-time cosmological evolution. Notably, positive values of $\phi'(z)^2$ which happen to occur near the upper $2\sigma$ contours can be used to predict $V\left(\phi\right)$. This should of course be taken with caution given that it hovers just around, and some times beyond, the $95\%$ confidence region. Nonetheless, the dark energy equation of state, $w_\phi = P_\phi / \rho_\phi$, can be obtained which in the quintessence model becomes
\begin{equation}\label{eq:w_de_quint}
    w_\phi = \dfrac{ \dot{\phi}^2/2 - V }{ \dot{\phi}^2/2 + V }\,.
\end{equation}
However, instead of $w_\phi$ it turns out to be as convenient to additionally sample over the compactified variable $\arctan\left( 1 + w_\phi \right)$. We refer to this as a \textit{compactified} dark energy equation of state. This alternative variable tames down the singularities which otherwise would make the MC procedure terribly unreliable, i.e., MC propagated errors in $w_\phi$ diverge at $z \sim 1$. The result of the additional sampling of the compactified dark energy equation of state is shown in Fig.~\ref{fig:w_de_quint}.

In Fig.~\ref{fig:w_de_quint}-(a), the median of the distribution and the $34.1 \%$ of the probability mass surrounding it on both sides are shown. To see why this is a more consistent statistic, compared with the usual Gaussian-anchored mean $\pm$ $1 \sigma$ statistic, we also show the resulting posterior distributions of $\arctan\left(1 + w_\phi(z)\right)$ at different redshifts in Figs.~\ref{fig:w_de_quint}-(b-d).

Clearly, for low redshifts, $z \lesssim 1$, the resulting distributions can be considered to be normally-distributed. However, at higher redshifts, it can be seen that the distribution not only starts to deviate away from normality, but even becomes bimodal and takes samples at $\arctan\left(1 + w_\phi\right) \sim \pm \pi/2$ where $w_\phi$ diverges. For this reason, the median and the $34.1\%$ probability mass surrounding it from both sides can be considered as a more consistent statistic as it agrees with the mean $\pm$ $1 \sigma$ when the distribution is normal, but it tracks the most concentration of probability mass even when the distribution is no longer normal. See Appendix \ref{sec:stat_compact_rv} for a supplementary illustration of this point. In addition, a naive use of the mean $\pm$ $1 \sigma$ statistic for a compactified variable may predict results outside of the domain of the compactified variable while the median $\pm$ $34\%$ probability mass will always only sample within this domain.

The results shown in Fig.~\ref{fig:w_de_quint} agree that dark energy may as well be vacuum energy, i.e., $w \approx -1$ or $\arctan\left(1 + w_\phi\right) = 0$, for low redshifts. On the other hand, the use of the compactified dark energy equation of state lets us understand a clearer picture of dark energy at earlier redshifts when the prediction of $w_\phi$ becomes spoiled by diverging uncertainties. For instance, at $z \gtrsim 1$, Figs.~\ref{fig:w_de_quint}-(b-d) reveal that the onset of the divergence in the uncertainty of $w_\phi$ is alternatively due to the compactified variable starting take values with $ | \arctan(x \sim \pm \pi/2) | \gg 1 $. This sampling procedure on a compactified random will be utilized further to study dark energy in the two following Horndeski models.

To end this section, we recall that $\Omega_m$ is a free parameter which we fixed to its P18 value (anchored on $\Lambda$CDM) in order to avoid an otherwise ad hoc choice. The above reconstruction is therefore bias to this information. We briefly describe the reconstruction for other values of $\Omega_m$. Smaller $\Omega_m$ leads to higher $V(\phi)$ and $\phi'(z)^2$, while larger $\Omega_m$ leads to lower $V(\phi)$ and $\phi'(z)^2$. This can be also be deduced from Eqs. (\ref{eq:V_quint}) and (\ref{eq:phip2_quint}). In particular, for $\Omega_m \lesssim 0.1$, the reconstruction of $\phi'(z)^2$ can be positive throughout. This can be taken to mean that quintessence and late-time data favor a smaller $\Omega_m$ than that of $\Lambda$CDM. We encourage the reader to test this out using our codes \cite{reggie_bernardo_4810864}.

\subsection{Designer Horndeski}
\label{subsec:designer_horndeski}

Another Horndeski construction method has been introduced recently in Ref.~\cite{Arjona:2019rfn}. This is referred to as \textit{designer Horndeski} (HDES) and is built on top of kinetic gravity braiding. To describe this method, we start with the cosmological field equations given by
\begin{equation}
\label{eq:Feq_hdes}
    3H^2 = \rho - K(X) + 2 X K_X + 3 H \dot{\phi}^2 G_X\,,
\end{equation}
\begin{equation}
\label{eq:Peq_hdes}
    2 \dot{H} + 3 H^2 = -P - K(X) + 2 X \ddot{\phi} G_X\,,
\end{equation}
and
\begin{equation}
\label{eq:Seq_hdes}
\ddot{\phi} \left(-\dot{\phi} \left(3 H \left(G_{XX} \dot{\phi}^2+2 G_X\right)+K_{XX} \dot{\phi}\right)-K_X\right)-3 \dot{\phi} \left(G_X \dot{H} \dot{\phi}+3 G_X H^2 \dot{\phi}+H K_X\right) = 0\,,
\end{equation}
where a subscript in the potentials $K(X)$ and $G(X)$ denote differentiation with respect to $X$. These are the Friedmann and scalar field equations. Also, a noteworthy feature in these shift symmetric models is that the scalar field equation can be written in the form
\begin{equation}
\label{eq:Seq_ss}
    \dot{J} + 3 H J = 0\,,
\end{equation}
where $J$ is the shift current given by
\begin{equation}
\label{eq:Seq_J}
    J = \dot{\phi} K_X + 3 H \dot{\phi}^2 G_X\,.
\end{equation}
Therefore, instead of the scalar field equation, one can consider as a surrogate the exact solution of Eq.~(\ref{eq:Seq_ss}), namely
\begin{equation}
\label{eq:J_hdes}
\dot{\phi} K_X + 3 H \dot{\phi}^2 G_X = \dfrac{\mathcal{J}}{a^3} \,,
\end{equation}
where $\mathcal{J}$ is an integration constant which we refer to as a \textit{shift charge}. The designer approach recognizes the constraint equations (Eqs.~(\ref{eq:Feq_hdes}) and (\ref{eq:J_hdes})) as two independent equations of the system but with three unknowns, i.e., $\left( H(a), K(X), G(X) \right)$. The system is then closed by \textit{a priori} assuming $H(X)$. In this case, the exact solution of Eqs.~(\ref{eq:Feq_hdes}) and (\ref{eq:J_hdes}) can be shown to be
\begin{equation}
\label{eq:K_hdes}
K(X) = -3 H_0^2 \Omega_\Lambda + \dfrac{\mathcal{J} \sqrt{2X} H(X)^2}{ H_0^2 \Omega_{m0} } - \dfrac{ \mathcal{J} \sqrt{2X} \Omega_\Lambda }{ \Omega_{m0} }\,,
\end{equation}
and
\begin{equation}
\label{eq:GX_hdes}
G_X(X) = - \dfrac{ 2 \mathcal{J} H'(X) }{ 3 H_0^2 \Omega_{m0} }\,.
\end{equation}
Eqs.~(\ref{eq:K_hdes}) and (\ref{eq:GX_hdes}) fleshes out the designer approach. By complementing this with the GP, we can then make a predictive analysis of the $k$-essence and braiding potentials.

Before proceeding, we clarify that the designer Horndeski approach relies on a mapping between $X$ and $a$ which leads to a Horndeski theory $\left( K(X), G(X) \right)$. This then means that the designer trajectory $X_\text{HDES}(a)$ must not have fixed points. However, the resulting theory $\left( K(X), G(X) \right)$ may have other trajectories $X(a)$ and so in general fixed points are not prohibited in the designer approach.

\begin{figure}[h!]
\center
	\subfigure[ ]{
		\includegraphics[width = 0.45 \textwidth]{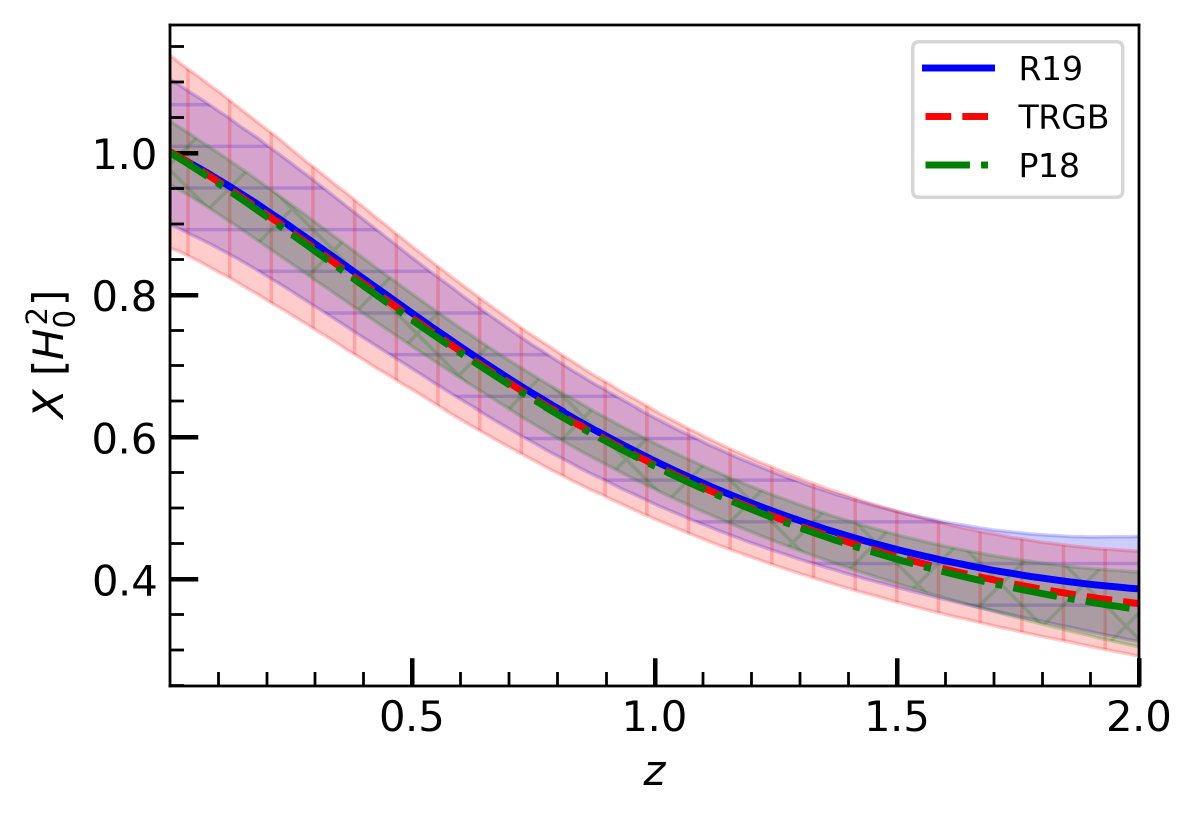}
		}
	\subfigure[ ]{
		\includegraphics[width = 0.45 \textwidth]{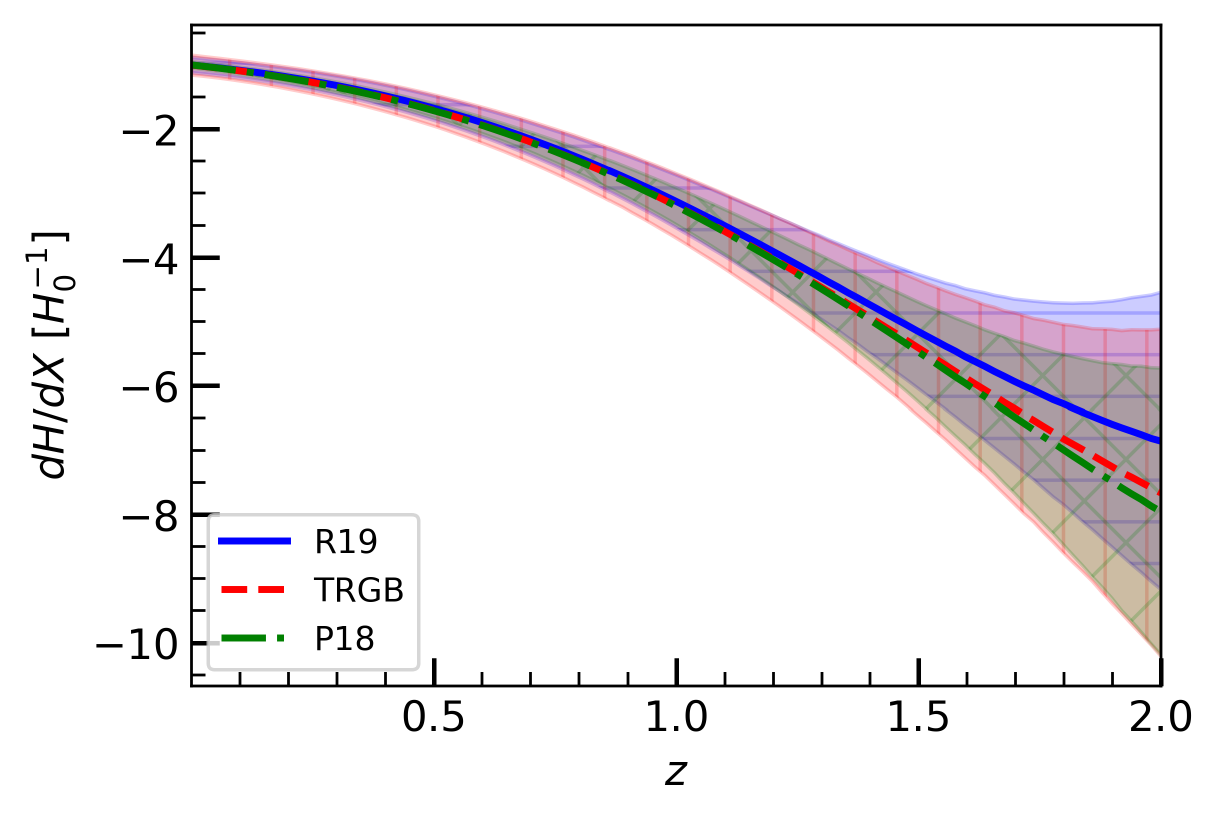}
		}
\caption{Reconstructed HDES kinetic density (a) $X$ and (b) $dH/dX$
as a function of the redshift $z$ for $c_0 = H_0^{n + 2}$ and $n = 1$ for different $H_0$ priors. The filled-hatched regions show the $2\sigma$ confidence intervals for each prior. Hatches used: $\left( '-' : H_0^\text{R19} \right)$ \cite{Riess:2019cxk}, $\left( '|' : H_0^\text{TRGB} \right)$ \cite{Freedman:2019jwv}, $\left( '\times' : H_0^\text{P18} \right)$ \cite{Aghanim:2018eyx}.}
\label{fig:X_dHdX_hdes}
\end{figure}

\begin{figure}[h!]
\center
	\subfigure[ ]{
		\includegraphics[width = 0.45 \textwidth]{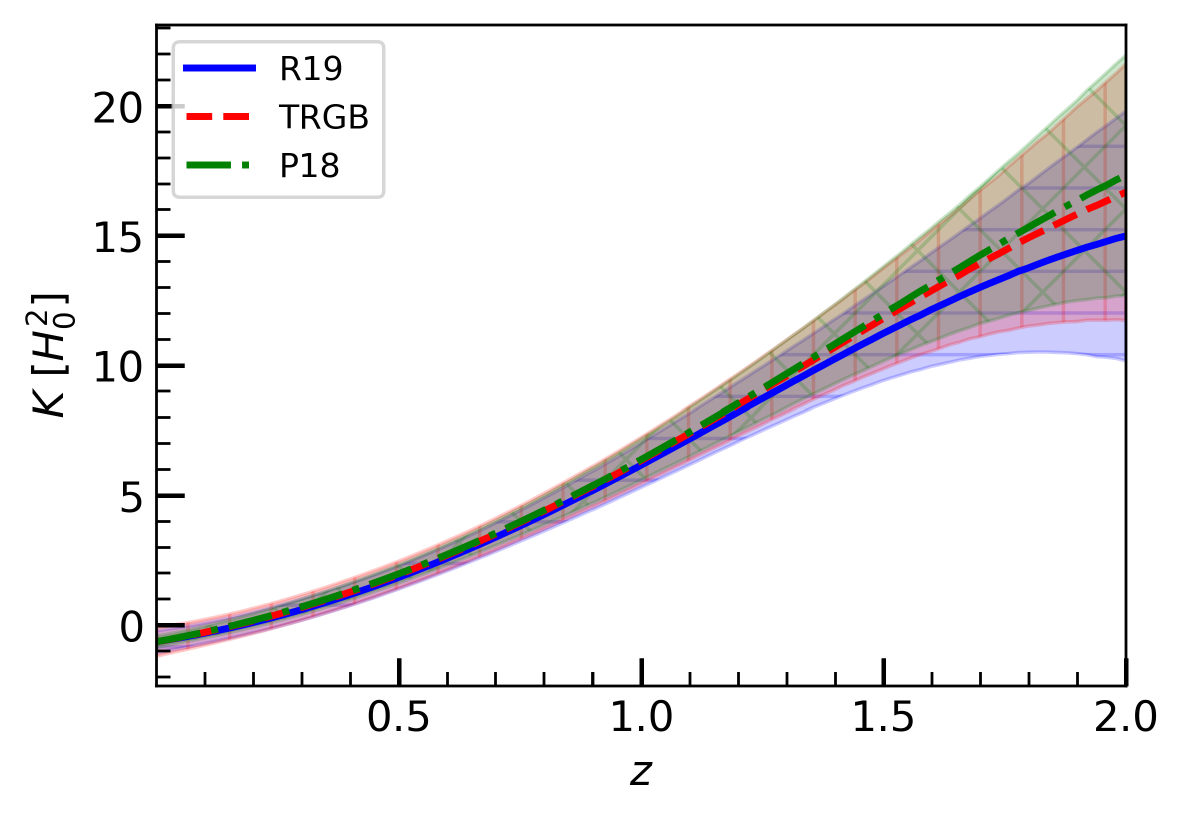}
		}
	\subfigure[ ]{
		\includegraphics[width = 0.45 \textwidth]{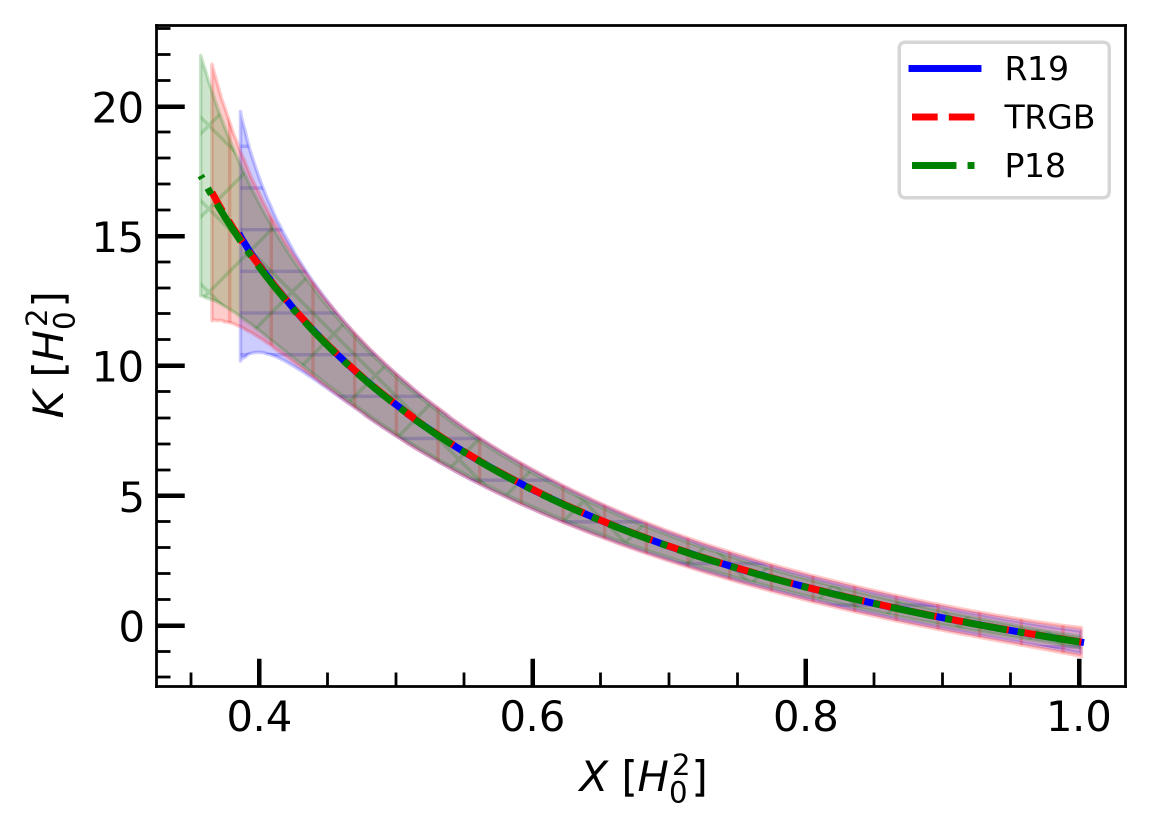}
		}
	\subfigure[ ]{
		\includegraphics[width = 0.45 \textwidth]{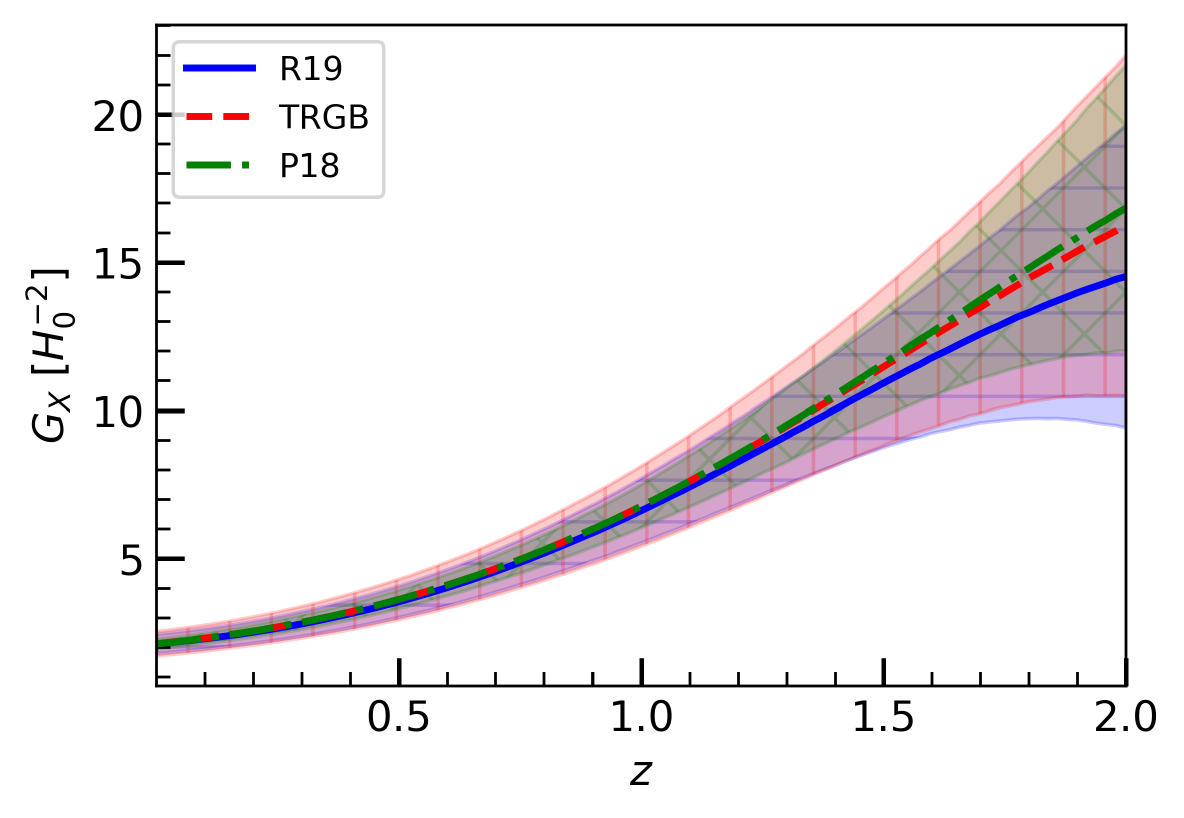}
		}
	\subfigure[ ]{
		\includegraphics[width = 0.45 \textwidth]{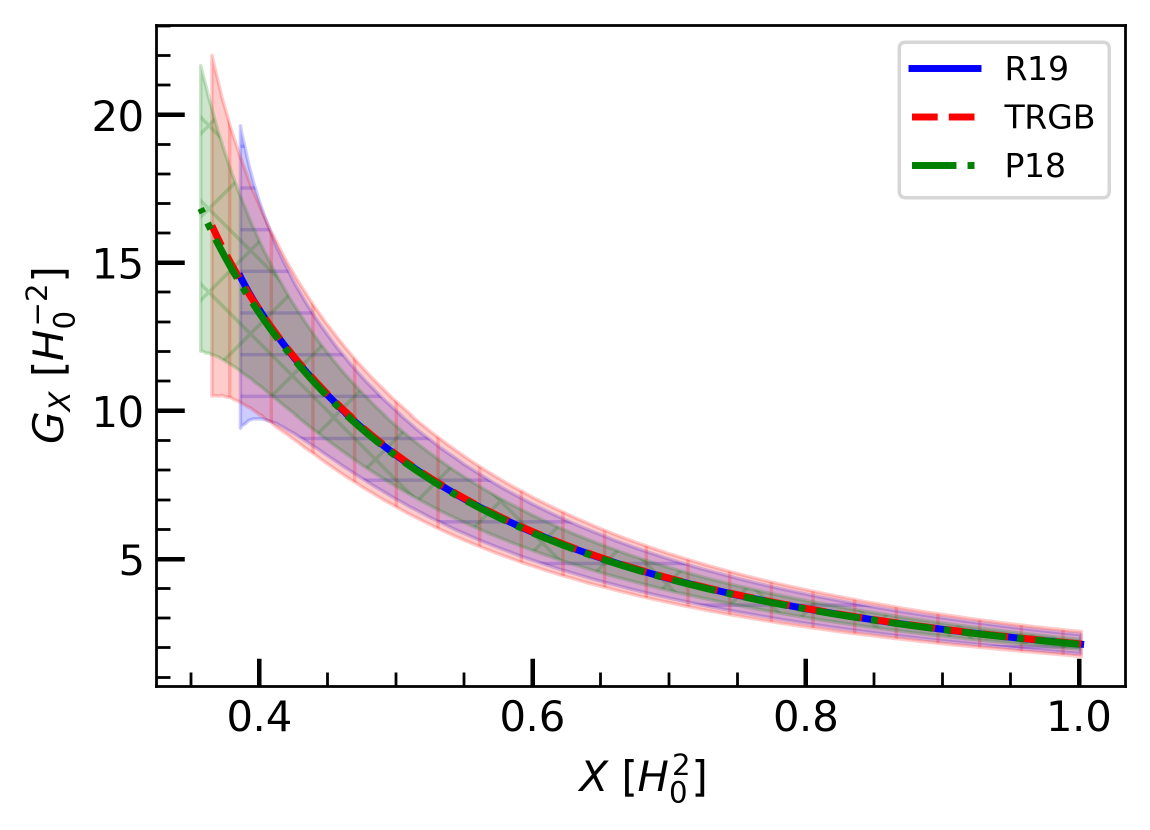}
		}
\caption{Reconstructed HDES potentials (a) $K\left(z\right)$, (b) $K(X)$, (c) $G_X\left(z\right)$, and (d) $G_X(X)$ for varying $H_0$ prior and fixed HDES parameters $c_0 = H_0^{n + 2}$, $n = 1$, and $\mathcal{J} = H_0$. The filled-hatched regions show the $2\sigma$ confidence intervals for each prior. Hatches used: $\left( '-' : H_0^\text{R19} \right)$ \cite{Riess:2019cxk}, $\left( '|' : H_0^\text{TRGB} \right)$ \cite{Freedman:2019jwv}, $\left( '\times' : H_0^\text{P18} \right)$ \cite{Aghanim:2018eyx}.}
\label{fig:hdes}
\end{figure}

As with Ref.~\cite{Arjona:2019rfn}, we take
\begin{equation}
\label{eq:X_hdes}
X = \dfrac{c_0}{H(X)^n}\,,
\end{equation}
where $c_0$ is a constant with units of $H_0^{n + 2}$. Using the GP reconstructed Hubble function, we then obtain the predictions for $X$ and $H'(X)$ shown in Fig.~\ref{fig:X_dHdX_hdes} for $n =1$ and $c_0 = H_0^{n + 2}$.
By expressing dimensionful physical quantities in units of the Hubble parameter $H_0$, we find that the mean $z = 0$ prediction for each $H_0$ prior is always within the $1\sigma$ contours of the two other priors. This holds for both $X$ and $H'(X)$ in Fig.~\ref{fig:X_dHdX_hdes} and, as we shall further see, also in the following figures. The precision of the $H_0$ prior also does play a role. In Fig.~\ref{fig:X_dHdX_hdes}, the prediction using the Planck prior turns out to be the most stringent for all redshifts. Moving forward, Fig.~\ref{fig:hdes} shows the predicted HDES potentials corresponding to Fig.~\ref{fig:X_dHdX_hdes} with $c_0 = H_0^{n + 2}$, $n = 1$, and $\mathcal{J}/H_0 = 1$. In this analysis, we also assume the P18 cosmological parameter values.
Fig.~\ref{fig:hdes} reveals the shape of the $k$-essence and braiding potentials. Most importantly, this shape appears to be consistent regardless of the choice of $H_0$ prior. 

\begin{figure}[h!]
\center
\subfigure[ ]{
		\includegraphics[width = 0.45 \textwidth]{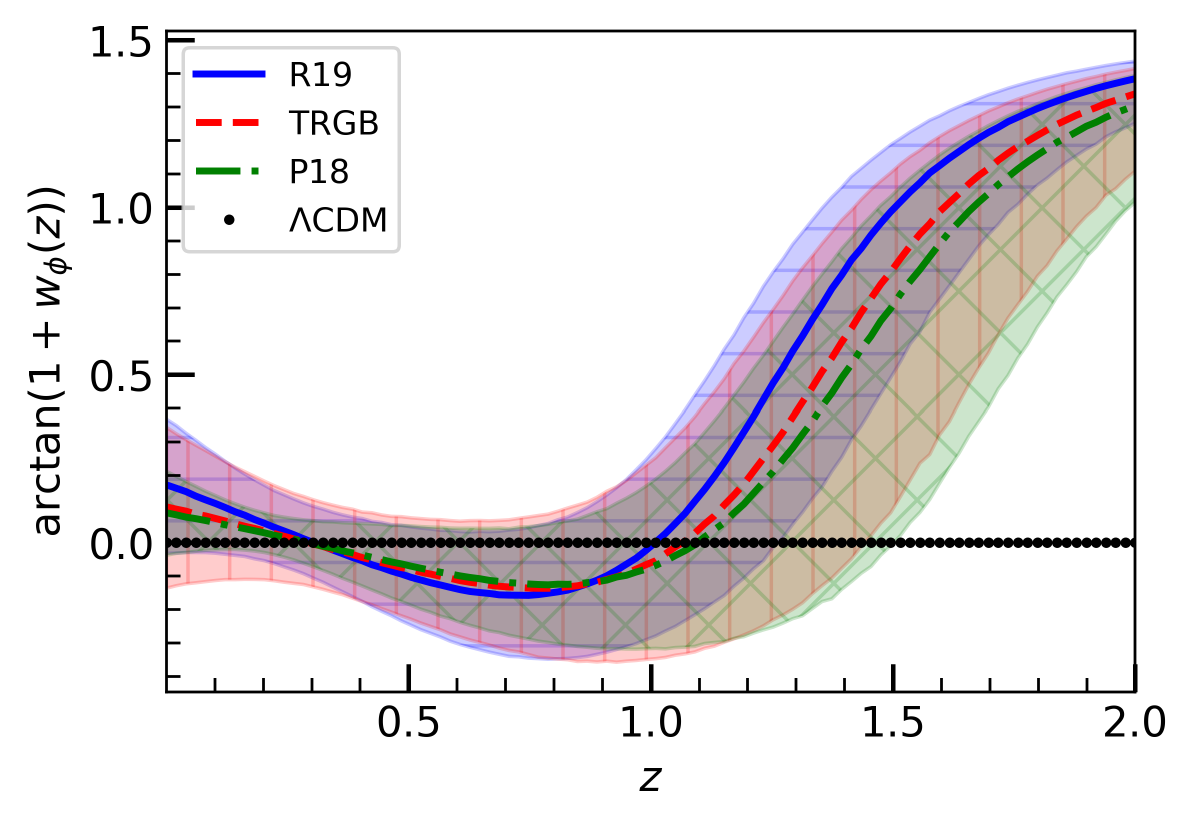}
		}
	\subfigure[ ]{
		\includegraphics[width = 0.4 \textwidth]{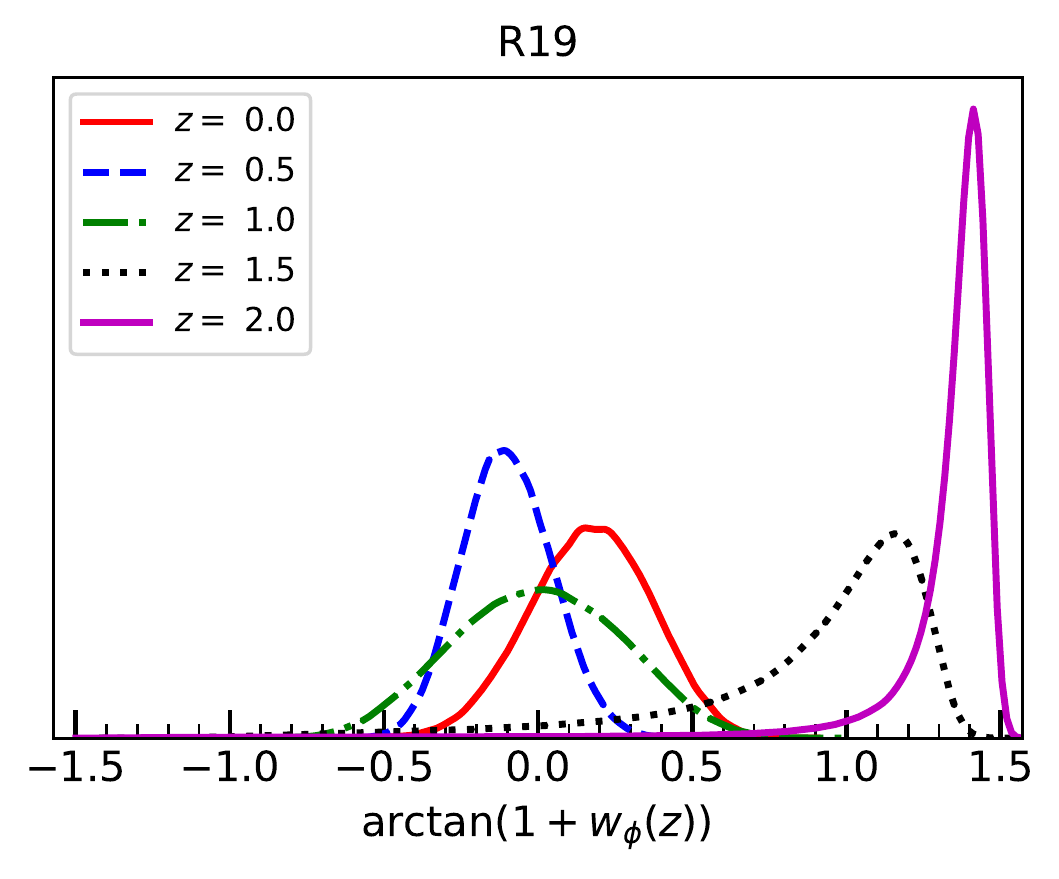}
		}
	\subfigure[ ]{
		\includegraphics[width = 0.4 \textwidth]{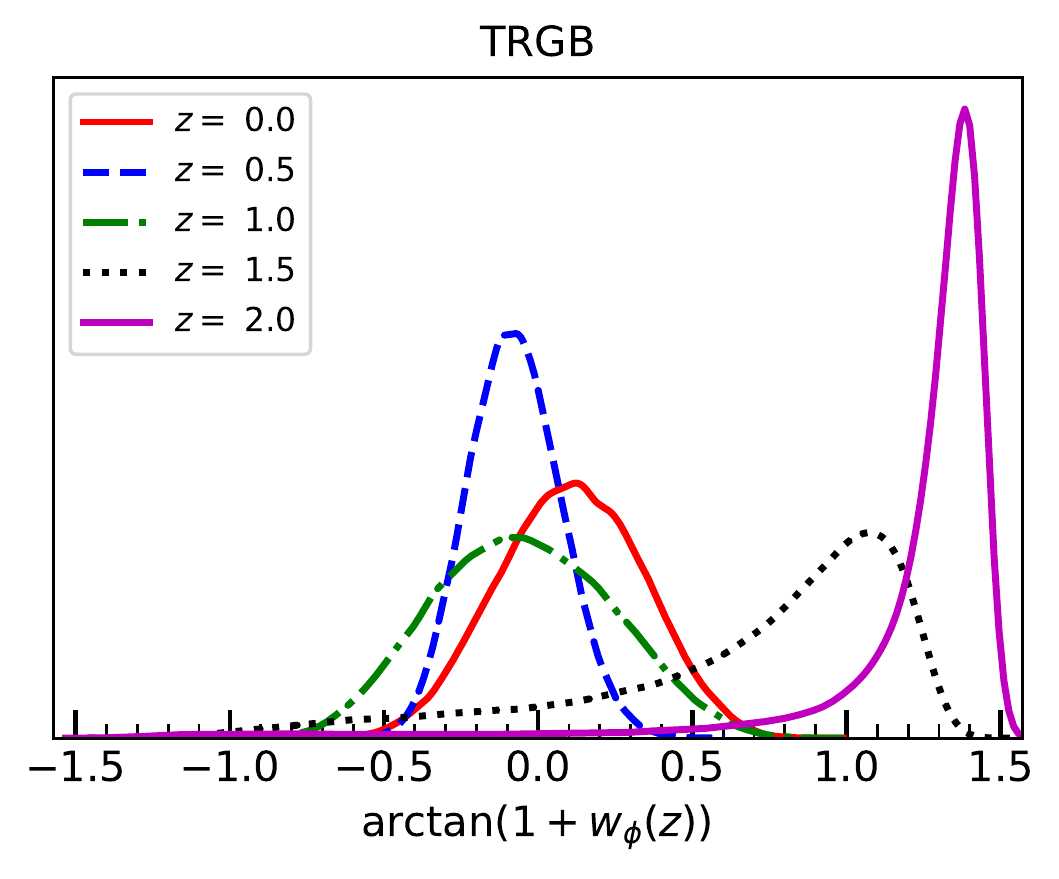}
		}
	\subfigure[ ]{
		\includegraphics[width = 0.4 \textwidth]{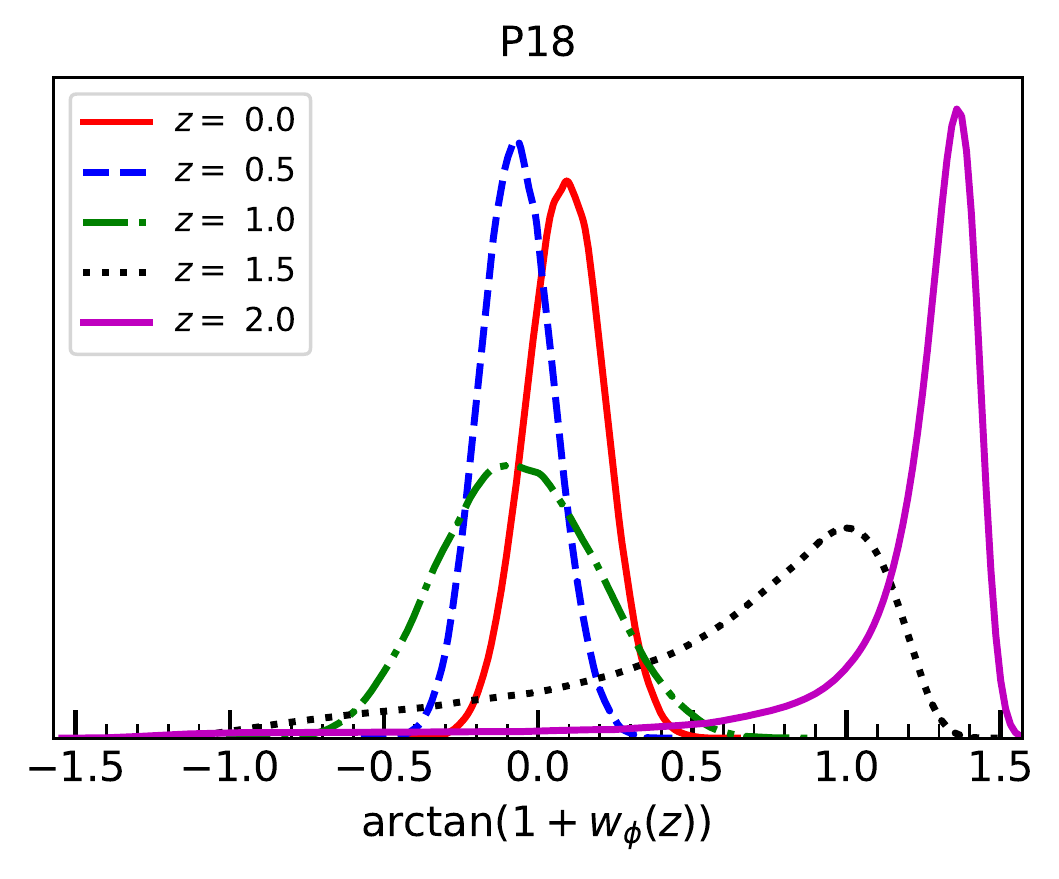}
		}
\caption{(a) The compactified HDES's dark energy equation of state, $\arctan\left(1 + w_\phi(z)\right)$, for varying $H_0$ prior as a function of the redshift. The solid, dashed, and dash-dotted lines represent the median of the distribution and the filled-hatched regions show the $34.1\%$ of the probability mass surrounding the median from both sides. Hatches used: $\left( '-' : H_0^\text{R19} \right)$ \cite{Riess:2019cxk}, $\left( '|' : H_0^\text{TRGB} \right)$ \cite{Freedman:2019jwv}, $\left( '\times' : H_0^\text{P18} \right)$ \cite{Aghanim:2018eyx}. Posteriors of the compactified dark energy equation of state at sample redshifts $z = 0, 0.5, 1, 1.5, 2$ for the (b) R19, (c) TRGB, and (d) P18 $H_0$ priors, respectively.}
\label{fig:w_de_hdes}
\end{figure}

The dark energy equation of state can also be computed. In general, in shift symmetric kinetic gravity braiding, this is given by
\begin{equation}
\label{eq:w_de_kgb}
    w_\phi = \dfrac{ - K + \sqrt{2X} \dot{X} G_X }{ K - 2 X \left( K_X + 3 \sqrt{2X} H(X) G_X \right)  }\,,
\end{equation}
where the potentials $K$ and $G$ and their derivatives are evaluated at $X$. By substituting the HDES solution to Eq.~(\ref{eq:w_de_kgb}), we obtain
\begin{equation}
\label{eq:w_de_hdes}
w_\phi = -1 + \dfrac{ \mathcal{J} \sqrt{2X} \left( H(z)^2 - H_0^2 \Omega_\Lambda \right) }{ 3 H_0^4 \Omega_{m0} \Omega_\Lambda } - \dfrac{ 2 \mathcal{J} \sqrt{2X} (1 + z) H(z) H'(z) }{ 9 H_0^4 \Omega_{m0} \Omega_\Lambda }\,.
\end{equation}
Using the reconstructed Hubble function and its first derivative, we can therefore obtain the dark energy equation of state. This is shown in compactified version, like we did with the quintessence case, in Fig.~\ref{fig:w_de_hdes}.
Interestingly, the compactified dark energy equation of state here retains single-modality for all redshifts, unlike its quintessence counterpart. Above all, this result is showing a strong deviation from $\Lambda$CDM at higher redshifts. This is shown in Fig.~\ref{fig:w_de_hdes}(a) and is also revealed more closely in Figs.~\ref{fig:w_de_hdes}-(b-d). In particular, the posteriors of the probability distribution for the $z = 1.5$ and $z = 2$ cases for all $H_0$ priors are strongly concentrated outside of the $\Lambda$CDM limit ($\arctan \left( 1 + w_\phi \right) = 0$). This may have been expected due to the $k$-essence and braiding potentials growing at higher redshifts (Figs.~\ref{fig:hdes}-(a) and (c)); nonetheless, this is worth highlighting as it points to potential future work with perturbations. Consider as an example the case of structure formation where the braiding potential may significantly affect.

To end the section, we recall that Fig. \ref{fig:w_de_hdes} shows the posterior of $\arctan \left(1 + w_\phi (z) \right)$ and not $w(z)$. So while the error bands do indeed become smaller in $\arctan \left( 1 + w_\phi (z) \right)$ for high redshifts, the corresponding uncertainty in $w(z)$ in fact becomes larger. In particular, Figs. \ref{fig:w_de_hdes}(b-d) show that the posterior for $z \sim 2$ is nearly localized at $\arctan \left( 1 + w(z) \right) \sim \pi/2$ which corresponds to very large mean values and uncertainties in $w(z)$. We clarify this further in Appendix \ref{sec:stat_compact_rv} where the posterior of a random variable and its compactified version are displayed side by side.

\subsection{Tailoring Horndeski}
\label{subsec:tailoring_horndeski}

\textit{Tailoring Horndeski} \cite{Bernardo:2019vln} is related to both quintessence potential and HDES models and is amenable to this reconstruction method. This can be considered as the $J(t) = 0$ limit of HDES but it works more similar with quintessence in the sense that one does not have to rely on an \textit{a priori} $H(X)$ in order to single out a particular potential \footnote{In Ref.~\cite{Arjona:2019rfn}, it was mentioned that the $J(t) = 0$ limit of HDES is $\Lambda$CDM.
This was based on naively making the shift charge $\mathcal{J}$ vanish in Eqs.~(\ref{eq:K_hdes}) and (\ref{eq:GX_hdes}). However, this limit should be merely realized to be just an artifact of the particular parametrization of the HDES solution (Eqs.~(\ref{eq:K_hdes}) and (\ref{eq:GX_hdes})). In general, with $\mathcal{J} = 0$, one should start again with Eqs.~(\ref{eq:Feq_hdes}) and (\ref{eq:J_hdes}), from which HDES was derived. This will inevitably lead to the tailoring Horndeski approach.}.

The field equations of the theory are given by
\begin{equation}
\label{eq:Feq_tail}
    3 H^2 = \rho + 2 \Lambda + \dfrac{\dot{\phi}^2}{2} + 3 H \dot{\phi}^3 G_X\,,
\end{equation}
\begin{equation}
\label{eq:Peq_tail}
    2 \dot{H} + 3 H^2 = -P + 2 \Lambda - \dfrac{\dot{\phi}^2}{2} + G_X \dot{\phi}^2 \ddot{\phi}\,,
\end{equation}
and
\begin{equation}
\label{eq:Seq_tail}
    \ddot{\phi} \left(-3 H \left(G_{XX} \dot{\phi}^3+2 G_X \dot{\phi}\right)-1\right)-3 \dot{\phi} \left(G_X \dot{ H } \dot{ \phi }+3 G_X H^2 \dot{\phi}+H \right) = 0\,.
\end{equation}
Note that the above equations can be obtained from Eqs.~(\ref{eq:Feq_hdes}), (\ref{eq:Peq_hdes}), and (\ref{eq:Seq_hdes}) by substituting $K(X) = X - 2 \Lambda$. The solution of the system will therefore also lie on the dynamical hypersurface determined by Eq.~(\ref{eq:J_hdes}). In contrast, the tailoring Horndeski approach recognizes the $J(t) = 0$ as a dynamical attractor and so builds the solution on top of this hypersurface. One can then write down
\begin{equation}
\label{eq:GX_tail}
G_X (X) = - \dfrac{1}{3 \sqrt{2 X} H(X)}\,.
\end{equation}
Since there is only one potential, i.e., $G_X$, Eq.~(\ref{eq:GX_tail}) can be used to obtain model independent necessary conditions by substituting into Eqs.~(\ref{eq:Feq_tail}) and (\ref{eq:Peq_tail}). The emerging necessary conditions are given by
\begin{equation}
\label{eq:X_tail}
X = 3 \left( H_0^2 \Omega_m(z) + H_0^2 \Omega_\Lambda - H^2 \right)\,,
\end{equation}
and
\begin{equation}
\label{eq:345_tail}
2 \dot{H} + 3 H^2 = -P + 3 H_0^2 \Omega_\Lambda - X - \dfrac{\dot{X}}{3 H}\,,
\end{equation}
where $\Omega_m(z) = \rho(z)/\left(3 H_0^2\right)$ and $\Omega_\Lambda = 2\Lambda/\left(3 H_0^2\right)$. The kinetic density can then be uniquely determined for a given Hubble function by using Eq.~(\ref{eq:X_tail}). Moreover, by eliminating $X$ in Eqs.~(\ref{eq:X_tail}) and (\ref{eq:345_tail}), one can recover the thermodynamic conservation law
\begin{equation}
\dot{\rho} + 3 H \left( \rho + P \right) = 0\,\
\end{equation}
which ensures the consistency of the method. In summary, the braiding potential and the scalar field's kinetic density can be uniquely assigned to a given Hubble function through Eqs.~(\ref{eq:GX_tail}) and (\ref{eq:X_tail}).

\begin{figure}[h!]
\center
\includegraphics[width = 0.45 \textwidth]{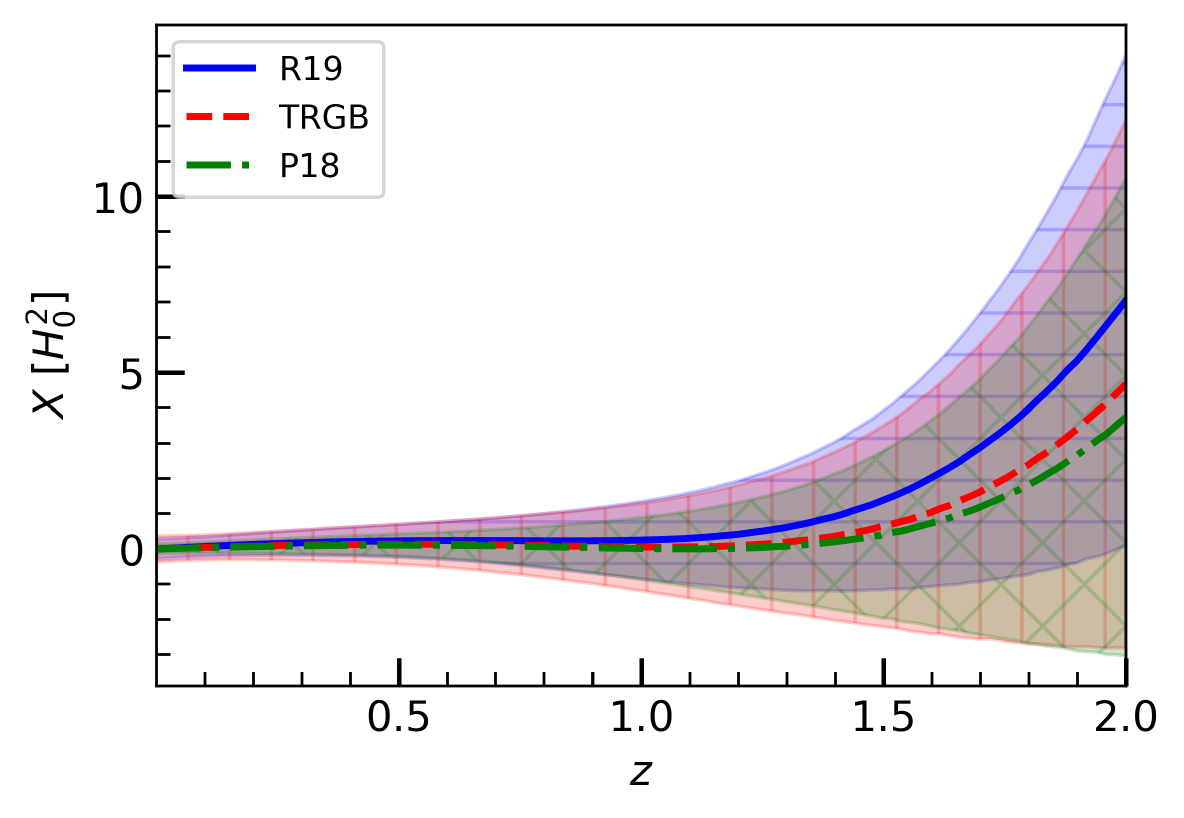}
\caption{Tailoring Horndeski's kinetic density $X\left(z\right)$ as a function of the redshift for varying $H_0$ prior. The filled-hatched regions show the $2\sigma$ confidence intervals for each prior. Hatches used: $\left( '-' : H_0^\text{R19} \right)$ \cite{Riess:2019cxk}, $\left( '|' : H_0^\text{TRGB} \right)$ \cite{Freedman:2019jwv}, $\left( '\times' : H_0^\text{P18} \right)$ \cite{Aghanim:2018eyx}.}
\label{fig:X_tail}
\end{figure}

\begin{figure}[h!]
\center
	\subfigure[ ]{
		\includegraphics[width = 0.45 \textwidth]{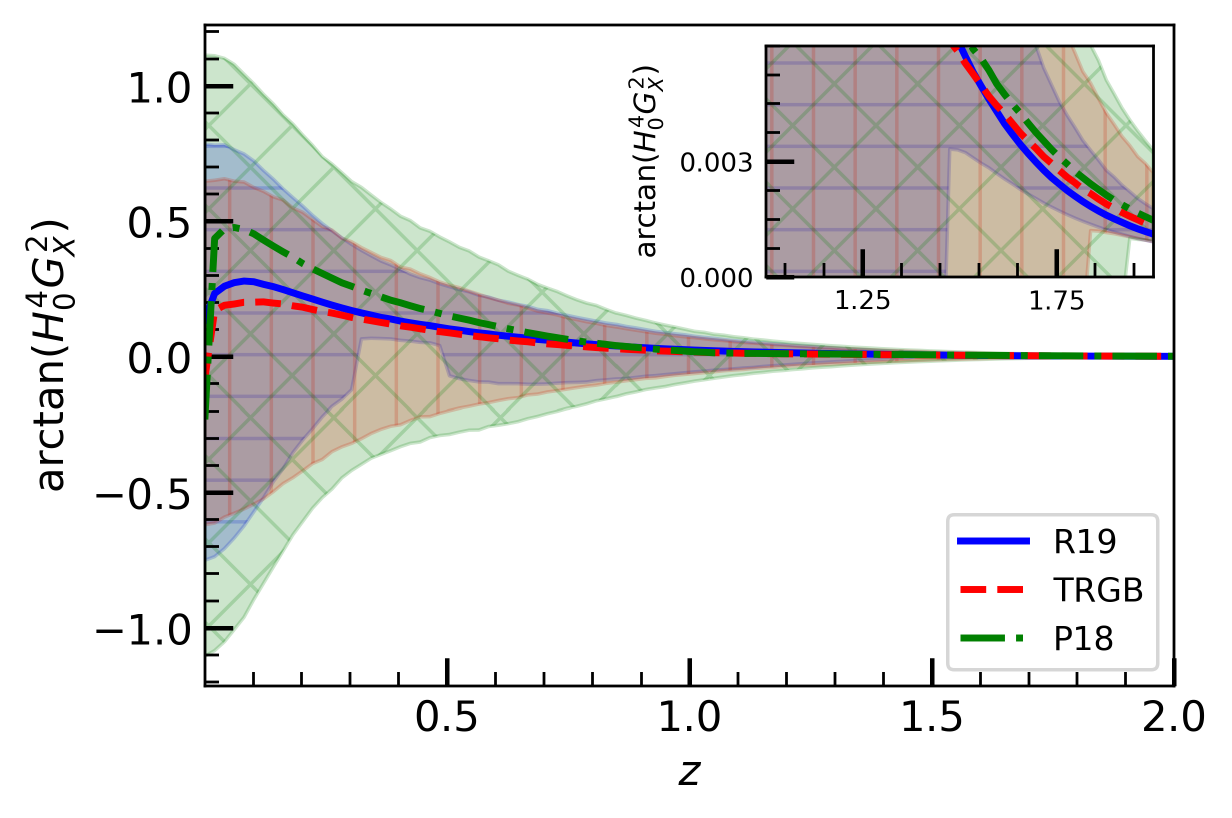}
		}
	\subfigure[ ]{
		\includegraphics[width = 0.45 \textwidth]{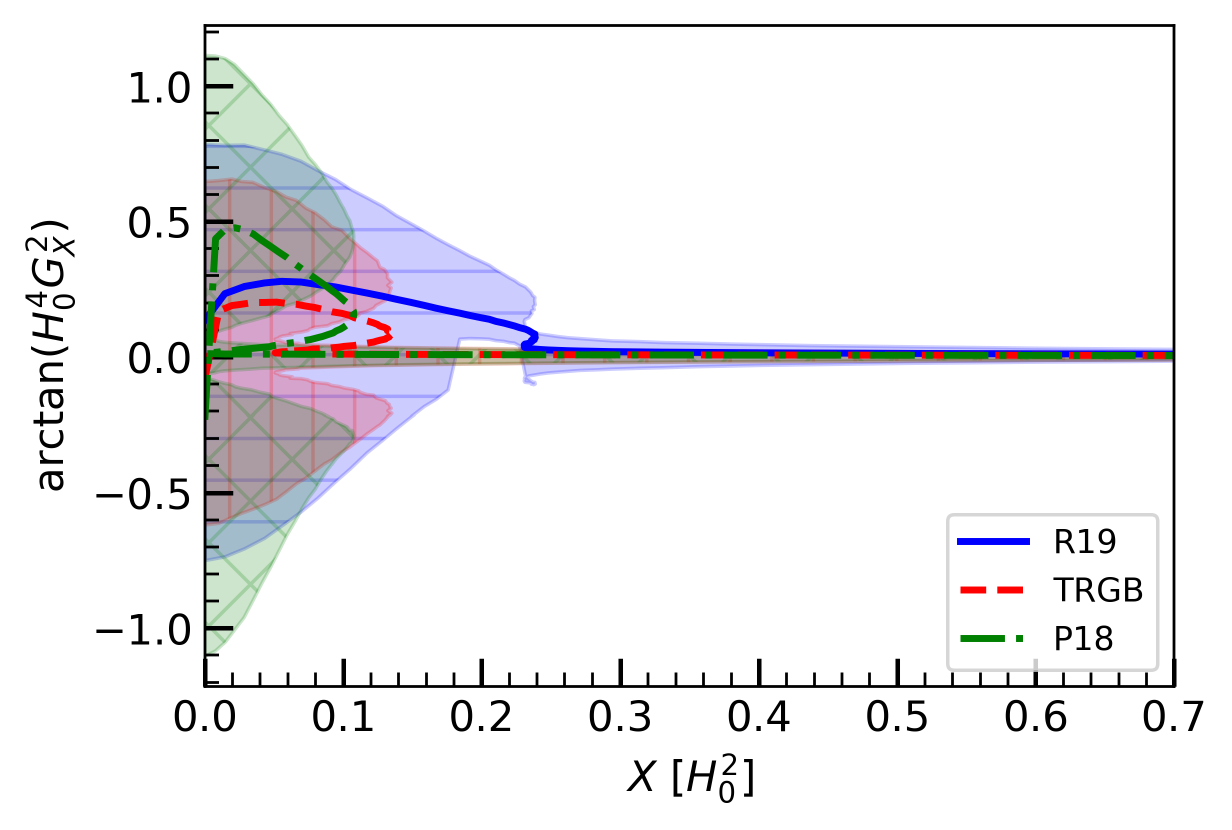}
		}
\caption{The compactified braiding potential, $\tilde{G}_X^2$ (Eq.~(\ref{eq:compactified_braiding})), in GP tailoring Horndeski as a function of the (a) redshift $z$ and (b) the kinetic density $X$. The solid, dashed, dash-dotted lines show the median of the distribution while the filled-hatched regions show the $34.1\%$ of the probability mass lying above and below the median. Hatches used: $\left( '-' : H_0^\text{R19} \right)$ \cite{Riess:2019cxk}, $\left( '|' : H_0^\text{TRGB} \right)$ \cite{Freedman:2019jwv}, $\left( '\times' : H_0^\text{P18} \right)$ \cite{Aghanim:2018eyx}. The inset of (a) is a closeup of the region $z \in \left(1, 2\right)$ and $\tilde{G}_X^2 \in \left(0, 6 \times 10^{-3}\right)$.}
\label{fig:GX_tail}
\end{figure}

\begin{figure}[h!]
\center
\subfigure[ ]{
		\includegraphics[width = 0.4 \textwidth]{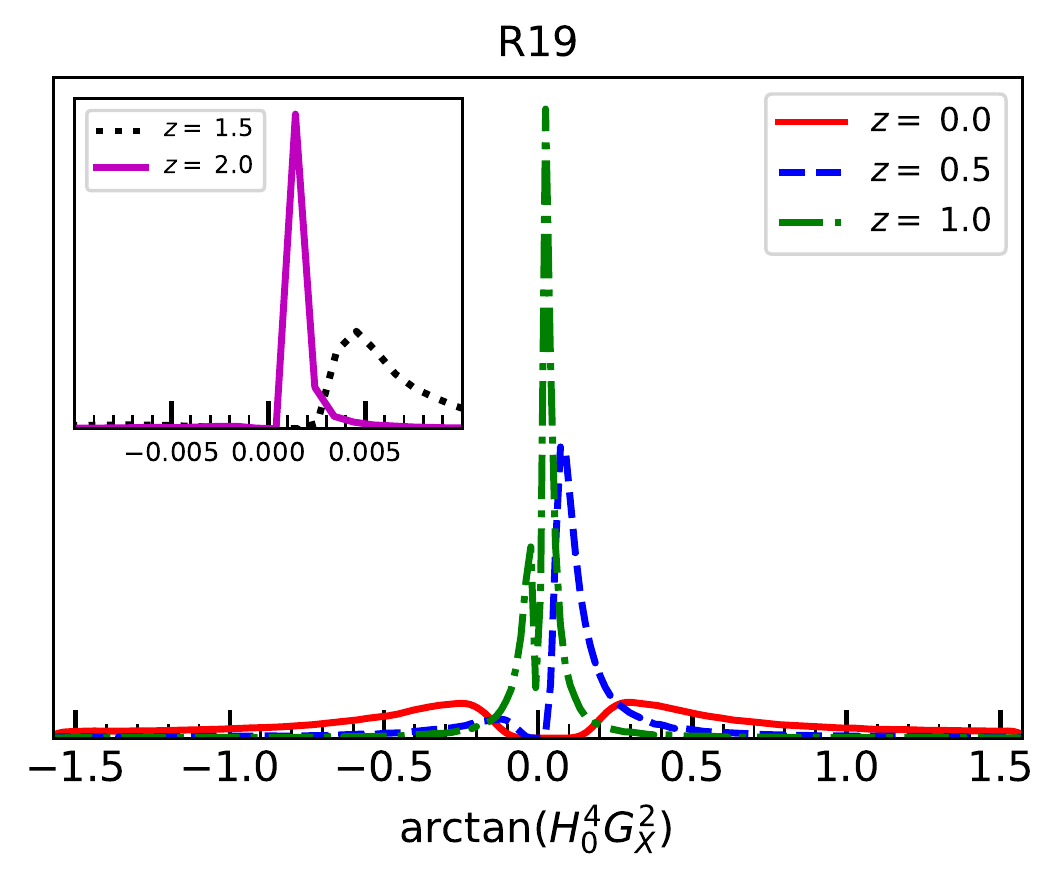}
		}
	\subfigure[ ]{
		\includegraphics[width = 0.4 \textwidth]{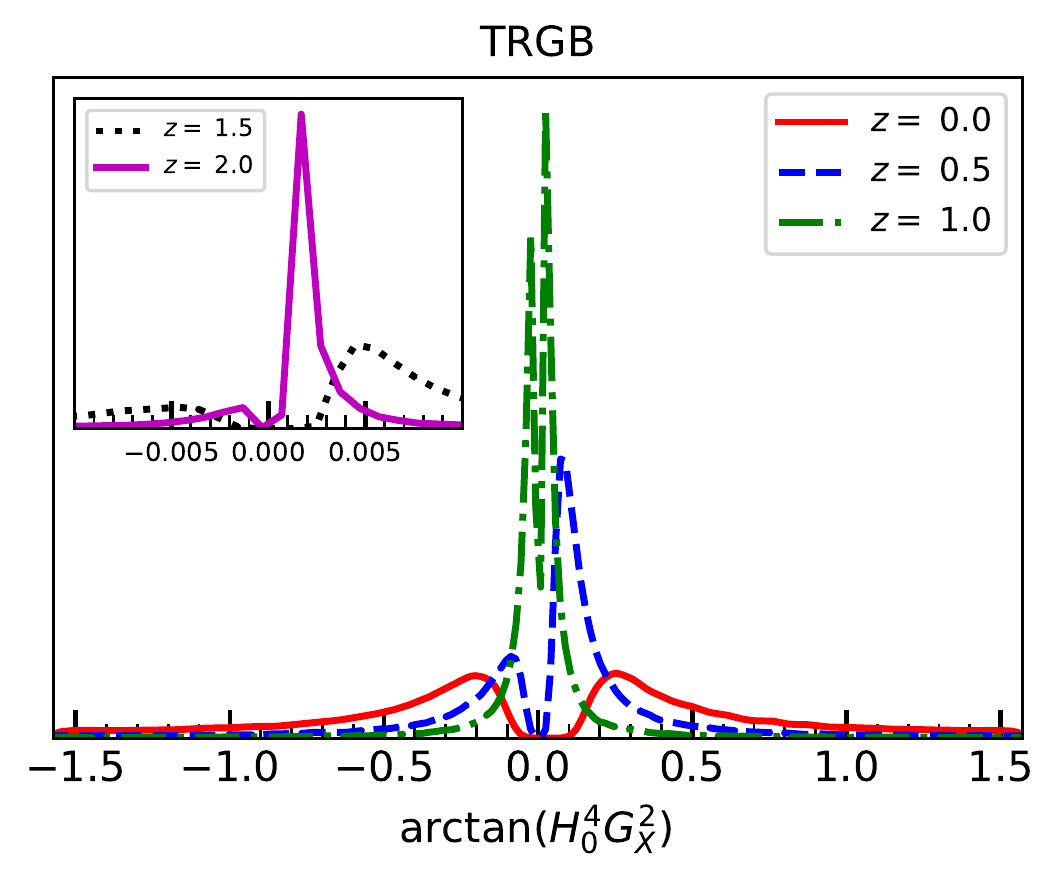}
		}
	\subfigure[ ]{
		\includegraphics[width = 0.4 \textwidth]{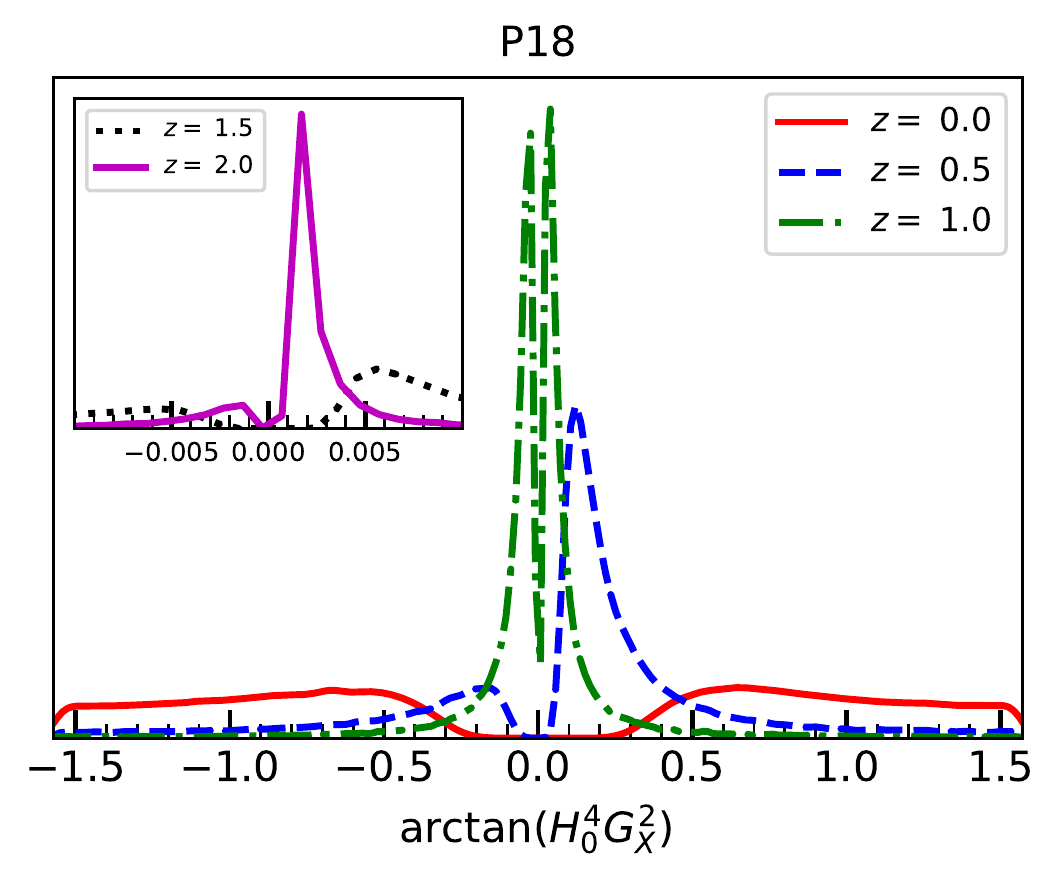}
		}
	\subfigure[ ]{
		\includegraphics[width = 0.4 \textwidth]{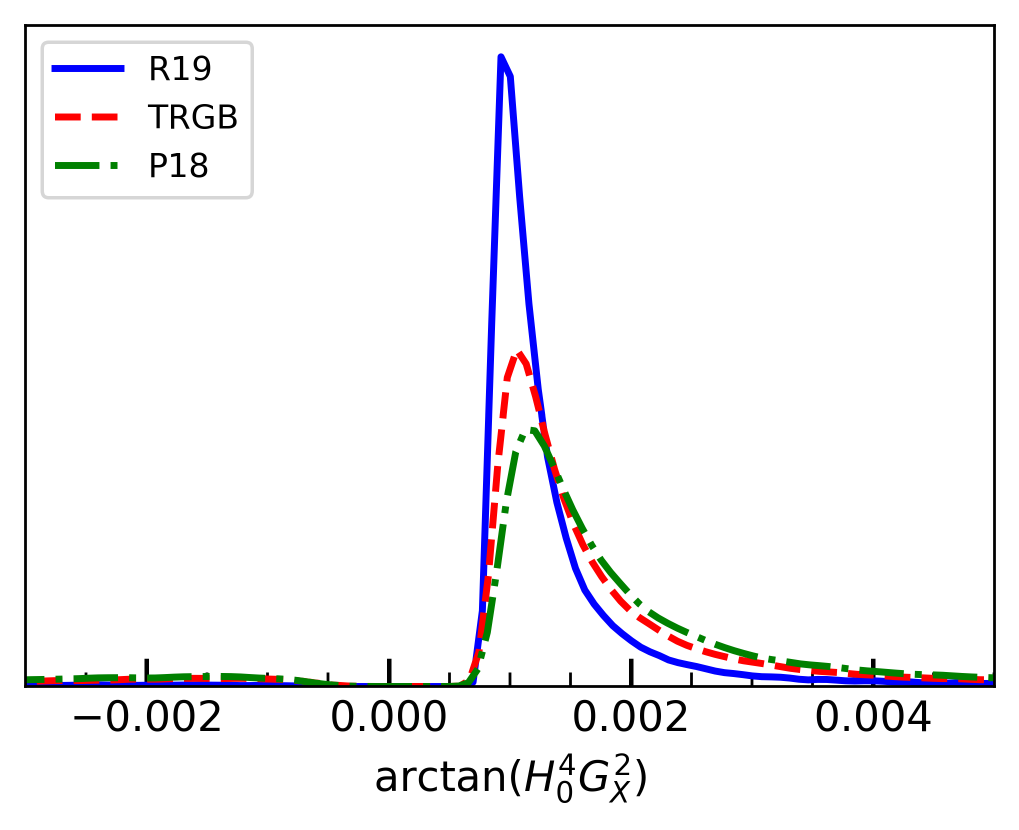}
		}
\caption{Snapshots of the posterior of the compactified braiding potential in tailoring Horndeski at sample redshifts $z = 0, 0.5, 1, 1.5, 2$ for the (a) R19, (b) TRGB, and (c) P18 $H_0$ priors. Inset plots separately show the $z = 1.5$ and $z = 2$ cases because the posterior for these two cases scale too short compared with the later redshift cases. (d) Posterior distribution at $z = 2$ for each $H_0$ prior.}
\label{fig:atGX2_samps_tail}
\end{figure}

By complementing the above outlined tailoring Horndeski method with the GP reconstructed Hubble function, we now obtain predictions of the theory. The resulting kinetic density is shown in Fig.~\ref{fig:X_tail}.
Here, we also consider the P18 cosmological parameter values, and so, once more, by expressing dimensionful quantities in units of $H_0$, we find consistent shapes of the predicted functions regardless of the choice of $H_0$ prior. It is noteworthy that the scalar field has spent most of its late-time dynamics with $ | X / H_0^2 | \ll 1$. This is important as the braiding potential (Eq.~(\ref{eq:GX_tail})) possesses a singularity at $X = 0$ and so MC sampling over $G_X$ can be expected to result to posteriors with very large, consequently unreliable, uncertainties. Fig.~\ref{fig:X_tail} also shows that an appreciable size of the samples fall to the region $X < 0$. This is problematic from a computational point of view when sampling $G_X$ because of the square root in Eq.~(\ref{eq:GX_tail}). To avoid the above mentioned issues, we proceed by additionally sampling over a compactified, real, random variable, $\tilde{G}_X^2$ given by
\begin{equation}\label{eq:compactified_braiding}
    \tilde{G}_X^2 = \arctan\left( \dfrac{H_0^4}{18 X H\left(X \right)^2} \right)\,.
\end{equation}
We refer to this as a compactified braiding potential. The results are shown in Fig.~\ref{fig:GX_tail}.
Here, it is shown that $\tilde{G}_X^2$ may also accept negative values at low redshifts (or low $X$), whenever $X < 0$ can be drawn with appreciable probability. In such cases, the braiding potential is going to be undefined, or rather that the model is disfavored by the data. However, at higher redshifts, the distribution tends to become more concentrated towards positive values. The inset of of Fig.~\ref{fig:GX_tail}-(a) proves this. As complementary to Fig.~\ref{fig:GX_tail}, snapshots of the posterior distribution of the distributions of $\tilde{G}_X^2$ at redshifts $z = 0, 0.5, 1, 1.5, 2$ are shown in Fig.~\ref{fig:atGX2_samps_tail}.
This shows that the distribution of $\tilde{G}_X^2$ is generally bimodal and supports the use of the median and its $34.1\%$ surrounding mass as a reasonable statistic. At $z = 0$, Figs.~\ref{fig:atGX2_samps_tail}-(a-c) show that $\tilde{G}_X^2$ may almost-equally be positive or negative and samples may even be drawn with appreciable probability near the boundaries, $\tilde{G}_X^2 \sim \pm \pi/2$, of the random variable. Interestingly, at higher, or earlier, redshifts, the probability mass at negative $\tilde{G}_X^2$ always decreases with an increase in the redshift, i.e., the weight of the samples drawn at negative $\tilde{G}_X^2$ becomes smaller for increasing $z$. This is a consistent result throughout the $H_0$ priors and so suggests that the braiding may have played a relevant earlier minor role in cosmic history. Fig.~\ref{fig:atGX2_samps_tail}-(d) supports this interpretation. Indeed, at $z = 2$, the posteriors of $\tilde{G}_X^2$ can already be considered to be practically single-modal and undeniably heavier at $\tilde{G}_X^2 > 0$, where the braiding potential $G_X$ can be computed. A crude estimate of $H_0^2 G_X$, based on Fig.~\ref{fig:atGX2_samps_tail}-(d) and $\arctan\left( x \ll 1 \right) \sim x$, leads to $H_0^2 G_X \sim 0.1$ at $z \sim 2$.

\begin{figure}[h!]
\center
\subfigure[ ]{
		\includegraphics[width = 0.45 \textwidth]{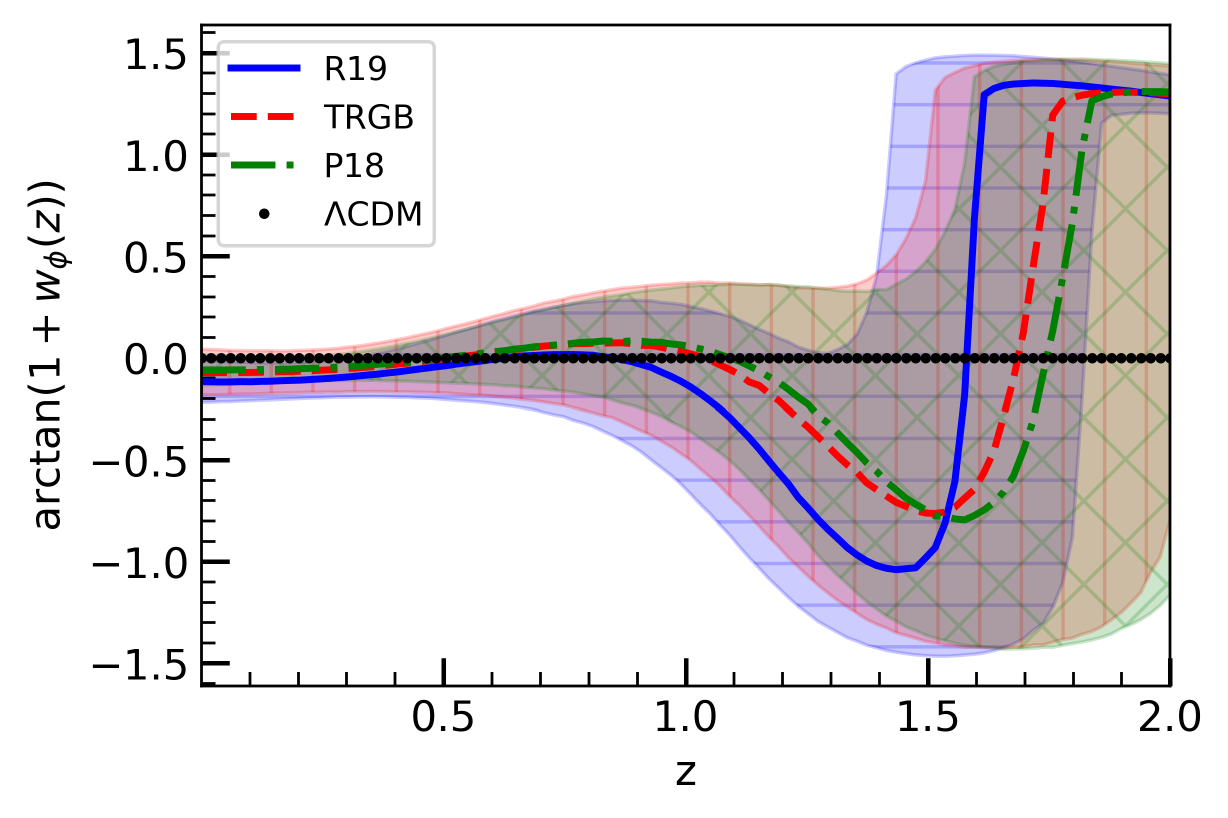}
		}
	\subfigure[ ]{
		\includegraphics[width = 0.4 \textwidth]{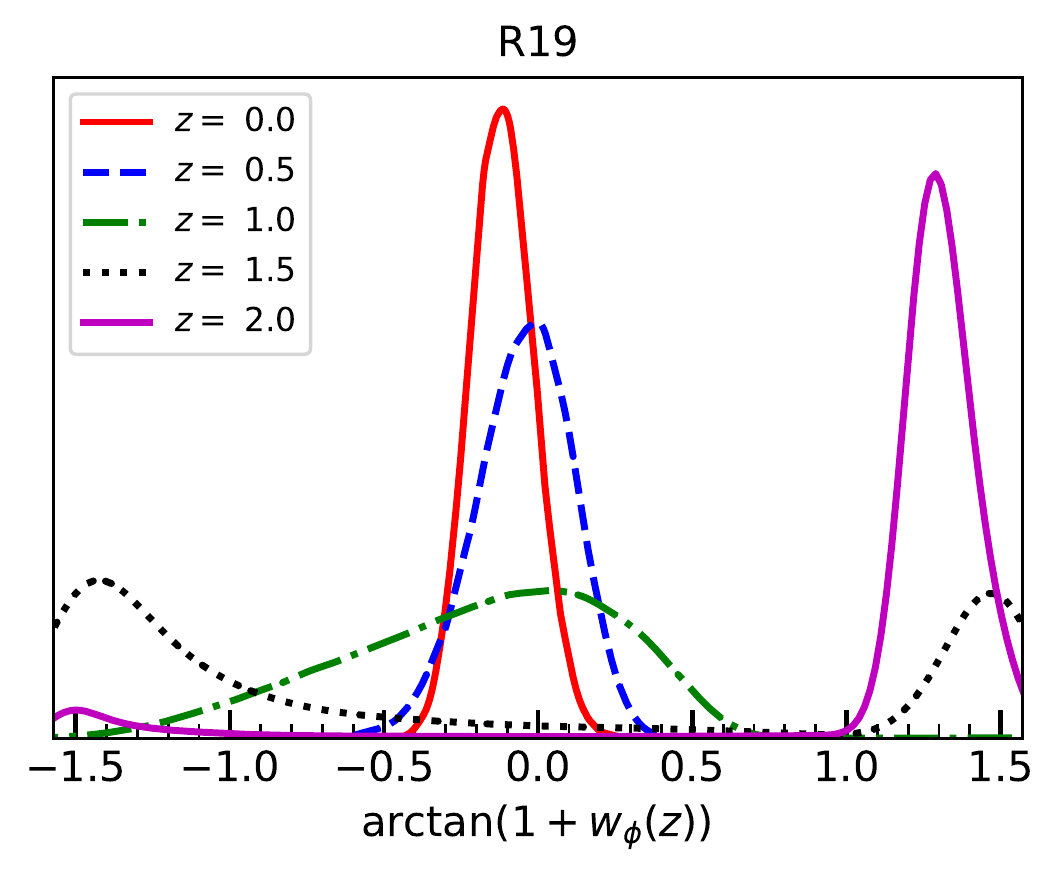}
		}
	\subfigure[ ]{
		\includegraphics[width = 0.4 \textwidth]{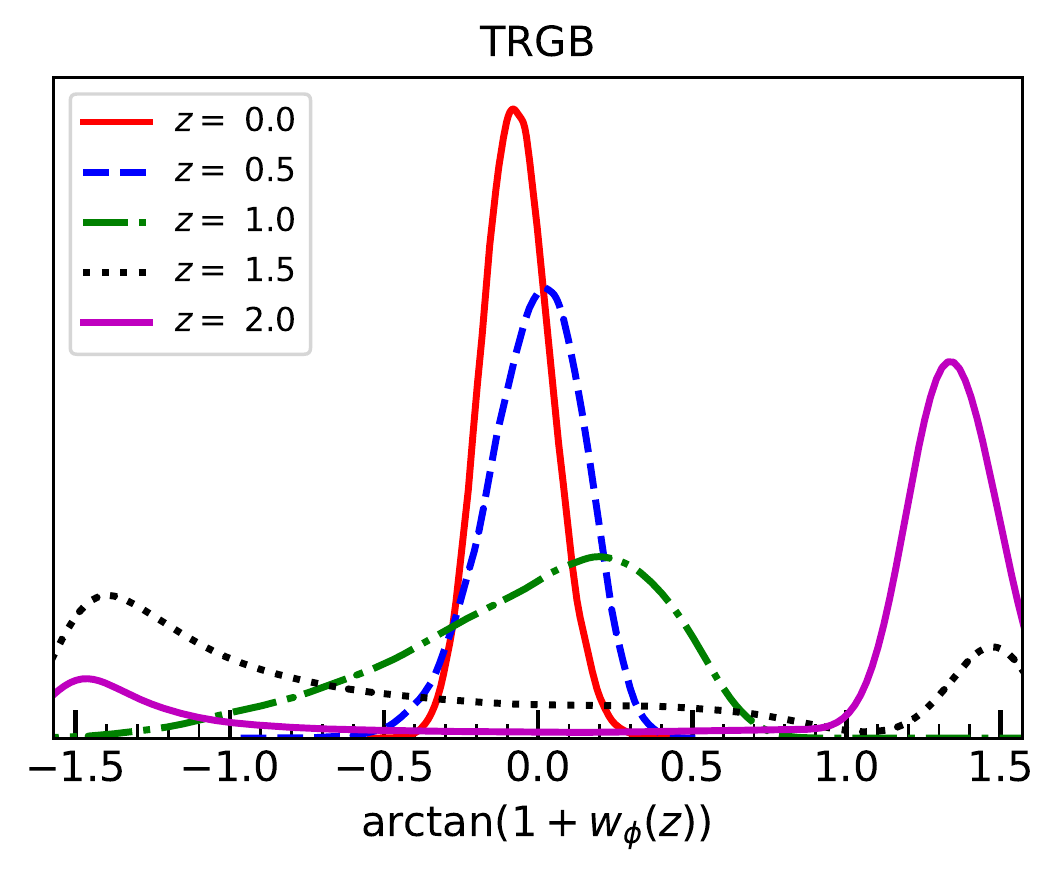}
		}
	\subfigure[ ]{
		\includegraphics[width = 0.4 \textwidth]{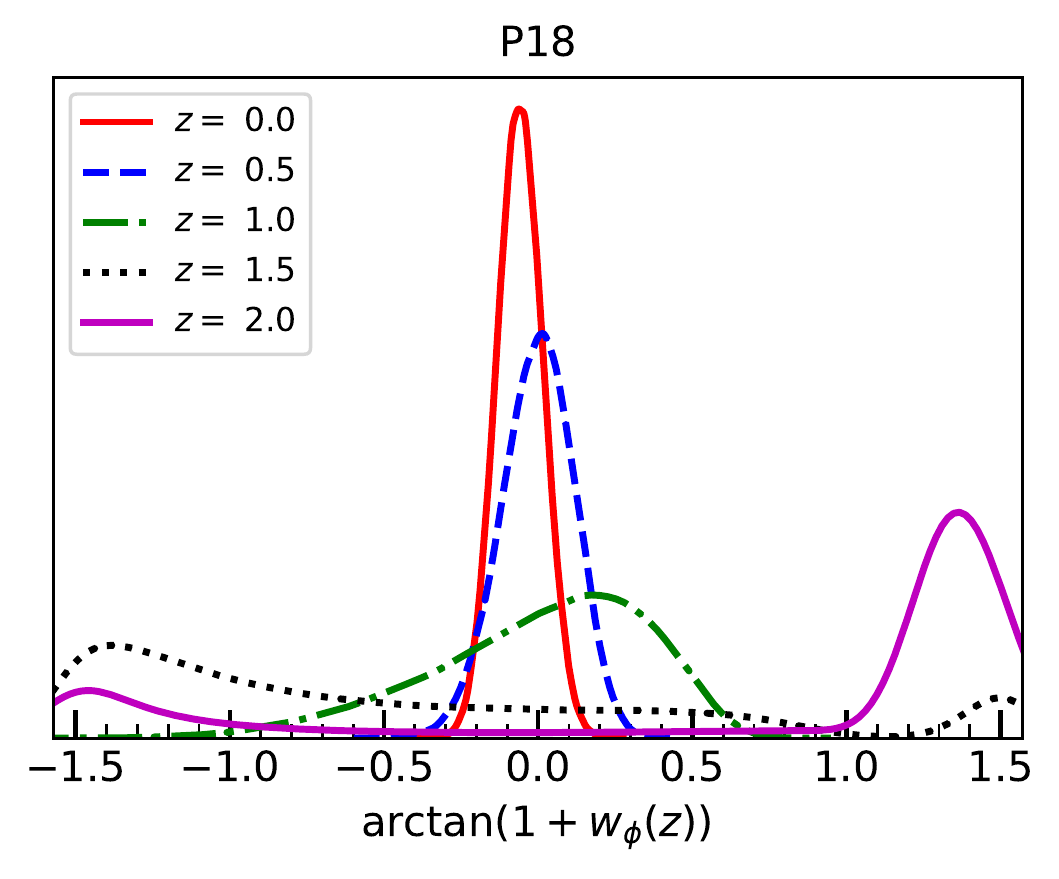}
		}
\caption{(a) The compactified dark energy equation of state, $\arctan\left(1 + w_\phi(z)\right)$, as a function of the redshift in tailoring Horndeski for varying $H_0$ prior. The solid, dashed, and dash-dotted lines represent the median of the distribution and the filled-hatched regions show the $34.1\%$ of the probability mass surrounding the median from both sides. Hatches used: $\left( '-' : H_0^\text{R19} \right)$ \cite{Riess:2019cxk}, $\left( '|' : H_0^\text{TRGB} \right)$ \cite{Freedman:2019jwv}, $\left( '\times' : H_0^\text{P18} \right)$ \cite{Aghanim:2018eyx}. Posteriors of the compactified dark energy equation of state at sample redshifts $z = 0, 0.5, 1, 1.5, 2$ for the (b) R19, (c) TRGB, and (d) P18 $H_0$ priors.}
\label{fig:w_de_tail}
\end{figure}

The dark energy equation of state in tailoring Horndeski can also be computed using Eq.~(\ref{eq:w_de_kgb}). Substituting $K(X) = X - 2\Lambda$ and then using the tailored solution (Eqs.~(\ref{eq:GX_tail}) and (\ref{eq:X_tail})), it is straightforward to obtain
\begin{equation}
\label{eq:w_de_tail}
w_\phi(z) = \dfrac{ H(z) \left( 3 H(z) - 2 (1 + z) H'(z) \right) }{ 3 \left( -H(z)^2 + H_0^2 \Omega_{m0} \left(1 + z\right)^3 \right) }\,.
\end{equation}
This is notably the exact same expression that can be obtained by assuming that dark energy is a perfect fluid with an equation of state $w(z)$. The results of the sampling over the compactified dark energy equation of state is shown in Fig.~\ref{fig:w_de_tail}.
It is most interesting to point out that this looks very similar to Fig.~\ref{fig:w_de_tail} even though the earlier analysis of quintessence did not take into account $\Omega_\Lambda$. Similar conclusions can therefore be drawn. First, at low redshifts, the posterior distributions of the compactified dark energy equation of state can be considered to be approximately Gaussian distributed. However, at higher redshifts, this cannot anymore be taken to be true as the distribution becomes bimodal and an appreciable fraction of samples fall too close to the boundaries $\sim \pm \pi/2$ of the compactified region. In particular, for $z = 2$ in Figs.~\ref{fig:w_de_tail}-(b-d), the probability mass appear to be mostly heavier at the positive side, i.e., $\arctan \left( 1 + w_\phi \right) \sim 1$, suggesting a deviation from $\Lambda$CDM. The implications of this should be further examined in a future work.

\subsection{Constraints on the dark energy equation of state}
\label{subsec:w_de_constraints}

We now summarize the constraints on the dark energy equation of state at $z = 0$ obtained in this work in Table~\ref{tab:w_de_constraints}.
\begin{table}
\center
\caption{Constraints on the dark energy equation of state in Horndeski cosmology. The columns $H_0^\text{R19}$, $H_0^\text{TRGB}$, and $H_0^\text{P18}$ stand for the GP analysis using the corresponding $H_0$ priors R19, TRGB, and P18. The P18 prior parameter values were assumed in this analysis \cite{Aghanim:2018eyx}. For designer Horndeski, $c_0 = H_0^{n + 2}$, $n = 1$, and $\mathcal{J} = H_0$ were additionally assumed.}
\begin{tabular}{| c | c | c | c |}
\hline
\phantom{ $\dfrac{1}{1}$ } \phantom{ $\dfrac{1}{1}$ } & \multicolumn{3}{c}{ $w_\text{DE}\,(z = 0)$ } \vline \\
\hline

\phantom{ $\dfrac{1}{1}$ } \textit{Theory} + parameters \phantom{ $\dfrac{1}{1}$ } & $H_0^\text{R19}$ & $H_0^\text{TRGB}$ & $H_0^\text{P18}$ \\ \hline \hline

\phantom{ $\dfrac{1}{1}$ } \textit{Quintessence} $+ \left( \Omega_{m0} \right)$ \phantom{ $\dfrac{1}{1}$ } & $-1.1 \pm 0.1$  & $-1.1 \pm 0.1$ & $-1.06 \pm 0.08$ \\ \hline

\phantom{ $\dfrac{1}{1}$ } \textit{Designer Horndeski} $+\left( \Omega_{m0}, \Omega_\Lambda, c_0, n, \mathcal{J} \right)$ \phantom{ $\dfrac{1}{1}$ } & $-0.8 \pm 0.2$ & $-0.9 \pm 0.3$ & $-0.9 \pm 0.1$ \\ \hline

\phantom{ $\dfrac{1}{1}$ } \textit{Tailoring Horndeski} $+ \left( \Omega_{m0}, \Omega_\Lambda \right)$ \phantom{ $\dfrac{1}{1}$ } & $-1.1 \pm 0.1$ & $-1.1 \pm 0.1$ & $-1.06 \pm 0.08$ \\ \hline \hline

\phantom{ $\dfrac{1}{1}$ } $\Lambda$CDM \phantom{ $\dfrac{1}{1}$ } & \multicolumn{3}{c}{-1} \vline \\ \hline

\phantom{ $\dfrac{1}{1}$ } $w_0$CDM (Planck + SNe + BAO) \phantom{ $\dfrac{1}{1}$ } & \multicolumn{3}{c}{ $-1.03 \pm 0.03$ \cite{Aghanim:2018eyx} } \vline \\ \hline

\phantom{ $\dfrac{1}{1}$ } $w_0 w_a$CDM (Planck + SNe + BAO) \phantom{ $\dfrac{1}{1}$ } & \multicolumn{3}{c}{ $-0.96 \pm 0.08$ \cite{Aghanim:2018eyx} } \vline \\ \hline
\end{tabular}
\label{tab:w_de_constraints}
\end{table}
The $w_\text{DE}\,(z = 0)$ constraints from quintessence and tailoring Horndeski differ only in their fifth significant digit which is well within the predictability region that can be reasonably expected from the reconstruction method. For both of these theories, the dark energy equation of state remains consistent with each other and regardless of the $H_0$ prior. In this background the priors on the Hubble data have little to no impact on the equation of state reconstructions since these fine differences will be suppressed by the precise formula that expresses the quantity. The situation is entirely different for the direct reconstruction of the Hubble diagram where priors not only have a significant impact on the diagrams that are produced but also on the reconstructed values of $H_0$ which is one of the most important results from this reconstruction method.

On the other hand, the $w_\text{DE}(z = 0)$ constraints coming from designer Horndeski should be taken more carefully as it strongly depends on the assumed shift charge $\mathcal{J}$ and so is not completely predictive with just the low-redshift data. Since the shift charge is a constant throughout the evolution, it would have been then highly coincidental if it scales with the Hubble parameter today. The shift charge can instead be constrained with the CMB as done in Ref.~\cite{Arjona:2019rfn}. Alternatively, combining the low redshift observations from Hubble data and $f \sigma_8$ can be considered to constrain the shift charge.

It is also interesting to point out the recent work of Ref. \cite{Banerjee:2020xcn} which suggests that quintessence is at odds with local determinations of $H_0$. Extending this perturbative analysis to broader Horndeski theories may lead to a better understanding of dark energy and the $H_0$ tension.

\section{Conclusion}
\label{sec:conclusions}

We first discuss briefly the related work of Ref.~\cite{Reyes:2021owe} which also focused on constraining Horndeski theory. Here, the novelty can be found in the GP reconstruction of $H(z)$ which exploited the use of observations of the function $d_p(z)$ and its first derivative $d_p'(z) = 1/H(z)$. The reconstructed Hubble function was then applied to $k$-essence and used to constrain a quintessence potential of the restricted form $V\left( \phi \right) = C \exp \left( \phi n \right)$.

In this paper, we combined the results from the GP approach to reconstruction and well-known inversions of the Friedmann equations in Horndeski theory in order to obtain predictions of completely \textit{unprescribed} Horndeski potentials. We particularly considered a combined data set from cosmic chronometers, supernovae, and baryon acoustic oscillations to construct the Hubble function using GP (Sec.~\ref{subsec:gp_application}) and later used this to single out Horndeski theories using three implementations: quintessence potential (Sec.~\ref{subsec:quintessence}), designer Horndeski (Sec.~\ref{subsec:designer_horndeski}), and tailoring Horndeski (Sec.~\ref{subsec:tailoring_horndeski}). In doing so, we obtained predictions of the potentials that are fully anchored on expansion history data. New constraints on the dark energy equation of state are presented in Table~\ref{tab:w_de_constraints}.

We also introduced a novel practical way of obtaining predictions of the dark energy equation of state $w_\text{DE}(z)$ for all redshifts $z$, i.e., by instead sampling over the compactified random variable $\arctan \left( 1 + w_\text{DE} (z) \right)$ (Figs.~\ref{fig:w_de_quint}, \ref{fig:w_de_hdes}, and \ref{fig:w_de_tail}). Through this, we were able to closely examine regions of the reconstructed function that would otherwise be spoiled by very large uncertainties because of the presence of a nearby singularity. The same technique was used to predict the braiding potential in tailoring Horndeski. We also argued that for a compactified random variable the median surrounded by $34.1\%$ of probability mass above and below it is a more reasonable statistic (Appendix \ref{sec:stat_compact_rv}). This is true for the dark energy equation of state which turned out to be generally bimodal. Improvements to this methodology are also interesting for future work.

Several more directions can be considered for future work. First, both quintessence and tailoring Horndeski theories have been completely specified by the Hubble data and prior values of $\Omega_m$ and $\Omega_\Lambda$, i.e., they have no more extra parameters left. Therefore, it is possible to put very tight constraints to both of these theories by comparing their predictions of the growth rate with observations. Similarly, the shift charge in designer Horndeski should be constrainable with additional information such as the existing $f \sigma_8$ data. Second, but not completely unrelated to the first, the perturbations of these theories should be examined for instabilities, e.g., ghost- and Laplace-type. In practice, the stability conditions are usually taken to be reasonable theoretical priors when sampling over parametrized dark energy theories and should also be integrable within GP reconstruction Horndeski implementations. Lastly, it would be interesting to further apply the GP inspired model reconstruction method to dark energy models outside of the Horndeski class such as degenerate higher-order scalar-tensor theories \cite{BenAchour:2016fzp} and vector-tensor theories \cite{DeFelice:2016yws, DeFelice:2016uil}.

\begin{acknowledgments}\label{sec:acknowledgements}
The authors would like to thank Eoin Colg\'ain for a helpful discussion on quintessence. JLS would like to acknowledge networking support by the COST Action CA18108 and funding support from Cosmology@MALTA which is supported by the University of Malta.
\end{acknowledgments}

\appendix

\section{Statistics for a compactified random variable}
\label{sec:stat_compact_rv}

\begin{figure}[h!]
\center
	\subfigure[ ]{
		\includegraphics[width = 0.4 \textwidth]{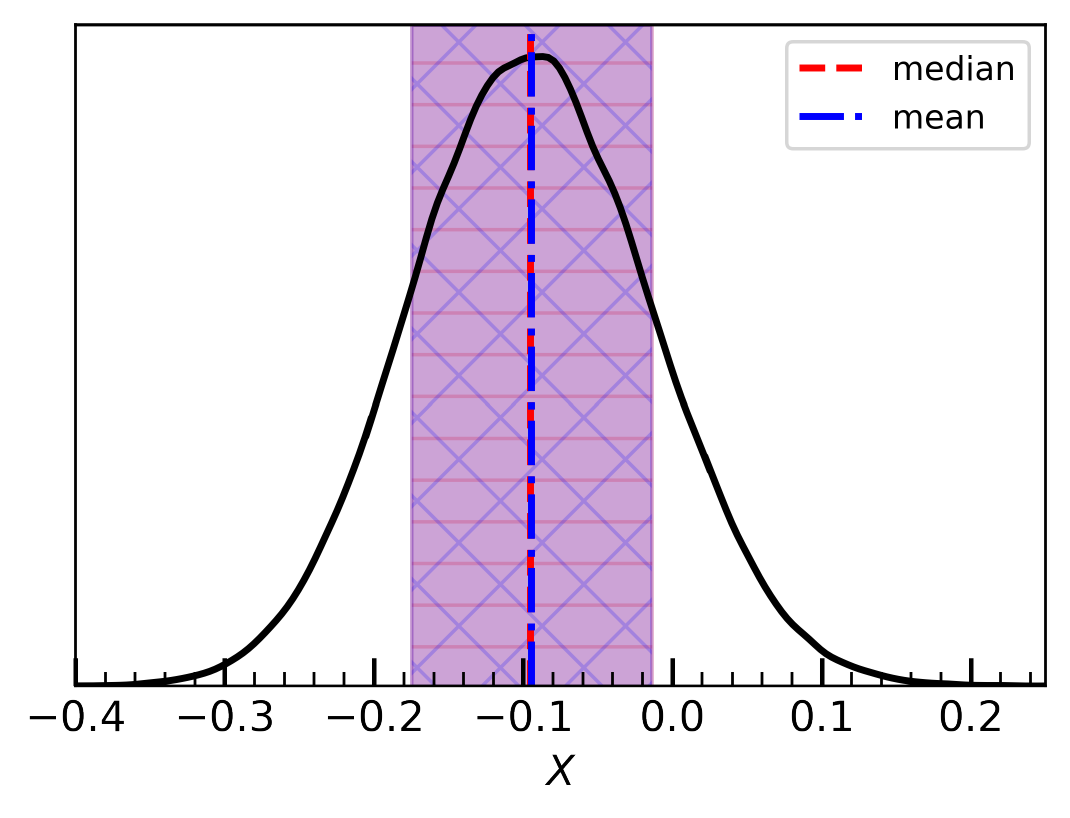}
		}
	\subfigure[ ]{
		\includegraphics[width = 0.4 \textwidth]{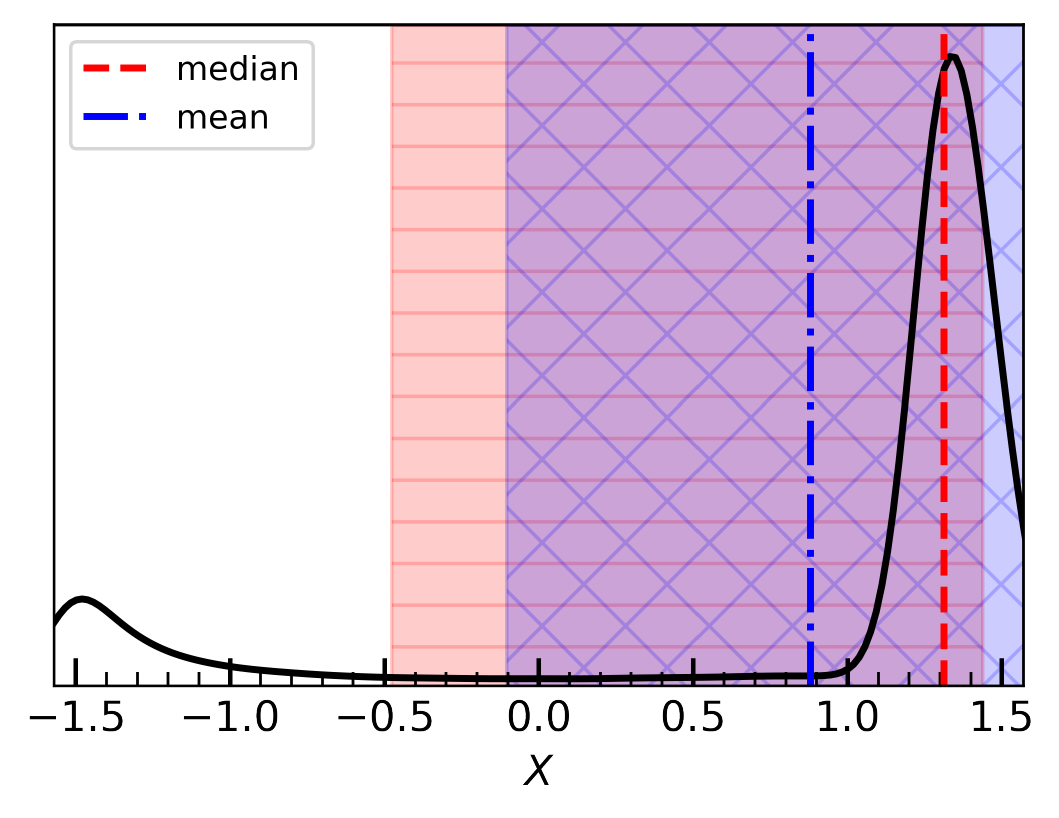}
		}
\caption{Distributions of a compactified random variable $X$ when it is (a) approximately Gaussian-distributed and (b) bimodal. The red-dashed and blue-dash-dotted lines show the median and the mean, respectively. The red-filled ($'-'$-hatched) region shows the 34.1\% probability mass above and below the median. The blue-filled ($'\times'$-hatched) region shows the $1\sigma$ confidence intervals from the mean, where $\sigma$ is the standard deviation.}
\label{fig:dist_examples}
\end{figure}

In Fig.~\ref{fig:dist_examples}, we show distributions of a compactified random variable $X = \arctan(x)$, where $x$ is a random variable with domain $x \in \left( - \infty, \infty \right)$, in cases where it is approximately Gaussian and bimodal.
Consider first Fig.~\ref{fig:dist_examples}-(a). In this case, $X$ is approximately Gaussian-distributed and so the median and mean of the distribution coincides. In addition, as shown by the colored-hatched regions in Fig.~\ref{fig:dist_examples}-(a), the $1\sigma$ confidence region around the mean also coincides with the $34.1\%$ probability mass above and below the median (50th percentile). On the other hand, Fig.~\ref{fig:dist_examples}-(b) shows an example where the compactified random variable is non-Gaussian, in particular bimodal. In this case, it can be seen that the locations of the median and the mean of the distribution are undisputably different. However, the mean also takes weight from the tails of a non-Gaussian-distribution; in this case, it therefore finds itself in a place where probability mass obviously is not localized. Moreover, a naive interpretation of the $1\sigma$ confidence intervals (blue-$'\times'$-hatched region of Fig.~\ref{fig:dist_examples}-(b)) will predict values supposedly outside the domain of the random variable. On the other hand, the median, by definition, is the place in a distribution where $50\%$ of the mass fall below it. This almost always ends up in a place where the mass is concentrated, as shown in Fig.~\ref{fig:dist_examples}-(b). Moreover, the $34.1\%$ of probability mass above and below the median will also always consistently predict samples within the domain of a compactified random variable, regardless of the shape of the distribution.

\begin{figure}[h!]
\center
	\subfigure[ ]{
		\includegraphics[width = 0.45 \textwidth]{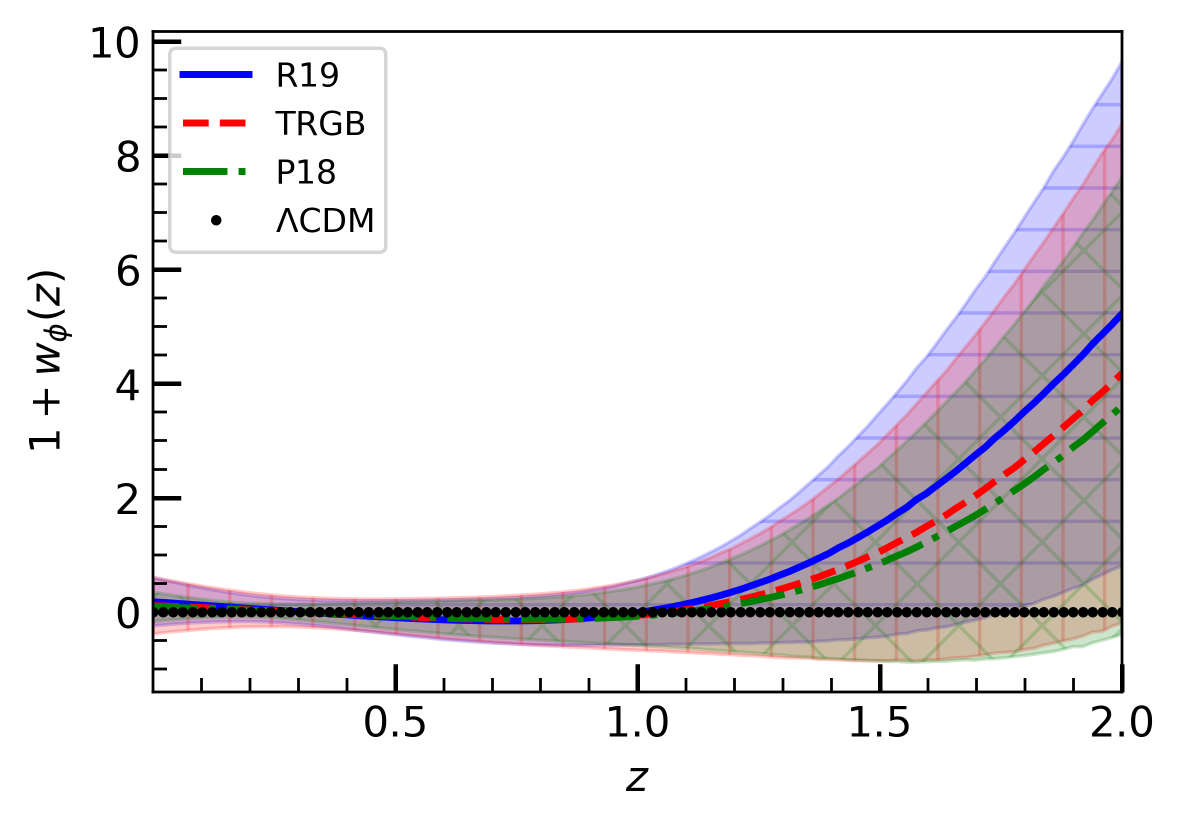}
		}
	\subfigure[ ]{
		\includegraphics[width = 0.45 \textwidth]{figs/atw_de_hdes.pdf}
		}
\caption{GP posteriors of (a) a random variable and (b) its compactified version.}
\label{fig:rv_vs_crv}
\end{figure}

We clarify one more statistical aspect of a compactified random variable. Fig. \ref{fig:rv_vs_crv} shows the GP posteriors of a random variable -- the dark energy equation of state (Sec. \ref{subsec:designer_horndeski}) -- and its compactified version.

Clearly, the posterior of a compactified random variable (Fig. \ref{fig:rv_vs_crv}(b)) may give the illusion that the error bars are getting smaller at high redshift where the data are sparse. This happens near the boundaries ($\sim \pm \pi/2$) of a compactified variable.


\begin{thebibliography}{100}

\bibitem{Riess:1998cb}
{\scshape Supernova Search Team} collaboration, \emph{{Observational evidence
  from supernovae for an accelerating universe and a cosmological constant}},
  \href{https://doi.org/10.1086/300499}{\emph{Astron.J.} {\bfseries 116} (1998)
  1009} [\href{https://arxiv.org/abs/astro-ph/9805201}{{\ttfamily
  astro-ph/9805201}}].

\bibitem{Perlmutter:1998np}
{\scshape Supernova Cosmology Project} collaboration, \emph{{Measurements of
  Omega and Lambda from 42 high redshift supernovae}},
  \href{https://doi.org/10.1086/307221}{\emph{Astrophys.J.} {\bfseries 517}
  (1999) 565} [\href{https://arxiv.org/abs/astro-ph/9812133}{{\ttfamily
  astro-ph/9812133}}].

\bibitem{dodelson2003modern}
S.~Dodelson, \emph{Modern Cosmology}. Academic Press, 2003.

\bibitem{Clifton:2011jh}
T.~Clifton, P.~G. Ferreira, A.~Padilla and C.~Skordis, \emph{{Modified Gravity
  and Cosmology}},
  \href{https://doi.org/10.1016/j.physrep.2012.01.001}{\emph{Phys. Rept.}
  {\bfseries 513} (2012) 1} [\href{https://arxiv.org/abs/1106.2476}{{\ttfamily
  1106.2476}}].

\bibitem{RevModPhys.61.1}
S.~Weinberg, \emph{The cosmological constant problem},
  \href{https://doi.org/10.1103/RevModPhys.61.1}{\emph{Rev. Mod. Phys.}
  {\bfseries 61} (1989) 1}.

\bibitem{Bull:2015stt}
P.~Bull et~al., \emph{{Beyond $\Lambda$CDM: Problems, solutions, and the road
  ahead}}, \href{https://doi.org/10.1016/j.dark.2016.02.001}{\emph{Phys. Dark
  Univ.} {\bfseries 12} (2016) 56}
  [\href{https://arxiv.org/abs/1512.05356}{{\ttfamily 1512.05356}}].

\bibitem{Sahni:1999gb}
V.~Sahni and A.~A. Starobinsky, \emph{{The Case for a positive cosmological
  Lambda term}}, \href{https://doi.org/10.1142/S0218271800000542}{\emph{Int. J.
  Mod. Phys. D} {\bfseries 9} (2000) 373}
  [\href{https://arxiv.org/abs/astro-ph/9904398}{{\ttfamily
  astro-ph/9904398}}].

\bibitem{Sahni:2006pa}
V.~Sahni and A.~Starobinsky, \emph{{Reconstructing Dark Energy}},
  \href{https://doi.org/10.1142/S0218271806009704}{\emph{Int. J. Mod. Phys. D}
  {\bfseries 15} (2006) 2105}
  [\href{https://arxiv.org/abs/astro-ph/0610026}{{\ttfamily
  astro-ph/0610026}}].

\bibitem{Copeland:2006wr}
E.~J. Copeland, M.~Sami and S.~Tsujikawa, \emph{{Dynamics of dark energy}},
  \href{https://doi.org/10.1142/S021827180600942X}{\emph{Int. J. Mod. Phys.}
  {\bfseries D15} (2006) 1753}
  [\href{https://arxiv.org/abs/hep-th/0603057}{{\ttfamily hep-th/0603057}}].

\bibitem{Capozziello:2011et}
S.~Capozziello and M.~De~Laurentis, \emph{{Extended Theories of Gravity}},
  \href{https://doi.org/10.1016/j.physrep.2011.09.003}{\emph{Phys. Rept.}
  {\bfseries 509} (2011) 167}
  [\href{https://arxiv.org/abs/1108.6266}{{\ttfamily 1108.6266}}].

\bibitem{DiValentino:2020zio}
E.~Di~Valentino et~al., \emph{{Cosmology Intertwined II: The Hubble Constant
  Tension}},  \href{https://arxiv.org/abs/2008.11284}{{\ttfamily 2008.11284}}.

\bibitem{DiValentino:2020vvd}
E.~Di~Valentino et~al., \emph{{Cosmology Intertwined III: $f \sigma_8$ and
  $S_8$}},  \href{https://arxiv.org/abs/2008.11285}{{\ttfamily 2008.11285}}.

\bibitem{Aghanim:2018eyx}
{\scshape Planck} collaboration, \emph{{Planck 2018 results. VI. Cosmological
  parameters}},
  \href{https://doi.org/10.1051/0004-6361/201833910}{\emph{Astron. Astrophys.}
  {\bfseries 641} (2020) A6}
  [\href{https://arxiv.org/abs/1807.06209}{{\ttfamily 1807.06209}}].

\bibitem{Bernal:2016gxb}
J.~L. Bernal, L.~Verde and A.~G. Riess, \emph{{The trouble with $H_0$}},
  \href{https://doi.org/10.1088/1475-7516/2016/10/019}{\emph{JCAP} {\bfseries
  10} (2016) 019} [\href{https://arxiv.org/abs/1607.05617}{{\ttfamily
  1607.05617}}].

\bibitem{Riess:2019cxk}
A.~G. Riess, S.~Casertano, W.~Yuan, L.~M. Macri and D.~Scolnic, \emph{{Large
  Magellanic Cloud Cepheid Standards Provide a 1\% Foundation for the
  Determination of the Hubble Constant and Stronger Evidence for Physics beyond
  $\Lambda$CDM}},
  \href{https://doi.org/10.3847/1538-4357/ab1422}{\emph{Astrophys. J.}
  {\bfseries 876} (2019) 85}
  [\href{https://arxiv.org/abs/1903.07603}{{\ttfamily 1903.07603}}].

\bibitem{Wong:2019kwg}
K.~C. Wong et~al., \emph{{H0LiCOW \textendash{} XIII. A 2.4 per cent
  measurement of H0 from lensed quasars: 5.3\ensuremath{\sigma} tension between
  early- and late-Universe probes}},
  \href{https://doi.org/10.1093/mnras/stz3094}{\emph{Mon. Not. Roy. Astron.
  Soc.} {\bfseries 498} (2020) 1420}
  [\href{https://arxiv.org/abs/1907.04869}{{\ttfamily 1907.04869}}].

\bibitem{Ade:2015xua}
{\scshape Planck} collaboration, \emph{Planck 2015 results. xiii. cosmological
  parameters},
  \href{https://doi.org/10.1051/0004-6361/201525830}{\emph{Astron.Astrophys.}
  {\bfseries 594} (2016) A13}
  [\href{https://arxiv.org/abs/1502.01589}{{\ttfamily 1502.01589}}].

\bibitem{Riess:2020sih}
A.~G. Riess, \emph{{The Expansion of the Universe is Faster than Expected}},
  \href{https://doi.org/10.1038/s42254-019-0137-0}{\emph{Nature Rev. Phys.}
  {\bfseries 2} (2019) 10} [\href{https://arxiv.org/abs/2001.03624}{{\ttfamily
  2001.03624}}].

\bibitem{Pesce:2020xfe}
D.~W. Pesce et~al., \emph{{The Megamaser Cosmology Project. XIII. Combined
  Hubble constant constraints}},
  \href{https://doi.org/10.3847/2041-8213/ab75f0}{\emph{Astrophys. J. Lett.}
  {\bfseries 891} (2020) L1}
  [\href{https://arxiv.org/abs/2001.09213}{{\ttfamily 2001.09213}}].

\bibitem{deJaeger:2020zpb}
T.~de~Jaeger, B.~E. Stahl, W.~Zheng, A.~V. Filippenko, A.~G. Riess and
  L.~Galbany, \emph{{A measurement of the Hubble constant from Type II
  supernovae}}, \href{https://doi.org/10.1093/mnras/staa1801}{\emph{Mon. Not.
  Roy. Astron. Soc.} {\bfseries 496} (2020) 3402}
  [\href{https://arxiv.org/abs/2006.03412}{{\ttfamily 2006.03412}}].

\bibitem{DiValentino:2019qzk}
E.~Di~Valentino, A.~Melchiorri and J.~Silk, \emph{{Planck evidence for a closed
  Universe and a possible crisis for cosmology}},
  \href{https://doi.org/10.1038/s41550-019-0906-9}{\emph{Nature Astron.}
  {\bfseries 4} (2019) 196} [\href{https://arxiv.org/abs/1911.02087}{{\ttfamily
  1911.02087}}].

\bibitem{Handley:2019tkm}
W.~Handley, \emph{{Curvature tension: evidence for a closed universe}},
  \href{https://doi.org/10.1103/PhysRevD.103.L041301}{\emph{Phys. Rev. D}
  {\bfseries 103} (2021) L041301}
  [\href{https://arxiv.org/abs/1908.09139}{{\ttfamily 1908.09139}}].

\bibitem{Horndeski:1974wa}
G.~W. Horndeski, \emph{{Second-order scalar-tensor field equations in a
  four-dimensional space}},
  \href{https://doi.org/10.1007/BF01807638}{\emph{Int. J. Theor. Phys.}
  {\bfseries 10} (1974) 363}.

\bibitem{Sotiriou:2008rp}
T.~P. Sotiriou and V.~Faraoni, \emph{{f(R) Theories Of Gravity}},
  \href{https://doi.org/10.1103/RevModPhys.82.451}{\emph{Rev. Mod. Phys.}
  {\bfseries 82} (2010) 451} [\href{https://arxiv.org/abs/0805.1726}{{\ttfamily
  0805.1726}}].

\bibitem{DeFelice:2010aj}
A.~De~Felice and S.~Tsujikawa, \emph{{f(R) theories}},
  \href{https://doi.org/10.12942/lrr-2010-3}{\emph{Living Rev. Rel.} {\bfseries
  13} (2010) 3} [\href{https://arxiv.org/abs/1002.4928}{{\ttfamily
  1002.4928}}].

\bibitem{Nojiri:2006gh}
S.~Nojiri and S.~D. Odintsov, \emph{{Modified f(R) gravity consistent with
  realistic cosmology: From matter dominated epoch to dark energy universe}},
  \href{https://doi.org/10.1103/PhysRevD.74.086005}{\emph{Phys. Rev. D}
  {\bfseries 74} (2006) 086005}
  [\href{https://arxiv.org/abs/hep-th/0608008}{{\ttfamily hep-th/0608008}}].

\bibitem{Hu:2007nk}
W.~Hu and I.~Sawicki, \emph{{Models of f(R) Cosmic Acceleration that Evade
  Solar-System Tests}},
  \href{https://doi.org/10.1103/PhysRevD.76.064004}{\emph{Phys. Rev. D}
  {\bfseries 76} (2007) 064004}
  [\href{https://arxiv.org/abs/0705.1158}{{\ttfamily 0705.1158}}].

\bibitem{Appleby:2007vb}
S.~A. Appleby and R.~A. Battye, \emph{{Do consistent $F(R)$ models mimic
  General Relativity plus $\Lambda$?}},
  \href{https://doi.org/10.1016/j.physletb.2007.08.037}{\emph{Phys. Lett. B}
  {\bfseries 654} (2007) 7} [\href{https://arxiv.org/abs/0705.3199}{{\ttfamily
  0705.3199}}].

\bibitem{Starobinsky:2007hu}
A.~A. Starobinsky, \emph{{Disappearing cosmological constant in f(R) gravity}},
  \href{https://doi.org/10.1134/S0021364007150027}{\emph{JETP Lett.} {\bfseries
  86} (2007) 157} [\href{https://arxiv.org/abs/0706.2041}{{\ttfamily
  0706.2041}}].

\bibitem{Appleby:2009uf}
S.~A. Appleby, R.~A. Battye and A.~A. Starobinsky, \emph{{Curing singularities
  in cosmological evolution of F(R) gravity}},
  \href{https://doi.org/10.1088/1475-7516/2010/06/005}{\emph{JCAP} {\bfseries
  06} (2010) 005} [\href{https://arxiv.org/abs/0909.1737}{{\ttfamily
  0909.1737}}].

\bibitem{Sotiriou:2008ve}
T.~P. Sotiriou, \emph{{6+1 lessons from f(R) gravity}},
  \href{https://doi.org/10.1088/1742-6596/189/1/012039}{\emph{J. Phys. Conf.
  Ser.} {\bfseries 189} (2009) 012039}
  [\href{https://arxiv.org/abs/0810.5594}{{\ttfamily 0810.5594}}].

\bibitem{TheLIGOScientific:2017qsa}
{\scshape LIGO Scientific, Virgo} collaboration, \emph{{GW170817: Observation
  of Gravitational Waves from a Binary Neutron Star Inspiral}},
  \href{https://doi.org/10.1103/PhysRevLett.119.161101}{\emph{Phys. Rev. Lett.}
  {\bfseries 119} (2017) 161101}
  [\href{https://arxiv.org/abs/1710.05832}{{\ttfamily 1710.05832}}].

\bibitem{Ezquiaga:2018btd}
J.~M. Ezquiaga and M.~Zumalacárregui, \emph{{Dark Energy in light of
  Multi-Messenger Gravitational-Wave astronomy}},
  \href{https://doi.org/10.3389/fspas.2018.00044}{\emph{Front. Astron. Space
  Sci.} {\bfseries 5} (2018) 44}
  [\href{https://arxiv.org/abs/1807.09241}{{\ttfamily 1807.09241}}].

\bibitem{Nicolis:2008in}
A.~Nicolis, R.~Rattazzi and E.~Trincherini, \emph{{The Galileon as a local
  modification of gravity}},
  \href{https://doi.org/10.1103/PhysRevD.79.064036}{\emph{Phys. Rev.}
  {\bfseries D79} (2009) 064036}
  [\href{https://arxiv.org/abs/0811.2197}{{\ttfamily 0811.2197}}].

\bibitem{Deffayet:2009wt}
C.~Deffayet, G.~Esposito-Farese and A.~Vikman, \emph{{Covariant Galileon}},
  \href{https://doi.org/10.1103/PhysRevD.79.084003}{\emph{Phys. Rev.}
  {\bfseries D79} (2009) 084003}
  [\href{https://arxiv.org/abs/0901.1314}{{\ttfamily 0901.1314}}].

\bibitem{Martin-Moruno:2015bda}
P.~Martin-Moruno, N.~J. Nunes and F.~S.~N. Lobo, \emph{{Horndeski theories
  self-tuning to a de Sitter vacuum}},
  \href{https://doi.org/10.1103/PhysRevD.91.084029}{\emph{Phys. Rev.}
  {\bfseries D91} (2015) 084029}
  [\href{https://arxiv.org/abs/1502.03236}{{\ttfamily 1502.03236}}].

\bibitem{Charmousis:2011bf}
C.~Charmousis, E.~J. Copeland, A.~Padilla and P.~M. Saffin, \emph{{General
  second order scalar-tensor theory, self tuning, and the Fab Four}},
  \href{https://doi.org/10.1103/PhysRevLett.108.051101}{\emph{Phys. Rev. Lett.}
  {\bfseries 108} (2012) 051101}
  [\href{https://arxiv.org/abs/1106.2000}{{\ttfamily 1106.2000}}].

\bibitem{Gubitosi:2011sg}
G.~Gubitosi and E.~V. Linder, \emph{{Purely Kinetic Coupled Gravity}},
  \href{https://doi.org/10.1016/j.physletb.2011.07.066}{\emph{Phys. Lett.}
  {\bfseries B703} (2011) 113}
  [\href{https://arxiv.org/abs/1106.2815}{{\ttfamily 1106.2815}}].

\bibitem{Krssak:2018ywd}
M.~Kr{\v{s}}{\v{s}}{\'{a}}k, R.~van~den Hoogen, J.~Pereira, C.~B{\"{o}}hmer and
  A.~Coley, \emph{{Teleparallel theories of gravity: illuminating a fully
  invariant approach}},
  \href{https://doi.org/10.1088/1361-6382/ab2e1f}{\emph{Class. Quant. Grav.}
  {\bfseries 36} (2019) 183001}
  [\href{https://arxiv.org/abs/1810.12932}{{\ttfamily 1810.12932}}].

\bibitem{Bahamonde:2019shr}
S.~Bahamonde, K.~F. Dialektopoulos and J.~Levi~Said, \emph{{Can Horndeski
  Theory be recast using Teleparallel Gravity?}},
  \href{https://doi.org/10.1103/PhysRevD.100.064018}{\emph{Phys. Rev. D}
  {\bfseries 100} (2019) 064018}
  [\href{https://arxiv.org/abs/1904.10791}{{\ttfamily 1904.10791}}].

\bibitem{Bahamonde:2020cfv}
S.~Bahamonde, K.~F. Dialektopoulos, M.~Hohmann and J.~Levi~Said,
  \emph{{Post-Newtonian limit of Teleparallel Horndeski gravity}},
  \href{https://doi.org/10.1088/1361-6382/abc441}{\emph{Class. Quant. Grav.}
  {\bfseries 38} (2020) 025006}
  [\href{https://arxiv.org/abs/2003.11554}{{\ttfamily 2003.11554}}].

\bibitem{Bahamonde:2019ipm}
S.~Bahamonde, K.~F. Dialektopoulos, V.~Gakis and J.~Levi~Said, \emph{{Reviving
  Horndeski theory using teleparallel gravity after GW170817}},
  \href{https://doi.org/10.1103/PhysRevD.101.084060}{\emph{Phys. Rev. D}
  {\bfseries 101} (2020) 084060}
  [\href{https://arxiv.org/abs/1907.10057}{{\ttfamily 1907.10057}}].

\bibitem{Seikel2012}
M.~Seikel, C.~Clarkson and M.~Smith, \emph{Reconstruction of dark energy and
  expansion dynamics using gaussian processes},
  \href{https://doi.org/10.1088/1475-7516/2012/06/036}{\emph{JCAP} {\bfseries
  2012} (2012) 036} [\href{https://arxiv.org/abs/1204.2832}{{\ttfamily
  1204.2832}}].

\bibitem{Shafieloo:2012ht}
A.~Shafieloo, A.~G. Kim and E.~V. Linder, \emph{{Gaussian Process
  Cosmography}}, \href{https://doi.org/10.1103/PhysRevD.85.123530}{\emph{Phys.
  Rev. D} {\bfseries 85} (2012) 123530}
  [\href{https://arxiv.org/abs/1204.2272}{{\ttfamily 1204.2272}}].

\bibitem{Seikel:2013fda}
M.~Seikel and C.~Clarkson, \emph{{Optimising Gaussian processes for
  reconstructing dark energy dynamics from supernovae}},
  \href{https://arxiv.org/abs/1311.6678}{{\ttfamily 1311.6678}}.

\bibitem{Yennapureddy:2017vvb}
M.~K. Yennapureddy and F.~Melia, \emph{{Reconstruction of the HII Galaxy Hubble
  Diagram using Gaussian Processes}},
  \href{https://doi.org/10.1088/1475-7516/2017/11/029}{\emph{JCAP} {\bfseries
  11} (2017) 029} [\href{https://arxiv.org/abs/1711.03454}{{\ttfamily
  1711.03454}}].

\bibitem{Gomez-Valent:2018hwc}
A.~Gómez-Valent and L.~Amendola, \emph{{$H_0$ from cosmic chronometers and
  Type Ia supernovae, with Gaussian Processes and the novel Weighted Polynomial
  Regression method}},
  \href{https://doi.org/10.1088/1475-7516/2018/04/051}{\emph{JCAP} {\bfseries
  04} (2018) 051} [\href{https://arxiv.org/abs/1802.01505}{{\ttfamily
  1802.01505}}].

\bibitem{Li:2019nux}
E.-K. Li, M.~Du, Z.-H. Zhou, H.~Zhang and L.~Xu, \emph{{Testing the effect of
  $H_0$ on $f\sigma_8$ tension using a Gaussian process method}},
  \href{https://doi.org/10.1093/mnras/staa3894}{\emph{Mon. Not. Roy. Astron.
  Soc.} {\bfseries 501} (2021) 4452}
  [\href{https://arxiv.org/abs/1911.12076}{{\ttfamily 1911.12076}}].

\bibitem{Liao:2019qoc}
K.~Liao, A.~Shafieloo, R.~E. Keeley and E.~V. Linder, \emph{{A
  model-independent determination of the Hubble constant from lensed quasars
  and supernovae using Gaussian process regression}},
  \href{https://doi.org/10.3847/2041-8213/ab5308}{\emph{Astrophys. J. Lett.}
  {\bfseries 886} (2019) L23}
  [\href{https://arxiv.org/abs/1908.04967}{{\ttfamily 1908.04967}}].

\bibitem{Keeley:2020aym}
R.~E. Keeley, A.~Shafieloo, G.-B. Zhao, J.~A. Vazquez and H.~Koo,
  \emph{{Reconstructing the Universe: Testing the Mutual Consistency of the
  Pantheon and SDSS/eBOSS BAO Data Sets with Gaussian Processes}},
  \href{https://doi.org/10.3847/1538-3881/abdd2a}{\emph{Astron. J.} {\bfseries
  161} (2021) 151} [\href{https://arxiv.org/abs/2010.03234}{{\ttfamily
  2010.03234}}].

\bibitem{Renzi:2020fnx}
F.~Renzi and A.~Silvestri, \emph{{A look at the Hubble speed from first
  principles}},  \href{https://arxiv.org/abs/2011.10559}{{\ttfamily
  2011.10559}}.

\bibitem{Colgain:2021ngq}
E.~Colg\'ain and M.~M. Sheikh-Jabbari, \emph{{Elucidating cosmological model
  dependence with $H_0$}},  \href{https://arxiv.org/abs/2101.08565}{{\ttfamily
  2101.08565}}.

\bibitem{Benisty:2020kdt}
D.~Benisty, \emph{{Quantifying the $S_8$ tension with the Redshift Space
  Distortion data set}},
  \href{https://doi.org/10.1016/j.dark.2020.100766}{\emph{Phys. Dark Univ.}
  {\bfseries 31} (2021) 100766}
  [\href{https://arxiv.org/abs/2005.03751}{{\ttfamily 2005.03751}}].

\bibitem{Belgacem:2019zzu}
E.~Belgacem, S.~Foffa, M.~Maggiore and T.~Yang, \emph{{Gaussian processes
  reconstruction of modified gravitational wave propagation}},
  \href{https://doi.org/10.1103/PhysRevD.101.063505}{\emph{Phys. Rev. D}
  {\bfseries 101} (2020) 063505}
  [\href{https://arxiv.org/abs/1911.11497}{{\ttfamily 1911.11497}}].

\bibitem{Moore:2015sza}
C.~J. Moore, C.~P.~L. Berry, A.~J.~K. Chua and J.~R. Gair, \emph{{Improving
  gravitational-wave parameter estimation using Gaussian process regression}},
  \href{https://doi.org/10.1103/PhysRevD.93.064001}{\emph{Phys. Rev. D}
  {\bfseries 93} (2016) 064001}
  [\href{https://arxiv.org/abs/1509.04066}{{\ttfamily 1509.04066}}].

\bibitem{Canas-Herrera:2021qxs}
G.~Ca\~nas Herrera, O.~Contigiani and V.~Vardanyan, \emph{{Learning how to
  surf: Reconstructing the propagation and origin of gravitational waves with
  Gaussian Processes}},  \href{https://arxiv.org/abs/2105.04262}{{\ttfamily
  2105.04262}}.

\bibitem{Briffa:2020qli}
R.~Briffa, S.~Capozziello, J.~Levi~Said, J.~Mifsud and E.~N. Saridakis,
  \emph{{Constraining teleparallel gravity through Gaussian processes}},
  \href{https://doi.org/10.1088/1361-6382/abd4f5}{\emph{Class. Quant. Grav.}
  {\bfseries 38} (2020) 055007}
  [\href{https://arxiv.org/abs/2009.14582}{{\ttfamily 2009.14582}}].

\bibitem{Cai:2019bdh}
Y.-F. Cai, M.~Khurshudyan and E.~N. Saridakis, \emph{{Model-independent
  reconstruction of $f(T)$ gravity from Gaussian Processes}},
  \href{https://doi.org/10.3847/1538-4357/ab5a7f}{\emph{Astrophys. J.}
  {\bfseries 888} (2020) 62}
  [\href{https://arxiv.org/abs/1907.10813}{{\ttfamily 1907.10813}}].

\bibitem{Ren:2021tfi}
X.~Ren, T.~H.~T. Wong, Y.-F. Cai and E.~N. Saridakis, \emph{{Data-driven
  Reconstruction of the Late-time Cosmic Acceleration with f(T) Gravity}},
  \href{https://doi.org/10.1016/j.dark.2021.100812}{\emph{Phys. Dark Univ.}
  {\bfseries 32} (2021) 100812}
  [\href{https://arxiv.org/abs/2103.01260}{{\ttfamily 2103.01260}}].

\bibitem{LeviSaid:2021yat}
J.~Levi~Said, J.~Mifsud, J.~Sultana and K.~Z. Adami, \emph{{Reconstructing
  teleparallel gravity with cosmic structure growth and expansion rate data}},
  \href{https://arxiv.org/abs/2103.05021}{{\ttfamily 2103.05021}}.

\bibitem{Cai:2015zoa}
T.~Yang, Z.-K. Guo and R.-G. Cai, \emph{{Reconstructing the interaction between
  dark energy and dark matter using Gaussian Processes}},
  \href{https://doi.org/10.1103/PhysRevD.91.123533}{\emph{Phys. Rev. D}
  {\bfseries 91} (2015) 123533}
  [\href{https://arxiv.org/abs/1505.04443}{{\ttfamily 1505.04443}}].

\bibitem{Reyes:2021owe}
M.~Reyes and C.~Escamilla-Rivera, \emph{{Improving data-driven
  model-independent reconstructions and new constraints in Horndeski
  cosmology}},  \href{https://arxiv.org/abs/2104.04484}{{\ttfamily
  2104.04484}}.

\bibitem{Lovelock:1971yv}
D.~Lovelock, \emph{{The Einstein tensor and its generalizations}},
  \href{https://doi.org/10.1063/1.1665613}{\emph{J. Math. Phys.} {\bfseries 12}
  (1971) 498}.

\bibitem{Kobayashi:2019hrl}
T.~Kobayashi, \emph{{Horndeski theory and beyond: a review}},
  \href{https://doi.org/10.1088/1361-6633/ab2429}{\emph{Rept. Prog. Phys.}
  {\bfseries 82} (2019) 086901}
  [\href{https://arxiv.org/abs/1901.07183}{{\ttfamily 1901.07183}}].

\bibitem{Hou:2017bqj}
S.~Hou, Y.~Gong and Y.~Liu, \emph{{Polarizations of Gravitational Waves in
  Horndeski Theory}},
  \href{https://doi.org/10.1140/epjc/s10052-018-5869-y}{\emph{Eur. Phys. J. C}
  {\bfseries 78} (2018) 378}
  [\href{https://arxiv.org/abs/1704.01899}{{\ttfamily 1704.01899}}].

\bibitem{Brans:1961sx}
C.~Brans and R.~H. Dicke, \emph{{Mach's principle and a relativistic theory of
  gravitation}}, \href{https://doi.org/10.1103/PhysRev.124.925}{\emph{Phys.
  Rev.} {\bfseries 124} (1961) 925}.

\bibitem{Goldstein:2017mmi}
A.~Goldstein et~al., \emph{{An Ordinary Short Gamma-Ray Burst with
  Extraordinary Implications: Fermi-GBM Detection of GRB 170817A}},
  \href{https://doi.org/10.3847/2041-8213/aa8f41}{\emph{Astrophys. J. Lett.}
  {\bfseries 848} (2017) L14}
  [\href{https://arxiv.org/abs/1710.05446}{{\ttfamily 1710.05446}}].

\bibitem{Ezquiaga:2017ekz}
J.~M. Ezquiaga and M.~Zumalac\'arregui, \emph{{Dark Energy After GW170817: Dead
  Ends and the Road Ahead}},
  \href{https://doi.org/10.1103/PhysRevLett.119.251304}{\emph{Phys. Rev. Lett.}
  {\bfseries 119} (2017) 251304}
  [\href{https://arxiv.org/abs/1710.05901}{{\ttfamily 1710.05901}}].

\bibitem{Tsamis:1997rk}
N.~C. Tsamis and R.~P. Woodard, \emph{{Nonperturbative models for the quantum
  gravitational back reaction on inflation}},
  \href{https://doi.org/10.1006/aphy.1998.5816}{\emph{Annals Phys.} {\bfseries
  267} (1998) 145} [\href{https://arxiv.org/abs/hep-ph/9712331}{{\ttfamily
  hep-ph/9712331}}].

\bibitem{Deffayet:2010qz}
C.~Deffayet, O.~Pujolas, I.~Sawicki and A.~Vikman, \emph{{Imperfect Dark Energy
  from Kinetic Gravity Braiding}},
  \href{https://doi.org/10.1088/1475-7516/2010/10/026}{\emph{JCAP} {\bfseries
  10} (2010) 026} [\href{https://arxiv.org/abs/1008.0048}{{\ttfamily
  1008.0048}}].

\bibitem{Kobayashi:2011nu}
T.~Kobayashi, M.~Yamaguchi and J.~Yokoyama, \emph{{Generalized G-inflation:
  Inflation with the most general second-order field equations}},
  \href{https://doi.org/10.1143/PTP.126.511}{\emph{Prog. Theor. Phys.}
  {\bfseries 126} (2011) 511}
  [\href{https://arxiv.org/abs/1105.5723}{{\ttfamily 1105.5723}}].

\bibitem{Kase:2018aps}
R.~Kase and S.~Tsujikawa, \emph{{Dark energy in Horndeski theories after
  GW170817: A review}},
  \href{https://doi.org/10.1142/S0218271819420057}{\emph{Int. J. Mod. Phys. D}
  {\bfseries 28} (2019) 1942005}
  [\href{https://arxiv.org/abs/1809.08735}{{\ttfamily 1809.08735}}].

\bibitem{Arjona:2019rfn}
R.~Arjona, W.~Cardona and S.~Nesseris, \emph{{Designing Horndeski and the
  effective fluid approach}},
  \href{https://doi.org/10.1103/PhysRevD.100.063526}{\emph{Phys. Rev. D}
  {\bfseries 100} (2019) 063526}
  [\href{https://arxiv.org/abs/1904.06294}{{\ttfamily 1904.06294}}].

\bibitem{Bernardo:2019vln}
R.~C. Bernardo and I.~Vega, \emph{{Tailoring cosmologies in cubic
  shift-symmetric Horndeski gravity}},
  \href{https://doi.org/10.1088/1475-7516/2019/10/058}{\emph{JCAP} {\bfseries
  10} (2019) 058} [\href{https://arxiv.org/abs/1903.12578}{{\ttfamily
  1903.12578}}].

\bibitem{10.5555/971143}
D.~J.~C. MacKay, \emph{Information Theory, Inference \& Learning Algorithms}.
  Cambridge University Press, USA, 2002.

\bibitem{10.5555/1162254}
C.~E. Rasmussen and C.~K.~I. Williams, \emph{Gaussian Processes for Machine
  Learning (Adaptive Computation and Machine Learning)}. The MIT Press, 2005.

\bibitem{Wang:2017jdm}
D.~Wang and X.-H. Meng, \emph{{Improved constraints on the dark energy equation
  of state using Gaussian processes}},
  \href{https://doi.org/10.1103/PhysRevD.95.023508}{\emph{Phys. Rev. D}
  {\bfseries 95} (2017) 023508}
  [\href{https://arxiv.org/abs/1708.07750}{{\ttfamily 1708.07750}}].

\bibitem{Zhang:2018gjb}
M.-J. Zhang and H.~Li, \emph{{Gaussian processes reconstruction of dark energy
  from observational data}},
  \href{https://doi.org/10.1140/epjc/s10052-018-5953-3}{\emph{Eur. Phys. J. C}
  {\bfseries 78} (2018) 460}
  [\href{https://arxiv.org/abs/1806.02981}{{\ttfamily 1806.02981}}].

\bibitem{Mukherjee:2020vkx}
P.~Mukherjee and N.~Banerjee, \emph{{Revisiting a non-parametric reconstruction
  of the deceleration parameter from observational data}},
  \href{https://arxiv.org/abs/2007.15941}{{\ttfamily 2007.15941}}.

\bibitem{Aljaf:2020eqh}
M.~Aljaf, D.~Gregoris and M.~Khurshudyan, \emph{{Constraints on interacting
  dark energy models through cosmic chronometers and Gaussian process}},
  \href{https://arxiv.org/abs/2005.01891}{{\ttfamily 2005.01891}}.

\bibitem{Busti:2014aoa}
V.~C. Busti, C.~Clarkson and M.~Seikel, \emph{{The Value of $H_0$ from Gaussian
  Processes}}, \href{https://doi.org/10.1017/S1743921314013751}{\emph{IAU
  Symp.} {\bfseries 306} (2014) 25}
  [\href{https://arxiv.org/abs/1407.5227}{{\ttfamily 1407.5227}}].

\bibitem{Cai:2015pia}
R.-G. Cai, Z.-K. Guo and T.~Yang, \emph{{Null test of the cosmic curvature
  using $H(z)$ and supernovae data}},
  \href{https://doi.org/10.1103/PhysRevD.93.043517}{\emph{Phys. Rev. D}
  {\bfseries 93} (2016) 043517}
  [\href{https://arxiv.org/abs/1509.06283}{{\ttfamily 1509.06283}}].

\bibitem{Bernardo:2021mfs}
R.~C. Bernardo and J.~Levi~Said, \emph{{Towards a model-independent
  reconstruction approach for late-time Hubble data}},
  \href{https://doi.org/10.1088/1475-7516/2021/08/027}{\emph{JCAP} {\bfseries
  08} (2021) 027} [\href{https://arxiv.org/abs/2106.08688}{{\ttfamily
  2106.08688}}].

\bibitem{scikit-learn}
F.~Pedregosa, G.~Varoquaux, A.~Gramfort, V.~Michel, B.~Thirion, O.~Grisel
  et~al., \emph{Scikit-learn: Machine learning in {P}ython}, {\emph{Journal of
  Machine Learning Research} {\bfseries 12} (2011) 2825}.

\bibitem{2020arXiv200505290T}
J.~{Torrado} and A.~{Lewis}, \emph{{Cobaya: Code for Bayesian Analysis of
  hierarchical physical models}}, {\emph{arXiv e-prints} (2020)
  arXiv:2005.05290} [\href{https://arxiv.org/abs/2005.05290}{{\ttfamily
  2005.05290}}].

\bibitem{Lewis:2019xzd}
A.~Lewis, \emph{{GetDist: a Python package for analysing Monte Carlo samples}},
   \href{https://arxiv.org/abs/1910.13970}{{\ttfamily 1910.13970}}.

\bibitem{2020NumPy-Array}
C.~R. Harris et~al., \emph{Array programming with {NumPy}},
  \href{https://doi.org/10.1038/s41586-020-2649-2}{\emph{Nature} {\bfseries
  585} (2020) 357–362}.

\bibitem{2020SciPy-NMeth}
P.~Virtanen et~al., \emph{{{SciPy} 1.0: Fundamental Algorithms for Scientific
  Computing in Python}},
  \href{https://doi.org/10.1038/s41592-019-0686-2}{\emph{Nature Methods}
  {\bfseries 17} (2020) 261}.

\bibitem{Waskom2021}
M.~L. Waskom, \emph{seaborn: statistical data visualization},
  \href{https://doi.org/10.21105/joss.03021}{\emph{Journal of Open Source
  Software} {\bfseries 6} (2021) 3021}.

\bibitem{4160265}
J.~D. Hunter, \emph{Matplotlib: A 2d graphics environment},
  \href{https://doi.org/10.1109/MCSE.2007.55}{\emph{Computing in Science
  Engineering} {\bfseries 9} (2007) 90}.

\bibitem{jupyter}
T.~Kluyver, B.~Ragan-Kelley, F.~P{\'e}rez, B.~Granger, M.~Bussonnier,
  J.~Frederic et~al., \emph{Jupyter notebooks - a publishing format for
  reproducible computational workflows},  in \emph{Positioning and Power in
  Academic Publishing: Players, Agents and Agendas}, F.~Loizides and B.~Scmidt,
  eds., (Netherlands), pp.~87--90, IOS Press, 2016,
  \href{https://eprints.soton.ac.uk/403913/}{https://eprints.soton.ac.uk/403913/}.

\bibitem{reggie_bernardo_4810864}
R.~Bernardo, ``{ reggiebernardo/notebooks: dark energy research notebooks }.''
  \href{ https://doi.org/10.5281/zenodo.4810864 }{ 10.5281/zenodo.4810864 },
  2021.

\bibitem{Freedman:2019jwv}
W.~L. {Freedman} et~al., \emph{{The Carnegie-Chicago Hubble Program. VIII. An
  Independent Determination of the Hubble Constant Based on the Tip of the Red
  Giant Branch}},
  \href{https://doi.org/10.3847/1538-4357/ab2f73}{\emph{Astrophys. J.}
  {\bfseries 882} (2019) 34}
  [\href{https://arxiv.org/abs/1907.05922}{{\ttfamily 1907.05922}}].

\bibitem{Moresco:2016mzx}
M.~Moresco, L.~Pozzetti, A.~Cimatti, R.~Jimenez, C.~Maraston, L.~Verde et~al.,
  \emph{{A 6\% measurement of the Hubble parameter at $z\sim0.45$: direct
  evidence of the epoch of cosmic re-acceleration}},
  \href{https://doi.org/10.1088/1475-7516/2016/05/014}{\emph{JCAP} {\bfseries
  05} (2016) 014} [\href{https://arxiv.org/abs/1601.01701}{{\ttfamily
  1601.01701}}].

\bibitem{Moresco:2015cya}
M.~Moresco, \emph{{Raising the bar: new constraints on the Hubble parameter
  with cosmic chronometers at z \ensuremath{\sim} 2}},
  \href{https://doi.org/10.1093/mnrasl/slv037}{\emph{Mon. Not. Roy. Astron.
  Soc.} {\bfseries 450} (2015) L16}
  [\href{https://arxiv.org/abs/1503.01116}{{\ttfamily 1503.01116}}].

\bibitem{2014RAA....14.1221Z}
C.~{Zhang}, H.~{Zhang}, S.~{Yuan}, S.~{Liu}, T.-J. {Zhang} and Y.-C. {Sun},
  \emph{{Four new observational H(z) data from luminous red galaxies in the
  Sloan Digital Sky Survey data release seven}},
  \href{https://doi.org/10.1088/1674-4527/14/10/002}{\emph{Research in
  Astronomy and Astrophysics} {\bfseries 14} (2014) 1221}
  [\href{https://arxiv.org/abs/1207.4541}{{\ttfamily 1207.4541}}].

\bibitem{2010JCAP...02..008S}
D.~{Stern}, R.~{Jimenez}, L.~{Verde}, M.~{Kamionkowski} and S.~A. {Stanford},
  \emph{{Cosmic chronometers: constraining the equation of state of dark
  energy. I: H(z) measurements}},
  \href{https://doi.org/10.1088/1475-7516/2010/02/008}{\emph{JCAP} {\bfseries
  2010} (2010) 008} [\href{https://arxiv.org/abs/0907.3149}{{\ttfamily
  0907.3149}}].

\bibitem{2012JCAP...08..006M}
M.~{Moresco} et~al., \emph{{Improved constraints on the expansion rate of the
  Universe up to z \raisebox{-0.5ex}\textasciitilde 1.1 from the spectroscopic
  evolution of cosmic chronometers}},
  \href{https://doi.org/10.1088/1475-7516/2012/08/006}{\emph{JCAP} {\bfseries
  2012} (2012) 006} [\href{https://arxiv.org/abs/1201.3609}{{\ttfamily
  1201.3609}}].

\bibitem{Scolnic:2017caz}
D.~M. Scolnic et~al., \emph{{The Complete Light-curve Sample of
  Spectroscopically Confirmed SNe Ia from Pan-STARRS1 and Cosmological
  Constraints from the Combined Pantheon Sample}},
  \href{https://doi.org/10.3847/1538-4357/aab9bb}{\emph{Astrophys. J.}
  {\bfseries 859} (2018) 101}
  [\href{https://arxiv.org/abs/1710.00845}{{\ttfamily 1710.00845}}].

\bibitem{Riess:2017lxs}
A.~G. Riess et~al., \emph{{Type Ia Supernova Distances at Redshift $>$ 1.5 from
  the Hubble Space Telescope Multi-cycle Treasury Programs: The Early Expansion
  Rate}}, \href{https://doi.org/10.3847/1538-4357/aaa5a9}{\emph{Astrophys. J.}
  {\bfseries 853} (2018) 126}
  [\href{https://arxiv.org/abs/1710.00844}{{\ttfamily 1710.00844}}].

\bibitem{Alam:2016hwk}
{\scshape BOSS} collaboration, \emph{{The clustering of galaxies in the
  completed SDSS-III Baryon Oscillation Spectroscopic Survey: cosmological
  analysis of the DR12 galaxy sample}},
  \href{https://doi.org/10.1093/mnras/stx721}{\emph{Mon. Not. Roy. Astron.
  Soc.} {\bfseries 470} (2017) 2617}
  [\href{https://arxiv.org/abs/1607.03155}{{\ttfamily 1607.03155}}].

\bibitem{Bautista:2020ahg}
J.~E. Bautista et~al., \emph{{The Completed SDSS-IV extended Baryon Oscillation
  Spectroscopic Survey: measurement of the BAO and growth rate of structure of
  the luminous red galaxy sample from the anisotropic correlation function
  between redshifts 0.6 and 1}},
  \href{https://doi.org/10.1093/mnras/staa2800}{\emph{Mon. Not. Roy. Astron.
  Soc.} {\bfseries 500} (2020) 736}
  [\href{https://arxiv.org/abs/2007.08993}{{\ttfamily 2007.08993}}].

\bibitem{Gil-Marin:2020bct}
H.~Gil-Marin et~al., \emph{{The Completed SDSS-IV extended Baryon Oscillation
  Spectroscopic Survey: measurement of the BAO and growth rate of structure of
  the luminous red galaxy sample from the anisotropic power spectrum between
  redshifts 0.6 and 1.0}},
  \href{https://doi.org/10.1093/mnras/staa2455}{\emph{Mon. Not. Roy. Astron.
  Soc.} {\bfseries 498} (2020) 2492}
  [\href{https://arxiv.org/abs/2007.08994}{{\ttfamily 2007.08994}}].

\bibitem{Tamone:2020qrl}
A.~Tamone et~al., \emph{{The Completed SDSS-IV extended Baryon Oscillation
  Spectroscopic Survey: Growth rate of structure measurement from anisotropic
  clustering analysis in configuration space between redshift 0.6 and 1.1 for
  the Emission Line Galaxy sample}},
  \href{https://doi.org/10.1093/mnras/staa3050}{\emph{Mon. Not. Roy. Astron.
  Soc.} {\bfseries 499} (2020) 5527}
  [\href{https://arxiv.org/abs/2007.09009}{{\ttfamily 2007.09009}}].

\bibitem{deMattia:2020fkb}
A.~de~Mattia et~al., \emph{{The Completed SDSS-IV extended Baryon Oscillation
  Spectroscopic Survey: measurement of the BAO and growth rate of structure of
  the emission line galaxy sample from the anisotropic power spectrum between
  redshift 0.6 and 1.1}},
  \href{https://doi.org/10.1093/mnras/staa3891}{\emph{Mon. Not. Roy. Astron.
  Soc.} {\bfseries 501} (2021) 5616}
  [\href{https://arxiv.org/abs/2007.09008}{{\ttfamily 2007.09008}}].

\bibitem{Neveux:2020voa}
R.~Neveux et~al., \emph{{The completed SDSS-IV extended Baryon Oscillation
  Spectroscopic Survey: BAO and RSD measurements from the anisotropic power
  spectrum of the quasar sample between redshift 0.8 and 2.2}},
  \href{https://doi.org/10.1093/mnras/staa2780}{\emph{Mon. Not. Roy. Astron.
  Soc.} {\bfseries 499} (2020) 210}
  [\href{https://arxiv.org/abs/2007.08999}{{\ttfamily 2007.08999}}].

\bibitem{Hou:2020rse}
J.~Hou et~al., \emph{{The Completed SDSS-IV extended Baryon Oscillation
  Spectroscopic Survey: BAO and RSD measurements from anisotropic clustering
  analysis of the Quasar Sample in configuration space between redshift 0.8 and
  2.2}}, \href{https://doi.org/10.1093/mnras/staa3234}{\emph{Mon. Not. Roy.
  Astron. Soc.} {\bfseries 500} (2020) 1201}
  [\href{https://arxiv.org/abs/2007.08998}{{\ttfamily 2007.08998}}].

\bibitem{Agathe:2019vsu}
V.~de~Sainte~Agathe et~al., \emph{{Baryon acoustic oscillations at z = 2.34
  from the correlations of Ly$\alpha$ absorption in eBOSS DR14}},
  \href{https://doi.org/10.1051/0004-6361/201935638}{\emph{Astron. Astrophys.}
  {\bfseries 629} (2019) A85}
  [\href{https://arxiv.org/abs/1904.03400}{{\ttfamily 1904.03400}}].

\bibitem{Blomqvist:2019rah}
M.~Blomqvist et~al., \emph{{Baryon acoustic oscillations from the
  cross-correlation of Ly$\alpha$ absorption and quasars in eBOSS DR14}},
  \href{https://doi.org/10.1051/0004-6361/201935641}{\emph{Astron. Astrophys.}
  {\bfseries 629} (2019) A86}
  [\href{https://arxiv.org/abs/1904.03430}{{\ttfamily 1904.03430}}].

\bibitem{Banerjee:2020xcn}
A.~Banerjee, H.~Cai, L.~Heisenberg, E.~O. Colg\'ain, M.~M. Sheikh-Jabbari and
  T.~Yang, \emph{{Hubble sinks in the low-redshift swampland}},
  \href{https://doi.org/10.1103/PhysRevD.103.L081305}{\emph{Phys. Rev. D}
  {\bfseries 103} (2021) L081305}
  [\href{https://arxiv.org/abs/2006.00244}{{\ttfamily 2006.00244}}].

\bibitem{BenAchour:2016fzp}
J.~Ben~Achour, M.~Crisostomi, K.~Koyama, D.~Langlois, K.~Noui and G.~Tasinato,
  \emph{{Degenerate higher order scalar-tensor theories beyond Horndeski up to
  cubic order}}, \href{https://doi.org/10.1007/JHEP12(2016)100}{\emph{JHEP}
  {\bfseries 12} (2016) 100}
  [\href{https://arxiv.org/abs/1608.08135}{{\ttfamily 1608.08135}}].

\bibitem{DeFelice:2016yws}
A.~De~Felice, L.~Heisenberg, R.~Kase, S.~Mukohyama, S.~Tsujikawa and Y.-l.
  Zhang, \emph{{Cosmology in generalized Proca theories}},
  \href{https://doi.org/10.1088/1475-7516/2016/06/048}{\emph{JCAP} {\bfseries
  06} (2016) 048} [\href{https://arxiv.org/abs/1603.05806}{{\ttfamily
  1603.05806}}].

\bibitem{DeFelice:2016uil}
A.~De~Felice, L.~Heisenberg, R.~Kase, S.~Mukohyama, S.~Tsujikawa and Y.-l.
  Zhang, \emph{{Effective gravitational couplings for cosmological
  perturbations in generalized Proca theories}},
  \href{https://doi.org/10.1103/PhysRevD.94.044024}{\emph{Phys. Rev. D}
  {\bfseries 94} (2016) 044024}
  [\href{https://arxiv.org/abs/1605.05066}{{\ttfamily 1605.05066}}].

\end{thebibliography}

\providecommand{\href}[2]{#2}\begingroup\raggedright\endgroup

\end{document}